\documentclass{emulateapj}
\usepackage{amssymb,amsmath}
\def\bs{\boldsymbol}

\def\gsim{\;\rlap{\lower 2.5pt
\hbox{$\sim$}}\raise 1.5pt\hbox{$>$}\;}
\def\lsim{\;\rlap{\lower 2.5pt
\hbox{$\sim$}}\raise 1.5pt\hbox{$<$}\;}
\makeatletter
\newcommand{\vast}{\bBigg@{3}}
\newcommand{\Vast}{\bBigg@{5}}
\makeatother

\newcommand{\Rmnum}[1]{\expandafter\@slowromancap\romannumeral #1@}

\begin{document}

\title{Turbulence-Induced Relative Velocity of Dust Particles II: The Bidisperse Case}

\author{Liubin Pan\altaffilmark{1}, Paolo Padoan\altaffilmark{2}, \& John Scalo\altaffilmark{3}}
\altaffiltext{1}{Harvard-Smithsonian Center for Astrophysics,
60 Garden St., Cambridge, MA 02138;  {\tt lpan@cfa.harvard.edu}}
\altaffiltext{2} {ICREA \& ICC, University of Barcelona, Marti i Franqu\`{e}s 1, E-08028 Barcelona, Spain; \tt{ppadoan@icc.ub.edu}}
\altaffiltext{3} {Department of Astronomy, University of Texas, Austin, TX 78712; \tt{parrot@astro.as.utexas.edu}}

\begin{abstract}

We extend our earlier work on turbulence-induced relative velocity between equal-size particles (Pan \& Padoan, Paper I) to particles of arbitrarily different sizes. The Pan \& Padoan (PP10) model shows that the relative velocity between different particles has two contributions, named the generalized shear and acceleration terms, respectively. The generalized shear term represents the particlesÕ memory of the spatial flow velocity difference across the particle distance in the past, while the acceleration term is associated with the temporal flow velocity difference on individual particle trajectories. Using the simulation of Paper I, we compute the root-mean-square relative velocity, $\langle w^2 \rangle^{1/2}$, as a function of the friction times, $\tau_{\rm p1}$ and $\tau_{\rm p2}$, of the two particles, and show that the PP10 prediction is in satisfactory agreement with the data, confirming its physical picture. For a given $\tau_{\rm p1}$ below the Lagrangian correlation time of the flow, $T_{\rm L}$, $\langle w^2 \rangle^{1/2}$ as a function of $\tau_{\rm p2}$ shows a dip at $\tau_{\rm p2} \simeq  \tau_{\rm p1}$, indicating tighter velocity correlation between similar particles. Defining a ratio $f\equiv \tau_{\rm p,l}/\tau_{\rm p,h}$, with  $\tau_{\rm p,l}$ and $\tau_{\rm p,h}$ the friction times of the smaller and larger particles, we find that $\langle w^2 \rangle^{1/2}$ increases with decreasing $f$ due to the  generalized acceleration contribution, which dominates 
at $f\lsim 1/4$. At a fixed $f$, our model predicts  that $\langle w^2 \rangle^{1/2}$  scales as $\tau_{\rm p,h}^{1/2}$ for $\tau_{\rm p,h}$ in the inertial range of the flow, stays roughly constant for $T_{\rm L} \lsim \tau_{\rm p,h}  \lsim T_{\rm L}/f$, and finally decreases as $\tau_{\rm p,h}^{-1/2}$ for $\tau_{\rm p,h} \gg T_{\rm L}/f$. 
The acceleration term is independent of the particle distance, $r$, and reduces the $r-$dependence of $\langle w^2 \rangle^{1/2}$ in the bidisperse case.

\end{abstract}

\section{Introduction}

This paper is a follow-up to our earlier work on turbulence-induced relative velocity of 
dust particles (Pan \& Padoan 2013; Paper I hereafter). The study is mainly motivated by the 
problem of dust particle growth and planetestimal formation in protoplanetary disks (e.g., 
Dullemond and Dominik 2005; Zsom et al.\ 2010, 2011; Birnstiel et al.\  2011; 
Windmark et al.\ 2012; Garaud et al.\ 2013). In Paper I, we conducted an extensive statistical study 
of the relative velocity and the collision kernel of equal-size particles suspended in turbulent flows 
using both analytical and numerical methods. 
%In Paper I, we presented a detailed statistical 
%analysis of the relative velocity 
%including its probability distribution, 
%and explored the collision kernel as a function of the particle inertia in 
The case of equal-size particles, usually referred to as the monodisperse case, is  
of theoretical interest, but insufficient for astrophysical 
applications, as dust particles in protoplanetary disks have a size distribution. 
%Even if 
%all dust particles started from the same initial size, a size distribution would 
%have developed from collisional coagulation or fragmentation. 
%Therefore, modeling dust collisions requires the collision velocity 
%between different particles. 
The main goal of the current paper is to investigate turbulence-induced relative velocity in 
the general case of different particles of arbitrary sizes, known as the bidisperse case. 
%We refer the reader to Paper I for a more detailed discussion for the motivation of our study.  

Saffman and Turner (1956, hereafter S-T) derived a formula for the variance 
of the turbulence-induced relative velocity in the limit of small particles 
with friction time, $\tau_{\rm p}$, much smaller than the Kolmogorov timescale, 
$\tau_\eta$. This limit, known as the S-T limit, is usually expressed as 
$St \ll 1$, where the Stokes number, $St$, is defined as $St \equiv \tau_{\rm p}/\tau_{\rm \eta}$. 
The Saffman-Turner formula consists of two terms, named 
the shear and the acceleration term, respectively (e.g., Zhou et al.\ 2001). 
%In the S-T prediction, 
The shear term is determined solely by the flow velocity difference across the particle 
distance $r$. It is independent of $St$, but has a significant dependence on $r$ 
(see Paper I).
%the relative velocity variance scales as $r^2$ for $r$ below the Kolmogorov 
%length scale, $\eta$. 
The name of the acceleration term originates from its dependence 
on the acceleration, ${\bs a}$, of the flow velocity, and it contributes 
a 1D variance of $a^2 (\tau_{\rm p2} -\tau_{\rm p1})^2$ to the 
relative velocity, where $\tau_{\rm p1}$  and $\tau_{\rm p2}$ are the friction 
times of the two particles, and $a$ is the 1D rms of the flow acceleration. 
This effect of the flow acceleration on the relative velocity of small particles of different 
sizes was also found by Weidenschilling (1984).  
%$\propto \frac{\bar{\epsilon}}{\nu} r^2$ where $\bar{\epsilon}$ and ${\nu}$ 
%and the average dissipation rate and the kinematic viscosity of the flow 
In the monodisperse case, the acceleration term vanishes and 
only the shear term contributes. The shear term in the S-T prediction 
for equal-size particles has been discussed in detail in Paper I. 
Its validity, accuracy and  limitations have been 
systematically examined using a numerical simulation. 
%We refer the reader to Paper I for 

%the relative velocity at a small distance $r$. 
In the bidisperse case, the fundamental difference from the case of equal-size particles  
is the contribution of the acceleration term, which
tends to increase the particle collision velocity. The dependence of the 
acceleration term on the friction time difference, $\tau_{\rm p2} -\tau_{\rm p1}$, 
corresponds to the fact that particles of different sizes 
have different responses  to the flow velocities along 
their trajectories. Interestingly, unlike the shear term, the acceleration contribution in the 
S-T formula is independent of the particle distance, $r$. This observation is of particular 
importance for the application to dust particles in protoplanetary disks.  
Because the size of dust particles is typically much smaller than the Kolmogorov 
scale, $\simeq 1$ km, of protoplanetary turbulence, 
%and they should be viewed as nearly point particles (e.g., Hubbard 2012). 
one is  required to examine the collisional statistics in 
the $r\to0$ limit (e.g., Hubbard 2012) that is not accessible to 
numerical simulations due to their limited resolution.
Therefore, unless the measured statistics already 
converges at the resolution scale, an extrapolation to the $r\to 0$ limit is 
needed. 
%before applying it to dust particle collisions. 
Such an extrapolation was found to be challenging 
for small equal-size particles with $St\lsim 1$ due to the $r-$dependence 
of the shear effect\footnote{To evaluate the collision kernel of small equal-size particles 
in the $r\to 0$ limit, a method is developed in Paper I to isolate an $r-$independent contribution  
by splitting particle pairs at given distances into two types, named continuous 
(S-T) pairs and caustic (sling) pairs, respectively (Falkovich et al.\ 2002, Wilkson 
et al.\ 2006).}(Paper I).  
%As will be shown in this paper, 
In the bidisperse case, the presence of the %$r-$independent 
acceleration term reduces the $r-$dependence 
and makes it easier to achieve numerical convergence  for the relative 
velocity between small particles of different sizes.  
%in simulations. 
%makes it easier to the extrapolate 

Pan and Padoan (2010, PP10) developed a model for the rms relative 
velocity in the general bidisperse case, for arbitrarily 
different particles of any size. It was shown that the model prediction is in 
good agreement with the simulation data 
of Zhou et al.\ (2001) at low resolutions. The PP10 formulation for the relative 
velocity also consists of two contributions, named as the generalized 
shear and acceleration terms, as they reduce, respectively, to the shear and acceleration terms in the S-T 
formula in the small particle limit. It can thus be viewed as a generalized formulation that extends the S-T limit ($St_{1,2} \ll 1$) 
to particles of arbitrary sizes. The generalized shear term has
a similar form as the monodisperse model discussed in Paper I. 
It represents the particles' memory of the {\it spatial} flow velocity difference, $\Delta {\bs u} (R)$,  
across the separation, $R$, of the two particles at 
given times in the past.  The physical meaning 
of the generalized acceleration term will be clarified in the present paper, and we will 
show its connection with the {\it temporal} flow velocity difference, $\Delta_{\rm T} {\bs u}$, 
along individual trajectories of the two particles. An approximate relation for the 
acceleration term will be established in terms of  $\Delta_{\rm T} {\bs u}$ and the particle friction times. 
Using the simulation of Paper I,
%at a moderate resolution, 
we will systematically test the PP10 model for the relative velocity between different particles.      

%A variety of theoretical models were constructed for relative velocity variances in 
%the general bidisperse case with $\tau_{\rm p1,2}$ covering the entire scale range of the flow. 

A variety of models have been developed to predict the relative velocity of particles of different 
sizes, covering the entire scale range of the turbulent flow (e.g., V\"olk et al.\ 1980; Yuu 1984; Kruis \& Kusters,1997; Zhou et al.\ 2001; Zaichik et al.\ 2006, 2008; 
see PP10 and references therein). In the astrophysics literature, the model of choice has been
that by V\"olk et al.\ (1980) and its later refinements (e.g., Markiewicz, Mizuno \& V\"olk 1991, Cuzzi \& Hogan 2003, and Ormel \& Cuzzi 2007). 
%As discussed in PP10 and Paper I, the V\"olk et al.\ model has serious physical weaknesses and overestimates the 
%relative velocity of equal-size particles. To our knowledge, the accuracy of 
%the V\"olk et al.\ model  has not been carefully tested in the bidisperse case.  
In this paper, we only test the PP10 model, in an effort of providing an improved physical insight.
Other models, particularly V\"olk et et al.\ (1980) and Zaichik et al.\ (2008), will be tested and compared with PP10 in a separate work. 

Theoretical models only predict the rms or variance of the relative velocity, which, 
%As the simplest statistical measure, the rms   
however, is not sufficient to model 
collisions of dust particles (Paper I). In fact, the rms does not directly enter the 
estimate of the collision kernel, which is determined by the first-order moments, i.e., 
the average of the radial component or the mean 3D amplitude of the relative velocity (Wang et al.\ 2000).  
%$\langle |w_{\rm r}| \rangle$ (or $\langle |{\bs w}| \rangle$) rather than the rms, $\langle w_{\rm r}^2\rangle^{1/2}$. 
The variance of the relative velocity does not even
represent the average collision energy per collision. Instead, using a collision-rate 
weighting, the average collision energy depends on the third-order moment of the collision velocity (e.g., Hubbard 2012).  
Furthermore, an accurate coagulation model for  dust particles 
in protoplanetary disks requires the entire probability distribution of the collision velocity, as 
the outcome of each collision depends on the collision velocity (Windmark et al.\ 2012; Garaud et al.\ 2013).  
%We defer a systematic study of the PDF to a later work. 
Despite these limitations, the rms relative velocity still provides a rough approximation to the mean of the relative velocity, 
and it is therefore a useful tool to shed light on the physics of turbulence-induced particle collisions.  
The main purpose of the current work is to confirm the accuracy of  the PP10 model for the rms relative velocity, and hence 
to validate the physical picture revealed by that model. We will show in a separate paper that this physical picture provides
an understanding of the probability distribution of the collision velocity as well.

In addition to turbulent motions,  there are other effects, such as differential settling or radial drift, 
that can provide important  contributions to  the relative velocity between dust particles of different sizes 
in protoplanetary disks. In this work, we do not consider these contributions.
The numerical experiment used here employs a statistically stationary and isotropic 
turbulent flow, which is a further idealization relative to realistic protoplanetary disks 
with Keplerian rotation, stratifications, etc. However, the highly idealized simulation provides a useful tool 
to study the role of turbulence-induced collisions.

%The large-scale anisotropy may leave an imprint on the relative speed 
%of large particles whose memory covers the length scales at which the rotation effect is significant. 

In \S 2, we present a simple model for the relative velocity between inertial particles 
and the local flow velocity,  a special bidisperse case that 
provides a clean comparison between our model and the 
simulation. The PP10 formulation for the bidisperse relative velocity is reviewed in \S 3.  
A brief presentation of our numerical simulation is given in \S 4. In \S 5, we examine the statistics 
of the particle-flow relative velocity.  In \S 6, we show simulation results for the rms relative velocity, and 
test the prediction of the PP10 model. 
%The relative velocity PDF as a 
%function of the friction time or Stokes number pair is systematically explored and interpreted in 
%\S7. 
We summarize the main results and conclusions in \S 7. 
%For small particles,  we expect our simulations and theoretical model to apply for the 
%turbulence-induced collision speed of small dust particles in protoplanetary disks, because 
%the statistical isotropy is likely to be restored the scale of the particle size, which is below the 
%Kolmogorov scale of the turbulent flow, Finally, the idealized sitation isolates various 
%complexities in a protoplanetary disk, and  is  thus a useful tool to reveal the fundamental 
%physics of turbulence-induced relative speed of inertial particles.   

\section{The Particle-flow Relative Velocity}

We first consider a simple model for the relative velocity between an 
inertial particle and the local flow element. This is a special bidisperse 
case where one of the particle is a tracer with zero inertia. 
It provides a useful illustration for the general bidisperse case. 
The particle-flow relative velocity is also of interest for practical application, e.g., in the 
formation of  fine dust  rims  of chondrules via an accretion process 
(e.g., Paque \& Cuzzi 1997, Morfill et al.\ 1998, Cuzzi \& Hogan 2003, Ormel et al.\ 2008, Carballido 2011). 
The velocity, ${\bs v}(t)$, 
of an inertial particle with a friction timescale, $\tau_{\rm p}$, evolving in 
a turbulent velocity field, ${\bs u} \left( {\bs x}, t \right)$, is governed by the equation, 
\begin{equation}
\frac {d {\bs v} } {dt} = \frac { {\bs u} \left( {\bs X} (t), t \right) - {\bs v}} {\tau_{\rm p}}       
\label{particlemomentum}     
\end{equation}
where ${\bs X} (t)$ is the particle position, and ${\bs u} \left( {\bs X} (t), t \right)$ is
the flow velocity ``seen"  by the particle.  Eq.\  (\ref{particlemomentum}) is a 
stochastic differential equation in a similar form as the Langevin equation, 
with the flow velocity acting as an random force. However, it differs from the 
Langevin equation in that the correlation time of the ``force" is 
significant in comparison with the friction time, $\tau_{\rm p}$.  
Eq.\ (\ref{particlemomentum}) can be formally integrated, as if it were a 
deterministic equation. As shown in Paper I, the particle velocity at any 
given time, say $t=0$, can be evaluated as, 
\begin{equation}
{\bs v}(0) = \frac {1} {\tau_{\rm p} }
\int_{t_0}^0{\bs u} \left({\bs X} (\tau), \tau \right) \exp \left(\frac{\tau}{\tau_{\rm p}}\right) d\tau
\label{formalsolution}           
\end{equation}
where it is assumed that at $t=0$ the particle has already forgot its inertial velocity at $t_{0}$.  This is equivalent to assuming $t_0 \ll -\tau_{\rm p}$, 
which also allows to replace the lower limit  $t_{0}$ by $-\infty$ (Paper I).  

At time $t=0$, we define a flow-particle relative velocity at ${\bs x}$ as 
${\bs w}_{\rm f} ({\bs x}, 0) \equiv {\bs u}({\bs x}, 0) -  {\bs v}(0)$. 
Using the formal solution, we have,
\begin{equation}
{\bs w}_{\rm f}=  \frac {1} {\tau_{\rm p} }
\int_{-\infty}^0 \left[ {\bs u}({\bs x}, 0)  -  {\bs u} \left({\bs X} (\tau), \tau \right)\right]\exp \left( \frac{\tau}{\tau_{\rm p}}\right) d\tau
\label{pfrelative}           
\end{equation}
where ${\bs X} (\tau) $ satisfies the condition ${\bs X} (0) = \bs{x}$. 
The equation suggests that the particle-flow relative velocity depends on
the difference between the local flow velocity and the velocity 
the particle saw within a friction timescale in the past. If we 
define a flow velocity difference, $\Delta_{\rm T} {\bs u} (\Delta \tau)$, at a time lag,  $\Delta \tau$, along 
the particle trajectory,   ${\bs w}_{\rm f}$ can be roughly estimated 
as ${\bs w}_{\rm f} \simeq  \Delta_{\rm T} {\bs u} (\tau_{\rm p})$.

The variance of ${\bs w}_{\rm f}$ 
can be calculated as, 
%\begin{equation}
%\langle  w^{\rm f}_i w^{\rm f}_j  \rangle = \int_{-\infty}^0 \frac {d\tau } {\tau_{\rm p} } \int_{-\infty}^0 \frac {d\tau' } {\tau_{\rm p} } 
%S^{\rm f}_{ij} (\tau, \tau')\exp \left( \frac{\tau}{\tau_{\rm p}}\right) \exp \left( \frac{\tau'}{\tau_{\rm p}}\right),
%\label{pfrelative}           
%\end{equation}
%where the tensor $S^{\rm f}_{ij}$ is defined as, 
%the temporal structure tensor with respect to the flow velocity at time 0 along the trajectory of one particle in 
%the past, 
%\begin{equation}
%\langle  w_{{\rm f}i} w_{{\rm f}j}  \rangle = \int_{-\infty}^0 \frac {d\tau } {\tau_{\rm p} } \int_{-\infty}^0 \frac {d\tau' } {\tau_{\rm p} } 
%\left[B_{ij} (0, 0)  - B_{ij} (0, \tau') - B_{ij} ( \tau, 0) + B_{ij} (\tau, \tau')\right] \exp \left( \frac{\tau}{\tau_{\rm p}}\right) \exp \left( \frac{\tau'}{\tau_{\rm p}}\right),
%\label{pfrelative}           
%\end{equation}
\begin{gather}
\langle  w_{{\rm f}i} w_{{\rm f}j}  \rangle = \int_{-\infty}^0 \frac {d\tau } {\tau_{\rm p} } \int_{-\infty}^0 \frac {d\tau' } {\tau_{\rm p} } 
\Bigg[B_{ij} (0, 0)  - B_{ij} (0, \tau') \notag \\
- B_{ij} ( \tau, 0) + B_{ij} (\tau, \tau')\Bigg] \exp \left( \frac{\tau}{\tau_{\rm p}}\right) \exp \left( \frac{\tau'}{\tau_{\rm p}}\right), \hspace{-0.5cm}  
\label{pfrelative}           
\end{gather}
%\begin{equation}    
%D^{\rm f}_{ij} (\tau, \tau') = \big \langle \big[ u_i({\bs x}, 0)  -   u_i \left({\bs X} (\tau), \tau \right) \big] \left[ u_j({\bs x}, 0)  -   u_j \left({\bs X} (\tau'), \tau' \right)\right] \big \rangle.
%\label{sfij}
%\end{equation}
where the trajectory correlation tensor, $B_{ij} (\tau, \tau')  \equiv  \left\langle  u_i \left({\bs X} (\tau), \tau \right) u_j \left({\bs X} (\tau'), \tau' \right) \right \rangle$, 
corresponds to a two-time correlation of the flow velocity along the particle trajectory (Paper I).  %Clearly, 
%$S^{\rm f}_{ij} (\tau, \tau') = B_{{\rm T}ij} (0, 0)  - B_{{\rm T}ij} (0, \tau') - B_{{\rm T}ij} ( \tau, 0) + B_{{\rm T}ij} (\tau, \tau')$. 
%We insert the four terms into eq.\  (\ref{pfrelative}).  
%The integral for the first three terms in the square bracket are simple, 
%and 
The integral of the fourth term, $B_{ij} (\tau, \tau')$,
 %\exp[(\tau+\tau')/\tau_{\rm p}]$ 
can be simplified using the fact that $B_{ij} (\tau, \tau')$ is an even function 
of the time lag, $\Delta \tau = \tau'-\tau$, i.e., $B_{ij} (\tau, \tau') = B_{ij} (|\Delta \tau|)$, 
in statistically stationary turbulence. Making a variable change, $\xi = \tau+\tau'$ and $\zeta =\Delta \tau= \tau' -\tau$, 
for this term,  we find,
\begin{gather}
\langle  w_{{\rm f}i} w_{{\rm f}j}  \rangle = \frac {1} {\tau_{\rm p} } \int_{-\infty}^0  [ B_{ij}(0)  - B_{ij} (\Delta \tau)] \exp \left( \frac{\Delta \tau}{\tau_{\rm p}}\right)  d{\Delta \tau } \notag \\=  \frac {1} {2\tau_{\rm p} } \int_{-\infty}^0  D_{ij} (\Delta \tau) \exp \left( \frac{\Delta \tau}{\tau_{\rm p}}\right)  {d\Delta \tau }, \hspace{.9cm}  
\label{pfrelative2}           
\end{gather}
where $ D_{ij} (\Delta \tau) \equiv \big \langle  \Delta_{\rm T}  u_i (\Delta\tau)\Delta_{\rm T}  u_j (\Delta\tau)\big \rangle$ 
%\big \langle \big[ u_i({\bs x}, 0)  -   u_i \left({\bs X} (\Delta \tau), \Delta \tau \right) \big] \big[ u_j({\bs x}, 0)  -   u_j \left({\bs X} (\Delta \tau), \Delta \tau \right) \big] \big \rangle$ 
is the temporal structure function of the flow velocity along the particle trajectory. %We refer to  $D_{ij} (\Delta \tau)$ as 
%the trajectory temporal structure tensor. 

For isotropic turbulence, $B_{ij} (\Delta \tau) = u'^2 \Phi_1(\Delta\tau) \delta_{ij} $, 
where $u'$ is the 1D rms velocity of the turbulent flow, and  $\Phi_1$ is the trajectory correlation 
function. It follows that $ D_{ij} (\Delta \tau)=2 u'^2\delta_{ij} (1-\Phi_1(\Delta \tau))$.
%$S^{\rm f}_{ij} (\tau, \tau') = u'^2 [\Phi_1(0)  -  \Phi_1(\tau') - \Phi_1(-\tau) + \Phi_1(\tau'-\tau)] \delta_{ij}$. 
%Inserting this into eq.\ (\ref{pfrelative}) and using the fact that $\Phi_1$  only depends on the time lag, 
%$\Delta \tau = \tau'-\tau$, one can make a variable change that allows an analytical integration 
%of one of the double integrals.  
A common assumption for $\Phi_1$ is to approximate it by the Lagrangian temporal correlation function $\Phi_{\rm L}$ (see e.g., 
Zaichick et al.\ 2006, 2008, Derivich 2006). This is equivalent to approximating $D_{ij} (\Delta \tau)$ in eq.\ (\ref{pfrelative2}) by the Lagrangian 
structure tensor, $D_{{\rm L}ij} (\Delta \tau) = D_{\rm L}(\Delta \tau)\delta_{ij}$, with $D_{\rm L} = 2u'^2 (1-\Phi_{\rm L}(\Delta \tau))$. 
For very large particles, it may be a better assumption to set $D_{ij} (\Delta \tau)$ to the Eulerian 
temporal structure tensor $D_{{\rm E}ij} (\Delta \tau)$ (see the definition in Appendix C). %Due to the relevance of the Lagrangian structure tensor, 
%We analyze $D_{{\rm L, E}ij}$ in our simulated flow in Appendix C. 
%where we provide a justification for the assumption 
%$D_{ij} \simeq D_{{\rm L}ij}$ for all particles.

%We will denote the normalized Lagrangian structure function as ${\widetilde S}_{\rm L} = S_{\rm L} (\tau)/2u'^2 = 1- \Phi_{\rm L}(\tau)$.
In Paper I, we considered single- and bi- exponential forms for $\Phi_{\rm L}$.  
The exponential form is $\Phi_{\rm L}(\Delta \tau) = \exp(-|\Delta \tau|/T_{\rm L})$, 
where $T_{\rm L}$ is the Lagrangian correlation timescale. 
Setting $\Phi_1 (\Delta \tau) =  \exp(-|\Delta \tau|/T_{\rm L})$ 
and integrating eq.\ (\ref{pfrelative2}), we obtain $\langle  w_{{\rm f}i} w_{{\rm f}j}  \rangle = (w^{\prime}_{\rm f})^2 \delta_{ij}$, 
where the 1D rms,  $w^{\prime}_{ \rm f}$, of the flow-particle relative velocity is given by,   
\begin{equation}
w^{\prime}_{\rm f} = u'\left( \frac{\tau_{\rm p}}{ \tau_{\rm p} + T_{\rm L} } \right)^{1/2},
\label{pfrelativesingle}
\end{equation}
which indicates a $\tau_{\rm p}^{1/2}$ scaling for particles with $\tau_{\rm p} \ll T_{\rm L}$. 
In the large-particle limit ($\tau_{\rm p} \gg T_{\rm L}$), $w_{\rm f}^{\prime}$ approaches  $u'$, as the particle velocity becomes much 
smaller than the flow velocity. 

As discussed in Paper I,  $\Phi_{\rm L} (\Delta \tau)$ is better fit by a bi-exponential form (see Fig.\  2 of Paper I),  
%is preferred as it provides an excellent fit tomeasured from 
%\begin{gather}
%\Phi_{\rm L}(\Delta \tau)= \frac{1}{2 \sqrt {1-2z^2} } \Bigg\{ \big(1 + \sqrt{1-2z^2}\big) \exp \Bigg[-\frac{2 |\Delta \tau|}{ \big(1+ \sqrt{1-2z^2}\big) T_{\rm L} }  \Bigg]  \notag\\ %%\vspace{7mm} 
%\hspace{4.1cm}  
%-\big(1-\sqrt{1-2z^2}\big) \exp \Bigg[ - \frac{2 |\Delta \tau|}{ \big(1 - \sqrt{1-2z^2} \big) T_{\rm L}}  \Bigg] \Bigg\},  
%\label{biexponential}
%\end{gather}
\begin{gather}
\Phi_{\rm L}(\Delta \tau)= {\frac{1}{2 \sqrt {1-2z^2} }}
\vast\{ \big(1 + \sqrt{1-2z^2}\big)  \times \hspace{5cm}\notag\\
\hspace{1cm} \exp \Bigg[-\frac{2 |\Delta \tau|}{ \big(1+ \sqrt{1-2z^2}\big) T_{\rm L} }  \Bigg] -  
\big(1-\sqrt{1-2z^2}\big) \times \notag\\ 
 \exp \left [ {- \frac{2 |\Delta \tau|}{ \big(1 - \sqrt{1-2z^2} \big) T_{\rm L}} } \right] \vast\}, \hspace{1.8cm}  
\label{biexponential}
\end{gather}
where the parameter $z$ ($\equiv \tau_{\rm T}/T_{\rm L}$) is the ratio of  the Taylor micro 
timescale, $\tau_{\rm T}$,  to $T_{\rm L}$. The Taylor (Lagrangian) timescale is defined as 
$\tau_{\rm T} =(2u'^2/a^2)^{1/2}$ with $a$ the 1D rms of the acceleration field, ${\bs a}$, 
of the flow. The theoretical motivation of adopting a bi-exponential function is that, by accounting for the flow 
acceleration, it correctly reflects the smoothness of $\Phi_{\rm L} (\Delta \tau)$ at small $\Delta \tau$ ($\lsim \tau_{\rm T}$).  
For $\Delta \tau \ll  \tau_{\rm T}$, the bi-exponential function ensures that the temporal velocity difference along a Lagrangian trajectory 
scales linearly with $\Delta \tau$, and thus correctly describes the dissipation range in the Lagrangian frame 
(Zaichik  et al.\ 2006). In other words, the  bi-exponential function reflects the transition from  the 
dissipation-range scaling to the inertial-range scaling in the Lagrangian frame. On the other hand, the single-exponential 
ignores the existence of the dissipation range, and is thus physical inadequate.  
 The parameter $z$ in eq.\ (\ref{biexponential}) is essentially a measure of the width of the inertial range in the Lagrangian 
 coordinate.  Approximately, $z^2 \sim \tau_\eta/T_{\rm L}$ with $\tau_\eta$ the Kolmogorov time, 
 and it is  thus an indicator of the timescale separation between the dissipation and the driving scales. 
 Clearly, $z$ decreases with the flow Reynolds number ($Re$), and  it  roughly scales with $Re$ as $z \propto Re^{-1/4}$. 
%or $\propto Re_{\lambda}^{-1/2}$, where $Re_{\lambda}$ is the Taylor Reynolds number. 
%For convenience, 
%here we give the bi-exponential form again.
With eq.\ (\ref{biexponential}) for $\Phi_{\rm L}$, we find, 
\begin{equation}
w^{\prime}_{ \rm f} = u'\left( \frac{\Omega^2} { \Omega +\Omega^2 + z^2/2}  \right)^{1/2},
\label{pfrelativebi}
\end{equation}
where $\Omega \equiv \tau_{\rm p}/T_{\rm L}$. %The behavior of eq.\ (\ref{pfrelativebi}) is the same 
%as eq.\ (\ref{pfrelativesingle}) with the single-exponential correlation for $\Omega \gg z^2/2$, or equivalently for 
%$\tau_{\rm p} \gg \tau_{\rm T}^2/2T_{\rm L} \simeq \tau_\eta$.  
Eq.\ (\ref{pfrelativebi}) shows that  $w^{\prime}_{\rm f}$ approaches $u'$ for $\Omega \gg 1$, and scales 
with $\tau_{\rm p}$ as $\tau_{\rm p}^{1/2}$, for 
intermediate particles with $z^2/2 \ll \Omega \ll 1$. For small particles with $\Omega \ll z^2/2$ (i.e., 
$\tau_{\rm p} \ll \tau_{\rm T}^2/2T_{\rm L} \simeq \tau_\eta$), 
%$\tau_{\rm p} \ll \tau_\eta$ (so that $\Omega \ll z^2/2$), 
eq.\ (\ref{pfrelativebi}) predicts $w^{\prime}_{ \rm f} = \sqrt{2}u' \Omega/z \equiv a \tau_{\rm p}$ (see also Weidenschilling 1984). This is expected  
from eq.\ (\ref{particlemomentum}), assuming $d{\bs v}/dt \simeq d{\bs u}/dt = {\bs a}$. 
The linear scaling of  $w_{\rm f}^{\prime}$ with $\tau_{\rm p}$ for  $St \ll 1$ particles cannot be correctly captured by 
eq.\ (\ref{pfrelativesingle}) that uses a single-exponential form for $\Phi_1$ or $\Phi_{\rm L}$.  We will compare the prediction, eq.\ (\ref{pfrelativebi}), against
our simulation data in \S 5. 

%We note that the flow-particle relative velocity is related to the particle acceleration, ${\bs a}_{\rm p}$($\equiv  d{\bs v}/dt$), as ${\bs w}_{\rm f} = \tau_{\rm p} {\bs a}_{\rm p}$ (see eq. (\ref{particlemomentum})). Therefore, the statistics of ${\bs w}_{\rm f}$ is essentially equivalent to that of ${\bs a}_{\rm p}$. The 
%latter has been studied both numerically and experimentally (Ayyalasomayajula et al.\ 2006, Bec et al.\ 2006, Salazar \& Collins 2012). 

\section{ The Particle Relative Velocity in the Bidisperse Case}

%The main goal of the current work is to understand and evaluate the velocity at which two nearby 
%particles collide. The collision speed of two particles is essentially their relative 
%velocity as they come cross each other, i.e., as their separation, $r$, becomes equal to the sum 
%of the particle radii (Saffman and Turner 1956). For applications %, such as rain droplets in atmospheric clouds or 
%to dust particles in protoplanetary disks, the particle size is typically much smaller than 
%the Kolmorgorov length scale, $\eta$, corresponding to the size of the smallest  eddies in  the turbulent 
%flow. The Kolmorgorov scale is defined as $\eta = (\nu^3/\bar{\epsilon})^{1/4}$, where $\nu$ and $\bar{\epsilon}$ 
%are the kinematic viscosity and the average energy dissipation rate in the turbulent flow.  
%We will explore the relative speed of nearby particles as a function of  the separation for $r \lsim \eta$, 
%as the particle collisions occur at scales below $ \eta$. 
Following PP10 and Paper I, we use superscripts $(1)$ and $(2)$ to label 
two particles coming together. In the bidisperse case, we denote the friction 
times of the two particles as $\tau_{\rm p1}$ and $\tau_{\rm p2}$. %respectively. 
At a given time $t$, we examine the relative velocity, ${\bs w} \equiv {\bs v}^{(1)}(t) - {\bs v}^{(2)}(t)$, 
of particle pairs at given separations, ${\bs r}$. The particle positions at $t$ are
constrained by ${\bs X}^{(1)} (t)- {\bs X}^{(2)}(t) = {\bs r}$.  
We denote the flow velocities seen by the two particles as ${\bs u}^{(1)}(t)$ ($\equiv {\bs u}({\bs X}^{(1)}, t)$) and 
${\bs u}^{(2)}(t)$.
%The relative velocity of the two particles is then ${\bs w} ({\bs r})$$(\equiv {\bs v}^{(1)} -  {\bs v}^{(2)}$. 
%with ${\bs r}$ the separation vector of the two particles.
%We will use numerical simulations to explore the full statistics of ${\bs w}$, 
%including both its second-order moments and the probability distribution function, as functions of the particle size or friction timescale. 
We characterize the second-order statistics of ${\bs w}$ by a particle velocity structure tensor, 
\begin{equation}
S_{{\rm p}ij} ({\bs r}) = \left \langle w_i w_j \right \rangle =  \left \langle \left(v^{(1)}_i -v^{(2)}_i\right) \left(v^{(1)}_j -v^{(2)}_j\right) \right \rangle,             
\label{particlestructure}
\end{equation}  
where $\langle \cdot \cdot \cdot \rangle$ denotes the ensemble average. From 
statistical homogeneity and stationarity, $S_{{\rm p}ij}$ depends only on ${\bs r}$. 
Further assuming isotropy, it can be written as (Paper I), 
\begin{equation}
S_{{\rm p}ij} ({\bs r})  = \langle w_{\rm t}^2 \rangle \delta_{ij} + \left( \langle w_{\rm r}^2 \rangle - \langle w_{\rm t}^2 \rangle \right) \frac{r_i r_j}{r^2}, 
\label{particlestructure2}
\end{equation} 
where $\langle w_{\rm r}^2 \rangle$ and $\langle w_{\rm t}^2 \rangle$ are the variances of the radial
component, $w_{\rm r}$ ($\equiv w_i r_i/r$), and a tangential component, $w_{\rm t}$, of the relative velocity, 
respectively.  From eq.\ (\ref{particlestructure2}), we see that $\langle w_{\rm r}^2 \rangle  = S_{{\rm p}ij} {r_i r_j}/{r^2}$, 
and the 3D variance, $\langle w^2 \rangle= S_{{\rm p}ii} = \langle w_{\rm r}^2 \rangle + 2\langle w_{\rm t}^2 \rangle$. 
%We will model $S_{{\rm p}ij}$ using the formal solution, eq.\ (\ref{formalsolution}), of the particle velocity and 
%the properties of the carrier flow. 
\subsection{The Limits of Small and Large Particles}

Saffman \& Turner (1956) studied turbulence-induced relative velocity in 
the small-particle limit  with $\tau_{\rm p1}, \tau_{\rm p2} $ much smaller 
than the Kolmogorov timescale, $\tau_{\eta}$, of the turbulent flow, or with  
%defined as $\tau_{\eta} \equiv (\nu/\bar{\epsilon})^{1/2}$, where $\nu$ 
%and $\bar{\epsilon}$ are the viscosity and the average energy dissipation rate in the flow. 
%The S-T limit  is usually expressed as $St_1, St_2  \ll 1$, with the 
Stokes numbers $St_{1,2} \equiv \tau_{\rm p1,2} /\tau_{\eta} \ll 1$. 
%If the friction timescale $\tau_{\rm p}$ is much smaller than  the smallest timescale, $\tau_{\eta}$, in the flow,
%The velocity of particles in this limit can be approximated 
Using a Taylor expansion of eq.\ (\ref{particlemomentum})
%${\bs v} (t) \simeq {\bs u} ({\bs X}, t) - \tau_{\rm p} {\bs a} ({\bs X}, t)$, 
%where ${\bs a} = d{\bs u}/dt$ is the local flow acceleration.  
for these small particles to calculate $S_{{\rm p}ij}$ gives (Saffman and Turner1956), 
%gives ${\bs w} = \left( {\bs  u}^{(1)} - {\bs  u}^{(2)} \right)+ 
%\%left( {\bs  a}^{(1)} \tau_{\rm p1} - {\bs  a}^{(2)} \tau_{\rm p2}\right) $. 
%where  ${\bs u}^{(1,2)}$ 
%($\equiv  {\bs  u}({\bs X}^{(1,2)}, t)$) and ${\bs  a}^{(1,2)}$ ($\equiv  {\bs  a}({\bs X}^{(1,2)}, t)$) 
%are the flow velocity and acceleration at the particle position. 
%Saffman and Turner (1956) derived the variance of  ${\bs w}$ based on 
%two assumptions. The first one is that the flow velocity,
%Assuming statistical independence of ${\bs  u}^{(1, 2)}$ with ${\bs  a}^{(1,2)}$ 
%and a unity correlation coefficient between ${\bs  a}^{(1)}$ and ${\bs  a}^{(2)}$ 
%across a small ${\bs r}$ (Saffman and Turner1956), 
\begin{equation} 
S_{{\rm p}ij} ({\bs r}) =  \langle a_i a_j \rangle (\tau_{\rm p1} -\tau_{\rm p2})^2 + S_{ij}({\bs r}),
\label{sttensor} 
\end{equation}
where ${\bs a}$ is the local flow acceleration, and the flow structure tensor, $S_{ij} (\bs {r})$,  is 
defined as $S_{ij} (\bs {r})\equiv\left \langle  \Delta u_i   \Delta u_j \right \rangle $ with $ \Delta u_i$  
the velocity difference,  $u_i({\bs x} + {\bs r}) - u_i({\bs x})$, across ${\bs r}$. In an isotropic flow,  
$S_{ij} = S_{\rm nn} \delta_{ij}  + (S_{\rm ll} - S_{\rm nn})r_i r_j /r^2$, where $S_{\rm ll}(r)$ 
and $S_{\rm nn}(r)$ are the longitudinal and transverse structure functions (Paper I). For $r$ below the Kolmogorov
scale $\eta$ ($\equiv (\nu^3\bar{\epsilon})^{1/4}$) of an incompressible flow,  
$S_{\rm ll}(r)= \frac{1}{15} \frac{\bar {\epsilon}} {\nu} r^2$ and $S_{\rm nn}(r) = \frac{2}{15} \frac{\bar {\epsilon}} {\nu} r^2$, 
respectively (Monin and Yaglom 1975). Here $\nu$ and $\bar{\epsilon}$ are the viscosity and the average 
energy dissipation rate of the flow.

%For a scale $r$
%Similar to the particle velocity structure tensor, $S_{\rm ll} = \langle (\Delta u_{\rm r})^2 \rangle$ and   $S_{\rm nn} = \langle (\Delta u_{\rm t} )^2 \rangle$ 
%where $\Delta u_{\rm r}$ and $\Delta u_{\rm t}$ are the radial and tangential components of the flow velocity difference respectively.

With $\langle a_i a_j \rangle = a^2 \delta_{ij}$ from isotropy, we find by  
comparing eq.\ (\ref {sttensor}) with eq.\ (\ref{particlestructure2}), 
%\begin{gather}
%\langle w_{\rm r}^2 \rangle = (\tau_{\rm p1}-\tau_{\rm p2} )^2 a^2  + \frac{1}{15} \frac {\bar{\epsilon}}{\nu} r^2,\hspace{1cm} \langle w_{\rm t}^2 \rangle = (\tau_{\rm p1}-\tau_{\rm p2} )^2 a^2  + \frac{2}{15} \frac {\bar{\epsilon}}{\nu} r^2. 
%\label{saffmanturner}
%\end{gather} 
\begin{gather}
\langle w_{\rm r}^2 \rangle = (\tau_{\rm p1}-\tau_{\rm p2} )^2 a^2  + \frac{1}{15} \frac {\bar{\epsilon}}{\nu} r^2,\notag \\
\langle w_{\rm t}^2 \rangle = (\tau_{\rm p1}-\tau_{\rm p2} )^2 a^2  + \frac{2}{15} \frac {\bar{\epsilon}}{\nu} r^2, 
\label{saffmanturner}
\end{gather} 
where the first and second terms on the right hand sides are usually referred 
to as the acceleration and shear terms, respectively  (Wang et al.\ 2000, Zhou et al.\ 2001). 
In the monodisperse case, the acceleration terms vanish, and  only the shear 
terms contribute. The accuracy and the weakness of the shear terms in the S-T formula 
for equal-size particles have been discussed in Paper I% 
%From eq.\ (\ref{saffmanturner}), the 3D variance is $\langle w^2 \rangle =  3 (\tau_{\rm p1}-\tau_{\rm p2} )^2 a^2 + \frac {\bar{\epsilon}}{3\nu} r^2$. 
\footnote{For example, the S-T formula predicts that the shear term for  $\langle w_{\rm t}^2 \rangle$ is 
twice larger than that for $\langle w_{\rm r}^2 \rangle$. %This was not confirmed by our simulation data in 
%Paper I for equal-size particles with $St\gsim 0.1$. 
But our simulation shows that the radial and tangential rms relative speeds are nearly equal for 
equal-size particles with $St \gsim 0.1$. % $\langle w_{\rm t}^2 \rangle = $
%and the shear terms in both $\langle w_{\rm r}^2 \rangle$ and $\langle w_{\rm t}^2 \rangle$ 
%are better approximated by $\frac {\bar{\epsilon}}{9\nu} r^2$. 
We also found that the linear scaling of the rms relative speed with $r$ predicted by S-T for 
equal-size particles does not hold for $St \gsim 0.1$, due to the sling effect or 
caustic formation (Paper I). The S-T formula for shear terms may have higher accuracy at $St \ll 0.1$.}.%(Falkovich et al.\ 2002; Wilkinson et al.\ 2006).}.  
The acceleration terms depend on the friction time difference (see also Weidenschilling 1984), %corresponding to the different
%responses of different particles to the flow, 
and, unlike the shear terms, they are independent of $r$. 
%We will test the accuracy of the S-T prediction for the bidisperse case in this work.

We next consider the opposite limit of  large particles with $\tau_{\rm p1,2} \gg T_{\rm L}$  with $T_{\rm L}$ 
the Lagrangian correlation time of the flow. Motions of these particles are similar to Brownian motion, and 
the velocities of any such particles are uncorrelated. Therefore, $S_{{\rm p}ij} ({\bs r}) = \left[ \left(v'^{(1)}\right)^2 + \left(v'^{(2)}\right)^2 \right] \delta_{ij}$, 
where $v'^{(1)}$ and $v'^{(2)}$ are the 1-particle rms velocities (see Paper I). In this limit, we have, 
\begin{gather}
\langle w_{\rm r}^2 \rangle = \langle w_{\rm t}^2 \rangle = u'^2 \left( \frac{1}{\Omega_{\rm 1}} + \frac{1}{\Omega_{\rm 2} } \right), 
\label{largeparticles}
\end{gather} 
where $\Omega_{1,2} = \tau_{\rm p1,2}/T_{\rm L}$, and we used $v'^{(1,2)} = u' \Omega_{1,2}^{-1/2}$ 
for $\Omega_{\rm 1, 2} \gg 1$ (Abrahamson 1975). 

An interesting limiting case is that only one of the particles is very large with $\tau_{\rm p} \gg T_{\rm L}$. 
The velocity of this particle is small ($\ll u'$), and its relative velocity with the small 
particle would be approximately the 1-particle velocity of the small particle.  The 1-particle velocity has been examined 
in Paper I.  If the small particle has $\tau_{\rm p} \ll T_{\rm L}$, its velocity is close to 
the flow rms velocity, $u'$, and the relative speed with the large particle is expected to be $\simeq u'$. 
The flow-particle relative velocity discussed in \S 2 
is a special limiting case with one of the particle being a tracer ($St=0$).  

%and with a zero particle separation ($r=0$).  

%we point out that the flow-particle relative velocity considered in \S 2 may 
%be viewed as 

\begin{figure*}[t]
\centerline{\includegraphics[width=1.3\columnwidth]{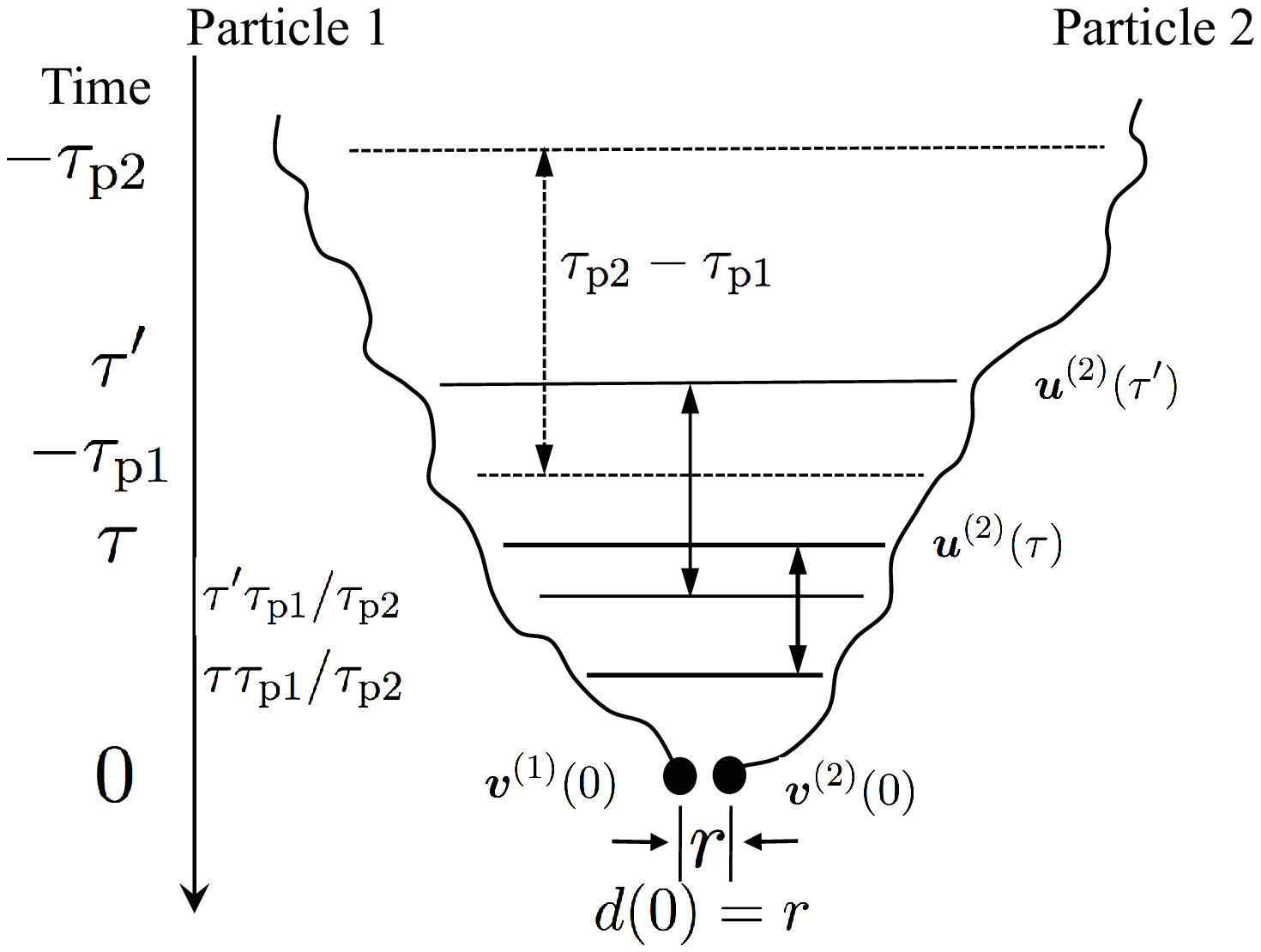}}
\caption{Schematic figure illustrating the generalized acceleration term, $\mathcal{A}_{ij}$, 
for the relative velocity of particles, (1) and (2), of different sizes at a separation $r$ at $t=0$. 
%The particle friction timescales are $\tau_{\rm p1}$ and $\tau_{\rm p2}$, respectively, and, 
Without loss of generality, it is assumed $\tau_{\rm p2}> \tau_{\rm p1}$. 
The figure is based on eq.\ (\ref{accelmean}), which suggests that  the effect of $\mathcal{A}_{ij}$ 
can be roughly viewed as due to the particle memory of the temporal flow velocity difference, $\Delta_{\rm T} {\bs u}$, seen by particle (2).
%at a time lag $\propto (\tau_{\rm p2} -\tau_{\rm p1})$. 
%One may also make a similar figure based 
%on the temporal velocity difference on the trajectory of particle (1) using eq.\ (\ref{accelmean2}).  
%Clearly, the term vanishes for equal-size particles. 
$\Delta_{\rm T} {\bs u}$ is thus crucial for understanding the generalized acceleration effect. 
%to the bidipserse relative velocity.
}
\label{cartoonaccel} 
\end{figure*}

\subsection{The Formulation of Pan and Padoan (2010)}

We briefly summarize the formulation of the model by Pan and Padoan (2010;  PP10) for the relative 
velocity variance in the bidisperse case. The monodisperse version of the model for equal-size 
particles was presented in Paper I. The model is derived by calculating $S_{{\rm p}ij} ({\bs r})$ in eq.\ (\ref{particlestructure}) 
using the formal solution eq.\ (\ref{formalsolution}) for the particle velocity. The derivation is reviewed  in Appendix A, 
where an assumption implicitly made in PP10 is pointed out and justified.  In the general PP10 model, 
$S_{{\rm p}ij} $ can be written as two terms, 
\begin{equation}
S_{{\rm p}ij} ({\bs r}) = \mathcal{A}_{ij} +\mathcal{S}_{ij} {\rm}, 
\label{pp10form}
\end{equation}
which reduce to the acceleration and shear terms in the S-T limit, eq.\ (\ref{saffmanturner}), 
respectively (see Appendix A and \S 3.2.1). The formulation may thus viewed as a generalization of the S-T formula 
for particles of any arbitrary sizes. We name $\mathcal{A}_{ij}$ and $\mathcal{S}_{ij}$ the generalized acceleration and shear terms, 
respectively (PP10). Equations for the two terms are given in Appendix A (eqs.\ (\ref{accel}) and (\ref{shear})). Note that the 
generalized shear term was denoted as $\mathcal{D}_{ij}$ in PP10.
  
%The tensor $\mathcal{T}$ is given by 
%\begin{equation}
% \mathcal{T}_{ij} =  \int_{-\infty}^0  \frac {d\tau}{\tau_{\rm p1}}\int_{-\infty}^0 \frac{d\tau'}{\tau_{\rm p2}}C^{(1)(2)}_{ij} ({\bs r}; \tau, \tau') \exp \left(\frac{\tau}{\tau_{\rm p1}} \right) \exp \left(\frac{\tau'}{\tau_{\rm p2}}\right)
%\end{equation}

Like the acceleration term in the S-T formula, $\mathcal{A}_{ij}$ vanishes for particles of equal size. 
$\mathcal{A}_{ij}$ relies  only on the flow velocity statistics along {\it individual} trajectories of the two particles, 
and is thus independent of ${\bs r}$.  The generalized shear term represents the contribution from the 
particles' memory of the flow velocity difference they saw in the past, and it can be modeled in a similar way as 
the case of equal-size particles discussed in Paper I. Due to the complexity of the problem, our formulation is 
complicated (see Appendix A) and not straightforward for applications. In further works, we will establish simple 
function fits  or  simplified forms that agree with our data and are practically easy to use.  
In this paper, we focus on testing the accuracy of the model. %and validating the physical picture. 

It is interesting to split the relative velocity, ${\bs w}$, into an acceleration component,  ${\bs w}_{\rm a}$
and a shear component ${\bs w}_{\rm s}$, such that $\langle w_{{\rm a}i} w_{{\rm a}j} \rangle = \mathcal{A}_{ij} $ 
and $\langle w_{{\rm s}i} w_{{\rm s}j} \rangle = \mathcal{S}_{ij}$. It is implicitly assumed that ${\bs w}_{\rm a}$ 
and ${\bs w}_{\rm s}$ are statistically independent. We discuss the modeling  and physical meaning 
of the generalized acceleration and shear  effects in the next two subsections. 

\subsubsection{The Generalized Acceleration Term}

To understand the physical meaning of the generalized acceleration term, we use an approximation 
for the equation (\ref{accel}) of $\mathcal{A}_{ij}$  derived in Appendix A. The term depends on 
the temporal flow velocity correlations, $\left \langle u_i^{(1)} (\tau) u_j^{(1)}(\tau') \right \rangle$ and $\left\langle u_i^{(2)} (\tau) u_j^{(2)}(\tau') \right \rangle$, 
along the individual trajectories of the two particles.  Assuming that  the  trajectory statistics of the two particles are equivalent, 
i.e., $\left \langle u_i^{(1)} (\tau) u_j^{(1)}(\tau') \right \rangle \simeq \left\langle u_i^{(2)} (\tau) u_j^{(2)}(\tau') \right \rangle$, 
%With this assumption, eq.\ (\ref{accel}) can be written as, 
eq.\ (\ref{accel}) can be rewritten as, 
%\begin{equation}
%\mathcal{A}_{ij} \simeq  \int_{-\infty}^0 \frac {d\tau}{\tau_{\rm p2}} \int_{-\infty}^0 \frac {d\tau'}{\tau_{\rm p2}} \left\langle \left[u_i^{(2)}(\tau) - u_i^{(2)}(f\tau)\right]\left[u_j^{(2)}(\tau') - u_j^{(2)}(f\tau')\right]  \right\rangle \exp \left(\frac{\tau}{\tau_{\rm p2}}\right) \exp \left(\frac{\tau'}{\tau_{\rm p2}}\right),
%\label{accelmean} 
%\end{equation}
\begin{gather}
\hspace{-1.5cm}\mathcal{A}_{ij} \simeq  \int_{-\infty}^0 \frac {d\tau}{\tau_{\rm p2}} \int_{-\infty}^0 \frac {d\tau'}{\tau_{\rm p2}} \Bigg\langle \left[ u_i^{(2)}(\tau) - u_i^{(2)}(f\tau)\right]
\notag \\ 
 \hspace{0.5cm} \times\left[u_j^{(2)}(\tau') 
- u_j^{(2)}(f\tau') \right]  \Bigg\rangle \exp \left(\frac{\tau}{\tau_{\rm p2}}\right) \exp \left(\frac{\tau'}{\tau_{\rm p2}}\right),
\label{accelmean} 
\end{gather}
%or 
%\begin{equation}
%\mathcal{A}_{ij} \simeq  \int_{-\infty}^0 \frac {d\tau}{\tau_{\rm p1}} \int_{-\infty}^0 \frac {d\tau'}{\tau_{\rm p1}} \left\langle \left[u_i^{(1)}(\tau) - u_i^{(1)}\left(\frac{\tau}{f}\right)\right]\left[u_i^{(1)}(\tau') - u_i^{(1)}\left(\frac{\tau'}{f}\right)\right]  \right\rangle \exp \left(\frac{\tau}{\tau_{\rm p1}}\right) \exp \left(\frac{\tau'}{\tau_{\rm p1}}\right),
%\label{accelmean2} 
%\end{equation}
where  $f \equiv \tau_{\rm p1}/\tau_{\rm p2}$ is the friction time ratio of the two particles. 
Without loss of generality, we assume $\tau_{\rm p1} \le \tau_{\rm p2}$ or $f\le1$.  
Throughout the paper, we define $f$ as the friction time (or Stokes number) ratio of the smaller particle to the larger one.
Note that eq.\ (\ref{accelmean}) depends on the flow velocity, ${\bs u}^{(2)}$, along the trajectory 
of particle (2) only\footnote{One could write an approximate equation for $\mathcal{A}_{ij}$ with ${\bs u}^{(1)}$ only. 
It would be similar to eq.\ (\ref{accelmean}), but with ${\bs u}^{(1)}$  replacing ${\bs u}^{(2)}$, 
$\tau_{\rm p1}$ replacing $\tau_{\rm p2}$, and $1/f$ replacing $f$.}.
If $\tau_{\rm p1}=0$,  $f = 0$, and eq.\ (\ref{accelmean}) becomes identical to eq.\ (\ref{pfrelative}) for the flow-particle 
relative velocity.  Based on eq.\ (\ref{accelmean}),  Fig.\ \ref{cartoonaccel} illustrates a 
physical picture for $\mathcal{A}_{ij}$.
%Eq.\ (\ref{accelmean}) suggests that the generalized acceleration term can roughly viewed as 
%due to the memory of the temporal flow velocity difference, $\Delta {\bs u}_{\rm T}$, 
%at a time lag of $(1-f) \tau$ %(or $(1/f-1)\tau$) 
%along the trajectory of the larger %(or smaller) 
%particle. This time lag is proportional to the friction time difference, $\tau_{\rm p2} - \tau_{\rm p1}$.  %Again, the flow-particle relative velocity is essentially a special case of 
%the generalized acceleration term with one of the particles being tracers.   
The ensemble average term in eq.\ (\ref{accelmean}) is expected to increases with $|\tau|$ and 
$|\tau'|$, and, together with the exponential cutoffs, it suggests that a major contribution to the integral is 
likely from $\tau$ and $\tau'$ values around $-\tau_{\rm p2}$. Therefore, a rough estimate for
the generalized acceleration effect is ${\bs w}_{\rm a} \sim {\bs u}^{(2)} (-\tau_{\rm p2})- {\bs u}^{(2)}(-\tau_{\rm p1})$, 
which is essentially the temporal flow velocity difference $\Delta {\bs u}_{\rm T} (\Delta \tau)$ 
along the trajectory of particle (2) at a time lag of $\Delta \tau \simeq |\tau_{\rm p2} -\tau_{\rm p1}|$. 
This establishes a relation between ${\bs w}_{\rm a}$ and 
the temporal statistics of the flow velocity along an individual trajectory. 
The approximation ${\bs w}_{\rm a} \simeq \Delta {\bs u}_{\rm T} (|\tau_{\rm p2} -\tau_{\rm p1}|)$ 
is crude especially for particles of similar sizes and for large particles with $\tau_{\rm p} \gsim T_{\rm L}$.
%It ignore the effect of the flow correlation time, which is important for large particles. Also it does not work for the case where the particle sizes are close. 
We will give a better expression for ${\bs w}_{\rm a}$  based on a quantitative calculation of  $\mathcal{A}_{ij}$ (see eq.\ (\ref{approxaccel}) below).    
%\footnote{This approximation is valid if $\tau_{\rm p1,2} \lsim T_{\rm L}$. 
%In that case, the term, $\left\langle \left[u_i^{(2)}(\tau) - u_i^{(2)}(f\tau)\right]\left[u_j^{(2)}(\tau') - u_j^{(2)}(f\tau')\right]  \right\rangle$, in eq.\ (\ref{accelmean})
%can be approximated by $\simeq \left\langle \left[u_i^{(2)}(\tau) - u_i^{(2)}(f\tau)\right] \left[u_j^{(2)}(\tau) - u_j^{(2)}(f\tau)\right]  \right\rangle$ for $-\tau_{\rm p2} \le\tau, \tau' \le 0$. 
%However, if $\tau_{\rm p2} \gsim T_{\rm L}$, this approximation breaks down, and the flow memory timescale, $T_{\rm L}$, starts to play a role in determining 
%the generalized acceleration term.}. 
%(or  $\sim {\bs u}\left(X^{(1)}(\tau_{\rm p2}), \tau_{\rm p2}\right) - {\bs u}\left(X^{(1)}(\tau_{\rm p1}), \tau_{\rm p1}\right)$ 
%if one makes use of  eq.\ (\ref{accelmean2})). 
%(or $u(X^{(1)}(\tau_{\rm p1}), \tau_{\rm p1}) - u(X^{(1)}(\tau_{\rm p2}), \tau_{\rm p2})$).
%and (\ref{accelmean2}) is 
%as it is based on the assumption of the statistical equivalence for the 
%trajectories of particles (1) and (2). 
Although the discussion here is qualitative, it does provide an insightful physical 
picture for the generalized acceleration effect. The discussion also suggests that, even if the trajectory statistics of two 
different particles were identical at all times in the past, their velocities at the current time would be 
different, as they have different memories of (or different responses to) the flow velocity. 
%which is a fundamental physical origin of the generalized acceleration effect.    

We now quantitatively evaluate the generalized acceleration term.  Assuming $B^{(1,2)}_{{\rm}ij}(\tau, \tau') =B^{(1,2)}_{{\rm}ij}(|\tau'-\tau|)$, we can 
simplify the double integrals in eq.\ (\ref{accel}) by a variable change, yielding, 
%For example, we may introduce $\xi = \tau + \tau_{\rm p1} \tau'/\tau_{\rm p2}$ and 
%$\zeta = \Delta \tau = \tau'-\tau$ 
%for the terms on the second line. 
%\begin{gather}
%\vspace {2mm}
%\mathcal{A}_{ij} = \frac{\tau_{\rm p2}- \tau_{\rm p1}}{\tau_{\rm p1}+\tau_{\rm p2}} 
%\int_{-\infty}^0   B^{+}_{ij} (\tau) \left[\frac {1}{\tau_{\rm p1}}  \exp \left (\frac {\tau}{\tau_{\rm p1}} \right)-\frac{1}{\tau_{\rm p2}} \exp \left (\frac {\tau}{\tau_{\rm p2}} \right) \right]  d\tau\notag\\
%+ \int_{-\infty}^0   B^{-}_{ij} (\tau) \left[\frac {1}{\tau_{\rm p1}}  \exp \left (\frac {\tau}{\tau_{\rm p1}} \right)-\frac{1}{\tau_{\rm p2}} \exp \left (\frac {\tau}{\tau_{\rm p2}} \right) \right]  d\tau
%\label{accel2}
%\end{gather}
%\begin{gather}
%\vspace {2mm}
%\mathcal{A}_{ij} = 
%\int_{-\infty}^0 \left [ \left(\frac{\tau_{\rm p2}- \tau_{\rm p1}}{\tau_{\rm p1}+\tau_{\rm p2}} \right)  B^{+}_{{\rm}ij} (\Delta \tau)+    B^{-}_{{\rm} ij} (\Delta \tau)\right]
%\left[\frac {1}{\tau_{\rm p1}}  \exp \left (\frac {\Delta \tau}{\tau_{\rm p1}} \right)-\frac{1}{\tau_{\rm p2}} \exp \left (\frac {\Delta \tau}{\tau_{\rm p2}} \right) \right]  d\Delta \tau
%\label{accel2}
%\end{gather}
\begin{gather}
%\vspace {2mm}
\hspace{-1.5cm}\mathcal{A}_{ij} = 
\int_{-\infty}^0 \left [ \left(\frac{\tau_{\rm p2}- \tau_{\rm p1}}{\tau_{\rm p1}+\tau_{\rm p2}} \right)  B^{+}_{{\rm}ij} (\Delta \tau)+    B^{-}_{{\rm} ij} (\Delta \tau)\right] \notag\\
\hspace{0.5cm} \times\left[\frac {1}{\tau_{\rm p1}}  \exp \left (\frac {\Delta \tau}{\tau_{\rm p1}} \right)-\frac{1}{\tau_{\rm p2}} \exp \left (\frac {\Delta \tau}{\tau_{\rm p2}} \right) \right]  d\Delta \tau, 
\label{accel2}
\end{gather}
where $B^{+}_{{\rm}ij} \equiv ( B^{(1)}_{{\rm}ij} + B^{(2)}_{{\rm}ij})/2$ and $B^{-}_{{\rm}ij} \equiv ( B^{(1)}_{{\rm}ij} - B^{(2)}_{{\rm}ij})/2$.  
%Note that $B^{-}_{{\rm}ij}$ corresponds to the difference in the flow velocity statistics along the trajectories of two different particles.  
%One can rewrite $\mathcal{A}_{ij}$ in terms of the temporal structure function,  $D_{{\rm}ij}^{(1,2)}$, of the flow velocity along each particle trajectory 
Using $D_{{\rm}ij}^{(1,2)} = 2(u'^2 \delta_{ij}- B^{(1,2)}_{{\rm }ij})$ for the temporal flow structure function, $D_{{\rm}ij}^{(1,2)}$, we can 
explicitly see that the acceleration term is connected to the temporal flow velocity difference, $\Delta_{\rm T} {\bs u} (\Delta \tau)$, on the particle trajectories. 
%Again we see that $\mathcal{A}_{ij} =0$ for equal-size particles. 
%\begin{gather}
%\mathcal{A}_{ij} =\frac{1}{2} 
%\int_{-\infty}^0 \left [ \left(\frac{\tau_{\rm p2}- \tau_{\rm p1}}{\tau_{\rm p1}+\tau_{\rm p2}} \right)  D^{+}_{{\rm }ij} (\Delta \tau)+    D^{-}_{{\rm} ij} (\Delta \tau)\right]
%\left[\frac {1}{\tau_{\rm p2}}  \exp \left (\frac {\Delta \tau}{\tau_{\rm p2}} \right)-\frac{1}{\tau_{\rm p1}} \exp \left (\frac {\Delta \tau}{\tau_{\rm p1}} \right) \right]  d\Delta \tau
%\label{accel3}
%\end{gather}
%where $D^{\pm}_{ij} = (D^{(1)}_{{\rm}ij} \pm D^{(2)}_{{\rm}ij})/2$. 
%The equation also indicates 

Following PP10, we approximate both $B^{(1)}_{{\rm}ij}$ and $B^{(2)}_{{\rm}ij}$ by the 
Lagrangian correlation tensor, $u'^2 \Phi_{\rm L}(\Delta \tau) \delta_{ij}$, %This assumption can be justified from the facts that the temporal series of the flow velocity seen by a particle with 
%finite inertial is in between Lagrangian and Eulerian and 
%Again, the assumption is based on the finding in Paper I that the Lagrangian and Eulerian temporal correlation 
%functions are close to each other. 
%With $B^{(1)}_{{\rm }ij}= B^{(2)}_{{\rm }ij} =u'^2 \Phi_{\rm L}(\Delta \tau) \delta_{ij}$, 
so that  $B^{+}_{{\rm }ij} = u'^2 \Phi_{\rm L}(\Delta \tau) \delta_{ij}$ 
and $B^{-}_{{\rm }ij} = 0$. Here, the possible asymmetry (i.e., $B^{-}_{{\rm }ij}$) between the trajectory statistics of different particles is neglected. Eq.\ (\ref{accel2}) then 
becomes,
\begin{equation}
\mathcal{A}_{ij} =A \delta_{ij}, 
\label{accelform}
 \end{equation}
with 
%\begin{gather}
%\vspace {2mm}
%A  = u'^2 \left(\frac{\tau_{\rm p2}- \tau_{\rm p1}}{\tau_{\rm p1}+\tau_{\rm p2}}\right) 
%\int_{-\infty}^0   \Phi_{\rm L}(\Delta \tau)  \left[\frac {1}{\tau_{\rm p1}}  \exp \left (\frac {\Delta \tau}{\tau_{\rm p1}} \right)-\frac{1}{\tau_{\rm p2}} \exp \left (\frac {\Delta \tau}{\tau_{\rm p2}} \right) \right]  d\Delta \tau.  
%\notag \\ =  \frac{\tau_{\rm p2}- \tau_{\rm p1}}{2(\tau_{\rm p1}+\tau_{\rm p2})}
%\int_{-\infty}^0   D_{\rm L}(\tau)  \left[\frac {1}{\tau_{\rm p2}}  \exp \left (\frac {\tau}{\tau_{\rm p2}} \right)-\frac{1}{\tau_{\rm p1}} \exp \left (\frac {\tau}{\tau_{\rm p1}} \right) \right]  d\tau. 
%\label{accel4}
%\end{gather}
\begin{gather}
\vspace {2mm}
\hspace{-3cm}A  = u'^2 \left(\frac{\tau_{\rm p2}- \tau_{\rm p1}}{\tau_{\rm p1}+\tau_{\rm p2}}\right) 
\int_{-\infty}^0   \Phi_{\rm L}(\Delta \tau) \notag \\ \hspace{1cm} \times\left[\frac {1}{\tau_{\rm p1}}  \exp \left (\frac {\Delta \tau}{\tau_{\rm p1}} \right)-\frac{1}{\tau_{\rm p2}} \exp \left (\frac {\Delta \tau}{\tau_{\rm p2}} \right) \right]  d\Delta \tau.  
%\notag \\ =  \frac{\tau_{\rm p2}- \tau_{\rm p1}}{2(\tau_{\rm p1}+\tau_{\rm p2})}
%\int_{-\infty}^0   D_{\rm L}(\tau)  \left[\frac {1}{\tau_{\rm p2}}  \exp \left (\frac {\tau}{\tau_{\rm p2}} \right)-\frac{1}{\tau_{\rm p1}} \exp \left (\frac {\tau}{\tau_{\rm p1}} \right) \right]  d\tau. 
\label{accel4}
\end{gather}
%One can derive an equivalent equation for $A$ from eq.\ (\ref{accel3}) assuming $D_{{\rm}ij}^{(1,2)}$ can be approximated 
%by the Lagrangian structure tensor, $D_{{\rm L}ij}$. 
Eq.\ (\ref{accelform}) indicates that $\mathcal{A}_{ij}$ has equal longitudinal (radial) and transverse (tangential) components.
% i.e., $\mathcal{A}_{\rm r} = \mathcal{A}_{\rm t} =A$.  

%In the resulting equation 
%eq.\ (\ref{accel3}) gives an equation similar 
%to eq.\ (\ref{accel4}), but with  $-D_{\rm L}/2$ replacing  $u'2 \Phi_{\rm L})$. %The equation indicates a connection of the acceleration 
%term to the Lagrangian velocity difference.  Again it may be better approximation to change $S_{\rm L}$ to $S_{\rm E}$ for large 
%particles. 
%\begin{gather}
%\vspace {2mm}
%A  = 
%\label{accel4}
%\end{gather}

If we adopt an exponential form for $\Phi_{\rm L}$ ($\equiv \exp(-|\Delta \tau|/T_{\rm L})$), 
%in eq.\ (\ref{accel4}), 
then a simple integration gives,  
 \begin{equation}
A = u'^2 (\Omega_2-\Omega_1)^2/\big[(\Omega_1 +\Omega_2)(1+\Omega_1) (1+\Omega_2)\big],
\label{accelsolutionsingle}
\end{equation}
where  $\Omega_{1, 2} \equiv  \tau_{\rm p 1,2}/T_{\rm L}$. With a biexponential $ \Phi(\Delta \tau)$ (eq.\ (\ref{biexponential})),  one obtains, 
%can also integrate eq.\ (\ref{accel4}) analytically,
\begin{equation}
A  = u'^2 \frac{(\Omega_2 -\Omega_1)^2 \left(\Omega_1 \Omega_2 + (\Omega_1 + \Omega_2) {\displaystyle \frac{z^2}{2} } \right)  }
{(\Omega_1 + \Omega_2) \left(\Omega_1 + \Omega_1^2 + {\displaystyle \frac{z^2}{2}} \right) \left(\Omega_2 + \Omega_2^2 + {\displaystyle \frac{z^2}{2}}\right) }, 
\label{accelsolution} 
\end{equation}
which reduces to eq.\ (\ref{accelsolutionsingle}) if $z=0$. Clearly, if both $\Omega_1$ and $\Omega_2$ are 
much larger than $z^2/2$, eq.\ (\ref{accelsolution}) is a good approximation for eq.\ (\ref{accelsolutionsingle}).  
In fact, eqs.\ (\ref{accelsolutionsingle}) and (\ref{accelsolution}) are close to each other if either 
of the two $\Omega$'s is much larger than $z^2/2$. 
%while the other particle has $\Omega \lsim z^2/2$. 
 A numerical comparison of eqs.\ (\ref{accelsolutionsingle}) and (\ref{accelsolution}) shows 
 that, if the large particle has 
$\Omega  \gsim 3 z^2$, the difference in the two equations is $\lsim 20\%$. For convenience, 
we will denote $\Omega$ of the larger and smaller particles as $\Omega_{h}$ and $\Omega_{\ell}$, respectively.
%and one can use either 
%of them to estimate the acceleration contribution. 
%can have considerable difference only when both $\Omega_1$ and $\Omega_2$ are 
%comparable or smaller than $z^2/2$. 
In the limit $\Omega_{h} \to \infty$, eq.\ (\ref{accelsolution}) approaches 
$u'^2 (\Omega_{\ell}+z^2/2)/(\Omega_{\ell}+ \Omega_{\ell}^2 + z^2/2)$, 
which is the 1-particle velocity variance of the smaller particle (see eq.\ (6) of Paper I).  

%Clearly, the generalized acceleration term vanishes for equal-size particles. 
For  small particles with $\Omega_{1,2} \ll z^2/2$, %(or equivalently $St_{1,2} \ll 1$), 
it is easy to show that eq.\ (\ref{accelsolution}) reduces to $ A= 2u'^2 (\Omega_2 -\Omega_1)^2/z^2 = (\tau_{\rm p2} -  \tau_{\rm p1})^2 a^2 $, 
meaning that the acceleration term in  the S-T formula, eq.\ (\ref{saffmanturner}), 
is correctly reproduced by eq.\ (\ref{accelsolution}).  
On the other hand, eq.\ (\ref{accelsolutionsingle}) from the single-exponential 
$\Phi_{\rm L}$ fails to recover the S-T acceleration term, %This suggests that, to 
%recover the acceleration terms in the S-T prediction, it is crucial to 
suggesting that the bi-exponential form (eq.\ (\ref{biexponential})) 
that accounts for the effect of the flow acceleration is a  
preferred choice for small particles. 
A numerical comparison with eq.\ (\ref{accelsolution}) 
shows that the S-T acceleration term is valid only if  both particles are quite 
small with  $\Omega  \lsim 0.08z^2$, and it becomes inaccurate if $\Omega_{h}  \gsim 0.08z^2$.  

In a short summary, if the larger particle has $\Omega_{h} \lsim 0.08z^2$, 
the S-T formula applies, and the acceleration effect is completely determined 
by the local flow acceleration, $a$. If $\Omega_{h} \gsim 3 z^2$,  
the generalized acceleration term is insensitive to $a$, %as it 
%is determined mainly by the temporal flow velocity difference in the inertial range. 
and one can estimate $A$ using either eq.\ (\ref{accelsolutionsingle}) or eq.\ (\ref{accelsolution}). 
In the intermediate case with $ 0.08 z^2 \lsim \Omega_{h} \lsim 3 z^2$, we need 
to use the general formula eq.\ (\ref{accelsolution}) for an accurate estimate. 
In this case,  $A$ depends on the temporal flow velocity difference at a time lag 
in between the dissipation and inertial ranges of the flow.

%In the large particle limit with $\Omega_1, \Omega_2 \gg 1$, we find $A \to u'^2 (\Omega_2 -\Omega_1)^2/[\Omega_1 \Omega_2 (\Omega_1 + \Omega_2)]$.  
If one of the particles, say particle (1), is a tracer particle, i.e., $\Omega_1=0$, then 
eq.\ (\ref{accelsolution}) becomes ${A}  = u'^2 \Omega_2^2/(\Omega_2 + \Omega_2^2 + z^2/2)$. 
It is  identical to the flow-particle relative velocity, ${\bs w}_{\rm f}$, derived in \S 2,  
which is a special bidisperse case where only the generalized acceleration term contributes. 

Using eq.\ (\ref{accelsolution}), we attempted to establish an approximate 
relation for ${\bs w}_{\rm a}$ with 
the temporal flow velocity difference, $\Delta {\bs u}_{\rm T}$, along the 
particle trajectory. 
The idea is to obtain an approximate expression for ${\bs w}_{\rm a}$ in terms of  
$\Delta {\bs u}_{\rm T}$, which satisfies the condition that the variance of 
${\bs w}_{\rm a}$ is consistent with eq.\ (\ref{accelsolution}). 
A good approximation is found to be,
\begin{equation}
{\bs w}_{\rm a} \simeq \frac{1-f}{(1+\Omega_{\ell} )^{1/2}} \Delta {\bs u}_{\rm T} (\tau_{\rm p,h}) 
\label{approxaccel}
\end{equation} 
%\begin{equation}
%{\bs w}_{\rm a} \simeq \frac{1-f}{[2(1+\Omega_{\rm l} )]^{1/2}} \Delta {\bs u}_{\rm T} (\tau_{\rm ph}) 
%\label{approxaccel}
%\end{equation} 
where $\tau_{\rm p,l} = \min(\tau_{\rm p1}, \tau_{\rm p2})$, $\tau_{\rm p,h} = \max(\tau_{\rm p1}, \tau_{\rm p2})$, 
$\Omega_{\ell, h} =  \tau_{\rm p,l,h}/T_{\rm L}$, and $f=\tau_{\rm p,l}/\tau_{\rm p,h}$. 
We computed the variance of ${\bs w}_{\rm a}$ from the above equation under the assumption 
that $\Delta {\bs u}_{\rm T} \simeq \Delta {\bs u}_{\rm L}$, 
%with  the assumption $\Delta {\bs u}_{\rm T} \simeq \Delta {\bs u}_{\rm L}$, 
and found that it agrees with eq.\ (\ref{accelsolution}) within a factor of 2 for $\tau_{\rm p1}$ and $\tau_{\rm p2}$ in any range.  
Note that here ${\bs w}_{\rm a}$ is related to $\Delta {\bs u}_{\rm T}$ at a time lag equal to the friction time, 
$\tau_{\rm p,h}$, of the larger particle.
%rather than the friction time difference, $\tau_{\rm ph}-\tau_{\rm pl}$.  
Eq.\ (\ref{approxaccel}) is useful for understanding 
our simulation result for the bidisperse relative velocity (see \S 6).

%where the relative speed is simply given by the generalized acceleration 
%term.  
%while the other particle is large with $\Omega_2 \gg1$.    

\begin{figure*}[t]
\centerline{\includegraphics[width=1.3\columnwidth]{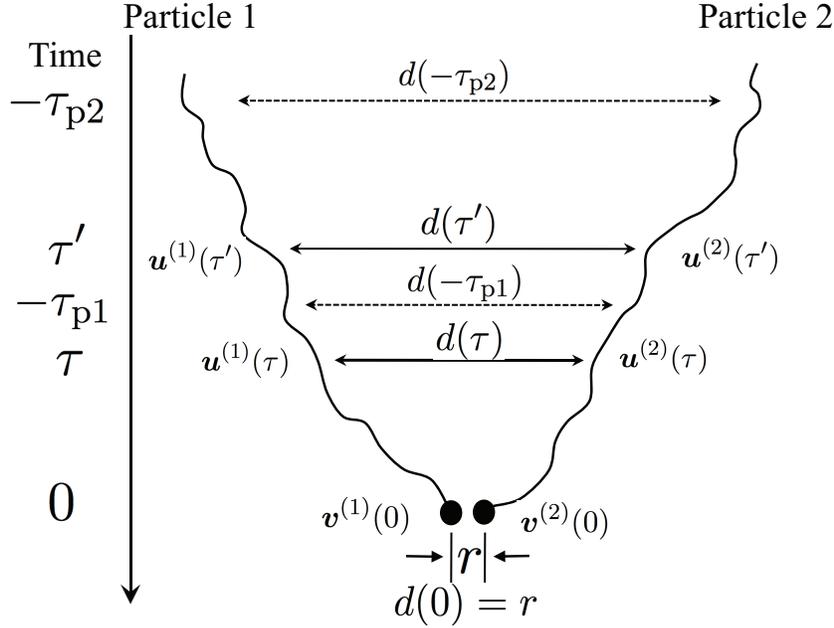}}
\caption{Schematic figure illustrating the generalized shear term for the relative velocity 
of two different particles, (1) and (2), % at a separation $r$ at time $0$. 
%The figure is similar to Fig.\ 1 of Paper I for the monodisperse case, except 
%that here the memory cutoffs occur at different times. 
which represents the memory of the spatial flow velocity difference  ``seen" 
by the two particles in the past. It depends on the particle separations, 
$d(\tau)$ and $d(\tau')$, backward in time, and also on the temporal decorrelation of 
the flow velocity structures across the particle distance %decorrelates temporally with a timescale associated with the lifetime of turbulent eddies 
%encountered by the two particles 
between $\tau$ and $\tau'$.}
\label{cartoonshear} 
\end{figure*}

\subsubsection{The Generalized Shear Term}

The generalized shear term is given by eq.\ (\ref{shear}) in Appendix A, and it represents the contribution 
to the relative velocity from the particles' memory of the {\it spatial} flow velocity differences along their trajectories 
in the past. The physical picture  is illustrated in Fig.\ \ref{cartoonshear}, which is similar to Fig.\ 1 of Paper I for identical 
particles, except that here the memory cutoffs occur at different times. 
The key of the PP10 model for this term is the evaluation of the trajectory 
structure tensor, $S_{{\rm T} ij} \equiv \langle [u_i^{(1)} (\tau) - u_i^{(2)} (\tau) ] [ u_j^{(1)} (\tau') - u_j^{(2)} (\tau') ]\rangle$, 
defined as the correlation of the flow velocity differences the particles saw at two times, 
$\tau$ and $\tau'$ (eq.\ (\ref{trajectstructuretensor}) in Appendix A). %Here we briefly 
%summarize the assumptions adopted for this tensor by PP10 and Paper I.  
%\begin{equation}            
%\mathcal{D}_{ij} = \int_{-\infty}^0  \frac {d\tau}{\tau_{\rm p1}}\int_{-\infty}^0 \frac{d\tau'}{\tau_{\rm p2}}
%S_{{\rm T}ij} ({\bs r}; \tau, \tau') \exp \left(\frac{\tau}{\tau_{\rm p1}} \right) \exp \left(\frac{\tau'}{\tau_{\rm p2}}\right).
%\end{equation}
%A detailed model for this term was described in Paper I for the monodisperse case.
We modeled it as, %a product of  the particle separation and time lag dependences,  
\begin{equation} 
S_{{\rm T}ij} ({\bs r}; \tau, \tau') \simeq \big\langle S_{ij} ({\bs R}) \Phi_2 \big(\tau' -\tau, R)\big\rangle_{\bs R}, 
\label{modeltrajectstructure}
\end{equation}
where ${\bs  R}(\tau, \tau')$ is the typical particle separation between $\tau$ and $\tau'$. The ensemble average  is over the 
probability distribution of the random vector ${\bs R}$. The Eulerian 
structure tensor $S_{ij} ({\bs R})$ is defined as $\langle \Delta u_i({\bs R}) \Delta u_j({\bs R}) \rangle$, with $\Delta {\bs u} ({\bs R}) \equiv  {\bs u} ({\bs x} + {\bs R}) -  {\bs u} ({\bs x})$, 
and $\Phi_2 \big(\tau' -\tau, R)$ is the temporal (de)correlation of the flow velocity structures at a time lag of $\Delta \tau = \tau'-\tau$.  
A simple assumption of $\Phi_2$ is $\Phi_2 \simeq \exp[-|\tau'-\tau|)/T(R)]$, where $T(R)$ is correlation time (or life time) 
of turbulent eddies of size $R$. A better approximation for $\Phi_2$  with a bi-exponetial form is 
given in Appendix B. Note that $\Phi_2$ differs from the temporal correlation, $\Phi_1$, on the trajectory of 
an individual particle. 
% which is associated with the lifetime of 
%turbulent eddies of size $R$.  %$\Phi_2$ only depends on the amplitude $R$, but not on 
%the direction. 
%%in the generalized shear term. %(see Paper I). 

%In a statistically isotropic flow, $S_{ij} ({\bs R}) = S_{\rm nn}(R) \delta_{ij}  + [S_{\rm ll}(R)- S_{\rm nn}(R)]\frac{R_i R_j} {R^2}$, 
%(see eq.\ (7) in Paper I), 
%which depends on both the direction ($R_i/R$) and the amplitude $R$ of 
%the separation.  
%On the other hand, $\Phi_2 \big(\tau -\tau', R)$ only depends on the amplitude $R$. In principle, to calculate  
%$S_{{\rm T} ij}$ from eq.\ (\ref{modeltrajectstructure0}), one needs to perform an average over the distributions 
%of both the amplitude and the direction of ${\bs R}$. 
 %the distribution of the amplitude $R$ is unknown, and thus, 
We neglect the fluctuations in the amplitude, $R$, of ${\bs R}$ and estimate $S_{{\rm T}ij}$ by simply using the 
rms value of $R$ (PP10). 
%This is equivalent 
%to assuming the PDF of the amplitude $R$ is a delta function at its rms value. 
%A justification for this assumption was given in PP10.
In the rest of the paper, $R$ denotes the rms particle distance.
%and it thus no longer represents a stochastic variable. 
%With these assumptions, 
We then have %\begin{equation}
$S_{{\rm T}ij} ({\bs r}; \tau, \tau') \simeq \big\langle S_{ij} ({\bs R}) \big\rangle_{\rm ang}\Phi_2 \big(\tau' -\tau, R)$,
%\label{modeltrajectstructure}
%\end{equation} 
where ``ang" denotes the angular average over the direction of ${\bs R}$. 
Inserting it into eq.\ (\ref{shear}) in Appendix A leads to, 
%\begin{equation}
%\mathcal{S}_{ij} =\int_{-\infty}^0  \frac {d\tau}{\tau_{\rm p1}}\int_{-\infty}^0 \frac{d\tau'}{\tau_{\rm p2}}
%\langle S_{ij} ({\bs R}) \big\rangle_{\rm ang} \Phi_2 \big(\tau' -\tau, R) \exp \left(\frac{\tau}{\tau_{\rm p1}} \right) \exp \left(\frac{\tau'}{\tau_{\rm p2}}\right). 
%\label{sij}
%\end{equation}
\begin{gather}
\mathcal{S}_{ij} =\int_{-\infty}^0  \frac {d\tau}{\tau_{\rm p1}}\int_{-\infty}^0 \frac{d\tau'}{\tau_{\rm p2}}
 \langle S_{ij} ({\bs R}) \big\rangle_{\rm ang} \Phi_2 \big(\tau -\tau', R) \notag \\ \hspace{2cm}\times \exp \left(\frac{\tau}{\tau_{\rm p1}} \right) \exp \left(\frac{\tau'}{\tau_{\rm p2}}\right). 
\label{sij}
\end{gather}
An approximate evaluation of $\langle S_{ij} ({\bs R}) \big\rangle_{\rm ang}$ in Appendix B gives, 
\begin{equation}
\langle S_{ij} ({\bs R}) \rangle_{\rm ang} = \frac{1}{3}\left[S_{\rm ll}(R) + 2 S_{\rm nn} (R) \right] \delta_{ij},
\label{randomdirection}
\end{equation}
where $S_{\rm ll}$, $S_{\rm nn}$ are the longitudinal and transverse structure functions of the flow.
The dependence of $S_{{\rm T}ij}$ on $R$ indicates the crucial role of the particle separation backward in time.  
We estimate $R(\tau, \tau')$ as the geometric average of the particle distances, $d(\tau)$ and $d(\tau')$, at $\tau$ 
and $\tau'$, i.e., $R(\tau, \tau') = \sqrt{d(\tau)d(\tau')}$. A two-phase separation behavior consisting of a ballistic phase 
($d^2(\tau) = r^2+ \langle w^2 \rangle \tau^2$ with $\langle w^2 \rangle$ the 3D variance of the particle 
relative velocity) and a Richardson phase ($d(\tau)^2 \propto g \bar{\epsilon} |\tau|^3$ with $g$ 
the Richardson constant) is used for $d(\tau)$ as a function of $\tau$ (Paper I). We connect the two 
behaviors at a transition time, $\tau_{\rm c} = -(\tau_{\rm p1} + \tau_{\rm p2})/2$. The justification for the assumptions is described in Appendix B.   

Based on eq.\ (\ref{sij}), we give an approximate estimate for the shear term  ${\bs w}_{\rm s}$ and relate 
it to the spatial flow velocity difference, $\Delta {\bs u} (\ell)$ ($\simeq u (x +\ell) - u (x) $).  As shown in \S 3.24 of Paper I, 
the shear term can be written as ${\bs w}_{\rm s} \simeq \Delta {\bs u} (R_{\rm p}) [T_{\rm p}/(T_{\rm p} + \tau_{\rm p})]^{1/2}$ 
for equal-size particles with a friction time $\tau_{\rm p}$. Here $R_{\rm p}$, named   
the primary distance in PP10 and Paper I, is defined as $R_{\rm p} \equiv R(-\tau_{\rm p}, -\tau_{\rm p})$, 
which is of particular interest because the memory cutoff sets in at $\tau, \tau' \lsim -\tau_{\rm p}$.  
In paper I, we set $R_{\rm p} \simeq \langle w^2 \rangle^{1/2} \tau_{\rm p}$, 
which assumes the duration of the ballistic separation phase is not shorter than $\tau_{\rm p}$.  
The timescale $T_{\rm p}$ is the  flow correlation (or eddy turnover) time at the scale 
$R_{\rm p}$, i.e., $T_{\rm p} = T(R_{\rm p})$,  and the factor $[T_{\rm p}/(T_{\rm p} + \tau_{\rm p})]^{1/2}$ 
is due to the $\Phi_2$ term in eq.\ (\ref{sij}) that gives a constraint, $|\tau' -\tau| \lsim T(R)$, on the memories 
of the two particles that can contribute (Paper I).

In the bidisperse case, an expression for ${\bs w}_{\rm s}$ is more complicated 
due to the different friction times. %The $\Phi_2$ term  in eq.\ (\ref{sij}) gives a constraint, 
%$|\tau' -\tau| \lsim T(R)$, on the memories of the two particles that can contribute. 
In this case,  the $\Phi_2$ term tends to limit or reduce the temporal 
range of the large particle's memory (around the memory time of the 
smaller particle) that can contribute to ${\bs w}_{\rm s}$. Roughly speaking, $R_{\rm p}$ 
is primarily determined by the smaller particle, and a simple assumption%to be close to that of the smaller one. 
%This makes the estimate of  the primary distance more complicated.  
\footnote{A more accurate evaluation of $R_{\rm p}$ can be obtained as 
follows. One may first compute $R_{\rm l,h} = R(-\tau_{\rm p,l}, -\tau_{\rm p,h})$, with $\tau_{\rm p,l,h}$ 
the friction times of smaller and larger particles, and then compare 
$T(R_{\rm l,h})$ with $\tau_{\rm p,h} -\tau_{\rm p,l}$. If $T(R_{\rm l,h})$ is larger,  we set $
R_{\rm p} = R_{\rm l,h}$ and  $T_{\rm p} = T(R_{\rm l,h})$. Otherwise, we 
define $R_{\rm p}$ such that $R_{\rm p} =  R(-\tau_{\rm p,l}, -\tau_{\rm p,l} -T_{\rm p})$. 
Combining this with $T_{\rm p} = T(R_{\rm p})$, one can solve $R_{\rm p}$ and $T_{\rm p}$.  
%This definition of $R_{\rm p}$ accounts for the fact that, due to the $\Phi_2$ term, 
%the larger particle's memory  before  $-\tau_{\rm pl} -T_{\rm p}$ cannot contribute to ${\bs w}_{\rm s}$.
Using these estimates of $R_{\rm p}$ and $T_{\rm p}$, we find that 
the variance of eq.\ (\ref{approxshear}) provides a satisfactory approximation for the shear contribution, $\mathcal{S}_{ij}$.} 
would be $R_{\rm p} \simeq R(-\tau_{\rm p,l}, -\tau_{\rm p,l})$, where $\tau_{\rm p,l}$ is the friction 
time of the smaller particle. 
In analogy with the monodisperese case, we then have, 
\begin{equation}
{\bs w}_{\rm s} \simeq \Delta {\bs u} (R_{\rm p}) \left(\frac{T_{\rm p}}{T_{\rm p} + \tau_{\rm p,h} }\right)^{1/2}, 
\label{approxshear} 
\end{equation} 
where the last term corresponds to the reduction in the time range of the larger 
particle's memory that can contribute when $T_{\rm p} < \tau_{\rm p,h}$. 
Although the assumption above is rough, it provides a useful picture to understand the generalized shear term.  %Note that for the bidisperse case the acceleration term makes a 
%contribution to $R_{\rm p}$ in the ballistic separation phase.  

\subsection{Summary}

We briefly summarize our model for the  general bidisperse case. Using eqs.\  (\ref{accelform}), (\ref{sij}), 
and (\ref{randomdirection}), the 3D rms relative velocity can be calculated from, 
\begin{equation}
\langle w^2 \rangle = 3A + \mathcal{S}_{ii}, 
\label{w2}
\end{equation} 
where $A$ is given by eq.\ (\ref{accelform}) and,  
%\begin{equation}
%\mathcal{S}_{ii} =\int_{-\infty}^0  \frac {d\tau}{\tau_{\rm p1}}\int_{-\infty}^0 \frac{d\tau'}{\tau_{\rm p2}}
%\left[S_{\rm ll}(R) + 2 S_{\rm nn} (R) \right]  \Phi_2 \big(\tau' -\tau, R) \exp \left(\frac{\tau}{\tau_{\rm p1}} \right) \exp \left(\frac{\tau'}{\tau_{\rm p2}}\right).
%\label{sii}
%\end{equation}
\begin{gather}
\mathcal{S}_{ii} =\int_{-\infty}^0  \frac {d\tau}{\tau_{\rm p1}}\int_{-\infty}^0 \frac{d\tau'}{\tau_{\rm p2}}
\left[S_{\rm ll}(R) + 2 S_{\rm nn} (R) \right]  \Phi_2 \big(\tau -\tau', R)  \notag\\ \hspace{2cm}
\times\exp \left(\frac{\tau}{\tau_{\rm p1}} \right) \exp \left(\frac{\tau'}{\tau_{\rm p2}}\right).
\label{sii}
\end{gather}
Here $S_{\rm ll}$, $S_{\rm nn}$, and the timescale $T$ in $\Phi_2$ are given by eqs.\ (\ref{sll}), (\ref{snn}), and (\ref{Tr}) in Appendix B.
%$\mathcal{S}_{ii}$ is the contraction of eq.\ (\ref{sii}). 
%\begin{equation}
%\mathcal{A}_{ii}  = 3A = 3u'^2 \frac{(\Omega_2 -\Omega_1)^2 \left(\Omega_1 \Omega_2 + (\Omega_1 + \Omega_2) {\displaystyle \frac{z^2}{2} } \right)  }
%{(\Omega_1 + \Omega_2) \left(\Omega_1 + \Omega_1^2 + {\displaystyle \frac{z^2}{2}} \right) \left(\Omega_2 + \Omega_2^2 + {\displaystyle \frac{z^2}{2}}\right) },
%\label{aii}
%\end{equation}
%and,
%In the last equation, $S_{\rm ll}$ and $S_{\rm nn}$ are given by eqs.\ (\ref{sll}) and (\ref{snn}), and $\Phi_2$ is 
%fixed by eqs.\ (\ref{biexponential2}) and (\ref{Tr}). $R$ is evaluated from eq.\ (\ref{distance}) 
%using the assumed backward separation behavior (eqs.\ (\ref {ballistic}) and (\ref{richardson})). 
Eq.\ (\ref{w2}) is an implicit equation of $\langle w^2 \rangle$, as %the particle separation 
$R$ in eq.\ (\ref{sii}) depends on $\langle w^2 \rangle$ in the ballistic separation phase (see \S 3.2.2 and eq.\ (\ref{ballistic}) in Appendix B). 
We solve eq.\ (\ref{w2}) by an iteration method. Unlike the monodisperse case of equal-size particles, the particle 
separation rate in the ballistic phase (which is given by $\langle w^2 \rangle^{1/2}$) has a contribution from the generalized 
acceleration term. 

The generalized acceleration and shear terms can recover the S-T 
prediction in the small particle limit with $St_{1,2} \ll 1$ (see \S 3.2.1 and Appendixes A and B). 
Here we show that eq.\ (\ref{w2}) also correctly reproduces the prediction, eq.\ (\ref{largeparticles}), for the large particle 
limit, i.e., $\tau_{\rm p1, p2} \gg T_{\rm L}$ (or $\Omega_{1,2} \gg1$). We start with an analysis of the generalized 
shear term, i.e., eq.\ (\ref{sii}). We first note that  $S_{\rm ll} (R)$, $S_{\rm nn}(R)$, and $T(R)$  increase as $R$ increases backward in time 
toward the integral scale, $L$, of the turbulent flow. %The exponential cutoffs then suggest that the main contrition to the double integral in eq.\ (\ref{dii}) is from  $\tau \simeq  -\tau_{\rm p1}$ 
%and $\tau' \simeq -\tau_{\rm p2}$ if $|\tau_{\rm p1} -\tau_{\rm p2}| \le T_{\rm L}$. 
At $|\tau|, |\tau'| \gsim T_{\rm L}$, $R$ is expected to exceed $L$ (see Paper I), and 
$S_{\rm ll}(R)$, $S_{\rm nn}(R)$, and $T (R)$ become constant, i.e., $S_{\rm ll} (R)= S_{\rm nn}(R) = 2u'^2$, and $T (R) = T_{\rm L}$ 
(see eqs.\ (\ref{sll}), (\ref{snn}), and (\ref{Tr})). We thus have $[S_{\rm ll}(R) + 2 S_{\rm nn} (R)] \Phi_2(\tau -\tau', R) = 6u'^2 \Phi_{\rm L}(\tau-\tau')$ for  $\tau, \tau' \lsim -T_{\rm L}$.   
%If  $|\tau_{\rm p1} -\tau_{\rm p2}| \ge T_{\rm L}$, the correlation function, $\Phi_2$, provides a constraint,  $|\tau -\tau'| \le T_{\rm L}$, for the pairs 
%of $\tau$ and $\tau'$ that considerably contribute to the integral. In either case, the main contribution is from $|\tau|, |\tau'| \gg T_{\rm L}$, 
%and in these ranges of $\tau$ and $\tau'$, the particle distance $R$ is expected to be much larger than the integral scale, $L$ of the flow, meaning that $S_{\rm ll} = S_{\rm nn} = 2u'^2$, and $T (R) = T_{\rm L}$. 
If the cutoff timescales, $\tau_{\rm p1}$ and $\tau_{\rm p2}$, in eq.\ (\ref{sii}) are much larger than $T_{\rm L}$,  
one may approximate $\mathcal{S}_{ii}$ by $\simeq 6 u'^2 \int_{-\infty}^0  \frac {d\tau}{\tau_{\rm p1}}\int_{-\infty}^0 \frac{d \tau'} {\tau_{\rm p2}}
\Phi_{\rm L} (\tau -\tau') \exp (\tau/\tau_{\rm p1}) \exp ({\tau'}/{\tau_{\rm p2}})$, which approaches  
%$\mathcal{S}_{ii} \to 
$12u'^2 /(\Omega_1 + \Omega_2)$ in the limit  $\Omega_{1,2} \gg 1$. Since $3A = 3u'^2(\Omega_1-\Omega_2)^2/[\Omega_1 \Omega_2 (\Omega_1+\Omega_2)]$ at $\Omega_{1,2} \gg1$, 
we have $\langle w^2 \rangle = 3A + \mathcal{S}_{ii}  \to 3u'^2(1/\Omega_1 + 1/\Omega_2)$, consistent with 
eq.\ (\ref{largeparticles}) in \S 3.1. This proves that  the large particle limit  is correctly recovered in our model.

\section{Numerical Simulation}

We use the same simulation data as in Paper I. We briefly describe the simulation, and refer the reader to Paper I for details. Using the Pencil code, 
we carried out the simulation in a periodic $512^3$ box with a length of $2\pi$ on each side. An isothermal equation of state was adopted. 
The flow was driven and maintained by a large-scale force using wave numbers $1\le k \le 2$. 
After reaching steady state, the 1D rms velocity of the flow is $u'\simeq 0.05$ 
in units of the sound speed. The 3D rms velocity of the flow 
corresponds to a Mach number of 0.085, suitable for applications to protoplanetary disks. 
At  a Mach number $\simeq 0.1$, the compressibility is very low, and the flow statistics 
of all orders are essentially identical to incompressible turbulence (Pan \& Scannapieco 2011).
%The simulated flow is fully developed and reaches a steady state at $t_{\rm dev} \simeq 5-10 T_{\rm eddy}$. 
The integral length, $L$, of the flow is about 1/6 box size, while the Kolmogorov length scale $\eta \equiv (\nu^3/\bar{\epsilon})^{1/4}$ is 
estimated to be 0.6 cell size of the grid, so that $L \simeq140\eta$. The Taylor Reynolds number of the 
flow is $Re_{\lambda}\simeq 200$, and the regular Reynolds number is $Re \simeq1000$.  
The Kolmogorov velocity, $u_\eta \equiv (\nu \bar{\epsilon})^{1/4}$, is related to the rms  
by $ u' = (Re_\lambda/\sqrt{15})^{1/2} u_\eta \simeq 7u_\eta$.
%in units of the sound speed.   

A computation of the large eddy turnover time, $T_{\rm eddy} \simeq L/u'$, and the Kolmogorov time, 
$\tau_\eta \equiv (\nu/\bar{\epsilon})^{1/2} $, shows that $T_{\rm eddy} \simeq 19.2 \tau_\eta$.
%and ($v_\eta$) velocity scales using the viscosity, 
%The Kolmogorov timescale is estimated to be $\tau_\eta \equiv (\nu/ \bar{\epsilon})^{1/2} = 1.04$. 
By integrating trajectories of tracer particles ($St=0$), we measured the 
Lagrangian correlation function, $\Phi_{\rm L}$, which is well fit by the 
bi-exponential form, eq.\ (\ref{biexponential}), with $T_{\rm L} = 14.4  \tau_\eta $ 
and $z=0.3$ (Paper I). The timescale, $T_{\rm E}$, for the Eulerian temporal 
correlation was estimated to be $15.9 \tau_\eta$. The near equality of  $T_{\rm E}$ with 
the Lagrangian correlation time $T_{\rm L}$ justifies the use of the Lagrangian correlation function 
for all particles  (Paper I). In Appendix C, we show the Lagrangian and Eulerian 
temporal structure functions, $D_{\rm L}$ and $D_{\rm E}$, which are of direct relevance 
for the particle-flow relative velocity and the generalized acceleration term.

%Both structure functions are expected to become smooth at $\Delta \tau \ll \tau_\eta$.
%The circles in the figure correspond to the Eulerian time structure function.  
%As discussed in \S 2, the particle-flow relative velocity can be roughly estimated as the 
%flow velocity difference seen by the particle at a temporal separation of $\tau_{\rm p}$. 
%Therefore, an examination of  $\Delta { \bs u}_{\rm L} (\Delta \tau)$ would be helpful to 
%understand the relative velocity between the flow and small particles, assuming that these particles more 
%or less follow Lagrangian trajectories. 

%For a large particle with $\tau_{\rm p} \gsim T_{\rm L}$, the temporal 
%series of the flow velocity along its trajectory may be better described as Eulerian 
%(Paper I). We thus also consider the temporal velocity difference, 
%as the contribution of the acceleration term is related to the 
%temporal flow velocity difference seen by the particles. 

We included 14 species of particles of different sizes in our simulated flow, 
each containing 33.6 million particles. The friction timescale, $\tau_{\rm p}$, 
of the smallest particles is $\simeq 0.1\tau_\eta$, while the largest particles have 
$ \tau_{\rm p}  =  54 T_{\rm L}$, corresponding to $St=795$.  %The friction timescales of the 
%14 species are equally spaced, increasing by a factor of two in each successive 
%species. 
Spanning 4 orders of magnitude, the $\tau_{\rm p}$ range covers the entire 
length scale range  of the simulated flow.  %Initially, the particles were randomly distributed in the 
%simulation box, and each velocity component of a particle was independently drawn from a 
%uniform distribution between -0.01 and 0.01 sound speed. 
When integrating the particle equation of motion (eq.\ \ref{particlemomentum}), we 
interpolated the flow velocity inside computational cells using the triangular-shaped-cloud (TSC) 
method %, already implemented in the Pencil code 
(Johansen and Youdin 2007).  An accurate integration of the particle trajectories 
requires an integration time step smaller than $\tau_{\rm p}$ by a factor of $\gsim 10-20$ (see, e.g., Ayala et al.\ 2008). 
The time step in our simulation is about $\simeq 0.01$ Kolmogorov time, which was set by the Courant condition for the flow 
evolution. The ratio of this time step to the friction time of the smallest particles in our simulation is  $\simeq 1/10$, 
close to the required value for accurate trajectory integration.         
%A justification for the 
%TSC interpolation method is given in Paper I.  
%The integration of the particle trajectories was computationally very expensive, and our run 
%costed 1.4 million CPU hours. 
%We evolved the hydrodynamic equations and integrated the particle trajectories together right 
%from the beginning of the simulation. This means that, when the particles were released, the turbulent 
%flow had not been developed. This is, however, not a problem because we choose to measure the particle 
%statistics after the steady state is reached.   
Our simulation run lasted $26 T_{\rm eddy}$, and, 
%we analyzed the temporal evolution of the 1-particle rms velocity, the number counts of 
%particle pairs at given distances ($r\lsim 1\eta$), and the rms relative velocity of the pairs. 
at the end of the run, all the statistical measures reached a quasi steady-state and 
the dynamics of all particles was relaxed (Paper I). 
%Even for the largest particles whose friction time ($\tau_{\rm p}  = 41 T_{\rm eddy}$ 
%or $54 T_{\rm L}$) is larger than the run time of the simulation, we found sufficient evidence 
%for their dynamical relaxation (Paper I). 
%As in Paper I,   
%At late times, even the largest particles are relaxed,  and the statistics of three well-separated snapshots are largely independent. 
%A detailed discussion and justification for the snapshot selection is given in Paper I. 
The simulation was computationally intensive. Using 4096 cores on the 
NASA/Ames Pleiades supercomputer, the run lasted 14 days, costing 1.4 million CPU hours. 
In our data analysis, we use several well-separated snapshots toward the end of the simulation.

\section{The Particle-Flow Relative Velocity}

\begin{figure}[h]
\centerline{\includegraphics[width=1.1\columnwidth]{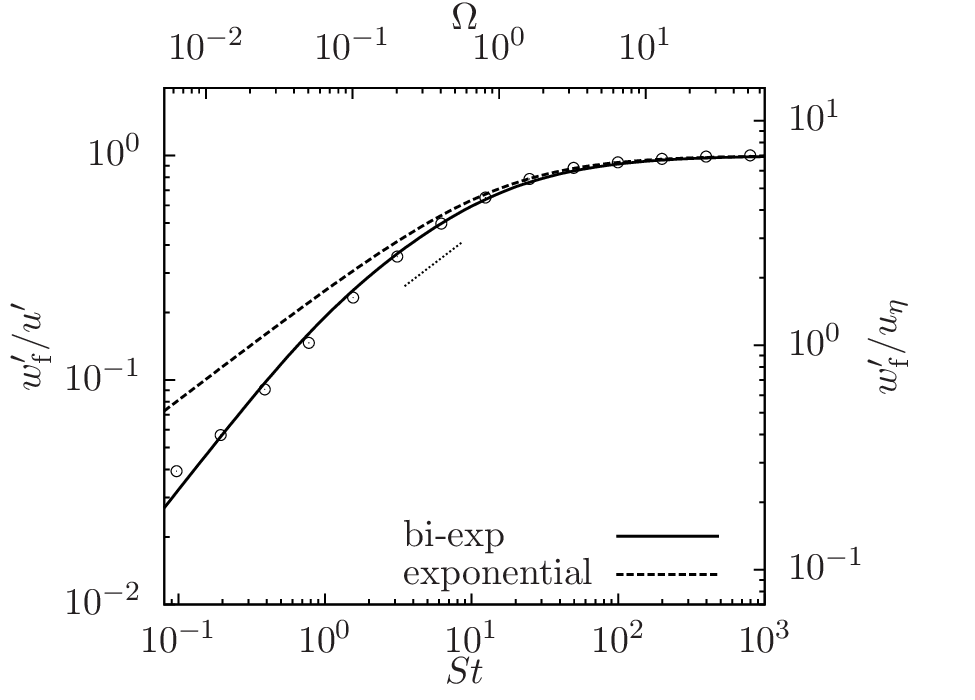}}
\caption{The 1D rms  of the particle-flow relative velocity as a function of the 
particle Stokes number, $St$.  The top X-axis normalizes $\tau_{\rm p}$ 
to the Lagrangian correlation time, $T_{\rm L}$. %The relative velocity is normalized to the rms ($u'$) and the Kolmorgrov velocity ($u_\eta$) of the 
%flow on the left and right Y-axises, respectively. 
The dash and solid lines are model predictions, eqs.\ (\ref{pfrelativesingle}) and (\ref{pfrelativebi}), 
using single- and bi- exponential forms for $\Phi_{\rm L}$ (and hence for $\Phi_1$), 
respectively. In both lines, $T_{\rm L}$ is taken to be 14.4 $\tau_{\eta}$. For the 
bi-exponential case, the parameter $z$ is set to 0.3. %The dashed line considerably overestimates $w^{\prime f}$ for $St \lsim 3$ particles.   
The dotted line segment denotes a $St^{1/2}$ scaling.}
\label{gaspartrms} 
\end{figure}

In Fig.\  \ref{gaspartrms}, we show the simulation results for the 1D rms of the particle-flow relative velocity, ${\bs w}_{\rm f}$. To compute ${\bs w}_{\rm f}$, 
we interpolated the flow velocity at the position of each particle with the same TSC method 
used in the simulation. The computed relative velocity is for zero particle-flow distance.  
The bottom and top axes normalize the particle friction time to the Kolmogorov and 
Lagrangian correlation times, respectively. The rms relative velocity, $w'_{\rm f}$, is 
normalized to the flow rms ($u'$) and the Kolmogorov velocity  ($u_\eta$) on the 
left and right Y-axises, respectively. Similar normalizations are used in most figures in the 
rest of the paper. 

Eq.\ (\ref{pfrelative2}) from our model suggests that ${\bs w}_{\rm f}$ is estimated by the 
temporal flow velocity difference $\Delta_{\rm T} {\bs u}$ along the particle trajectory at a time lag of $\tau_{\rm p}$. 
$\Delta_{\rm T} {\bs u}$ may be approximated by the Lagrangian ($\Delta_{\rm L} {\bs u}$) and Eulerian 
($\Delta_{\rm E} {\bs u}$) temporal velocity differences for particles in the $St\ll 1$ and $\tau_{\rm p} \gg T_{\rm L}$ 
limits, respectively. For a better understanding of  ${\bs w}_{\rm f}$,  we show the Lagrangian and Eulerian 
temporal structure functions ($D_{\rm L}$ and $D_{\rm E}$) in Fig.\ \ref{temporalstructure} of Appendix C.   
%that ${\bs w}_{\rm f}$ is estimated by $\Delta_{\rm T} {\bs u}$. which XXX under the assumption $D_{{\rm T}ij}\simeq D_{{\rm L}ij}$.
As argued in Appendix C, the behaviors of $\Delta_{\rm L} {\bs u}$ and $\Delta_{\rm E} {\bs u}$ 
suggest that one may approximate  $\Delta_{\rm T} {\bs u}$ by the Lagrangian 
velocity difference for particles of any sizes. 
Comparing with Fig.\ \ref{temporalstructure} in Appendix C, 
we see that  ${w'}_{\rm f}$ as a function of $St$ is similar to the Lagrangian 
velocity difference 
%$D_{\rm L}^{1/2}(\Delta \tau)$ 
as a function of the time lag, %$\Delta \tau$ 
confirming the physical picture of our model. 

The  solid line in Fig.\ (\ref{gaspartrms}) is our model prediction, eq.\ (\ref{pfrelativebi}), 
using a bi-exponential form for the Lagrangian correlation function $\Phi_{\rm L}$ (and hence 
for $\Phi_1$; see \S 2). In $\Phi_{\rm L}$, we set $T_{\rm L}$ to 14.4 $\tau_\eta$ and 
the parameter $z$ to  0.3. These values of $T_{\rm L}$ and $z$  were obtained 
from the best fit of eq.\ (\ref{biexponential}) to the Lagrangian correlation function measured 
in our simulated flow (see Fig. 2 of Paper I and Fig.\ \ref{temporalstructure} in Appendix C of the current paper). 
As mentioned in \S 2, the parameter $z$ has a dependence on 
the flow Reynolds number, $Re$, and, considering $Re\simeq 10^3$ in our flow, one can estimate $z$  in 
protoplantary turbulence using appropriate values of $Re$ and the scaling $z \simeq 0.3 (Re/10^3) ^{-1/4}$. 
%slightly larger than the directly measured value ($14.4 \tau_\eta$) in our simulated flow (Paper I). 
%as derived in Paper I from the best fits to the measured Lagrangian 
%correlation 
% from the Lagrangian structure functions. 
%using the bi-exponential form, eq.\ (\ref{biexponential}), of $\Phi_{\rm L}$. 
The solid line fits the data well except for the smallest particles.
 
The dashed line corresponds to eq.\ (\ref{pfrelativesingle}) based on a single 
exponential $\Phi_{\rm L}$.  Here $T_{\rm L}$ is also set to $14.4 \tau_\eta$. 
The fitting quality of eq.\ (\ref{pfrelativesingle}) cannot be improved 
by tuning $T_{\rm L}$. As discussed in \S 2, eq.\ (\ref{pfrelativesingle}) predicts 
a $St^{1/2}$ scaling for $w_{\rm f}^{\prime}$ for all particles with $\Omega \ll1$. 
This prediction is inaccurate, and the scaling of $w_{\rm f}^{\prime}$ is significantly 
steeper than $St^{1/2}$ at $St \lsim 3$. This discrepancy is due to the fact that the single exponential 
form does not reflect the smooth part of  $\Phi_{\rm L}$ (or $D_{\rm L}$) 
at time lags, $\Delta \tau$, below the Taylor micro timescale, $\tau_{\rm T}$. 
At  $\Delta \tau < \tau_{\rm T}$, $\Phi_{\rm L}$ is affected by the flow acceleration. 
The Taylor timescale was found to be 4.3 $\tau_\eta$ in our flow (Paper I), and this 
explains the deviation of the dotted line from the data points at $St \lsim 3$.   
Again, the steepening of $w_{\rm f}^{\prime}$ below $\tau_{\rm p} \lsim \tau_{\rm T}$ 
corresponds to the transition from the inertial range of the temporal flow velocity 
difference, $\Delta {\bs u}_{\rm T}$, to the dissipation range, which is not  captured 
by the single-exponential correlation.

The $St^{1/2}$ scaling applies only to inertial-range particles.  
%as predicted by our model (eq.\ (\ref{pfrelativebi})) with a bi-exponential $\Phi_{\rm L}$.  
The dotted line segment in Fig.\ \ref{gaspartrms} represents such a
scaling at intermediate $St$. The data points do not show a clear 
$St^{1/2}$ range probably due to the limited inertial range of the simulated 
flow. A confirmation of the predicted $St^{1/2}$ scaling would require 
larger resolutions. For small particles in the $St \ll 1$ limit, 
$w'_{\rm f}$ is predicted to be equal to 
$a \tau_{\rm p}$ (see \S 2), suggesting a linear increase of 
$w_{\rm f} ^{\prime}$ with $St$.  The linear scaling is not observed 
even at the smallest $St$($=0.1$) in our simulation. 
%From eq.\ \ref{}, we see that 
The linear scaling of $w_{\rm f}^{\prime}$ is based on the assumptions that tiny 
particles closely follow Lagrangian trajectories and that  the Lagrangian structure 
function, $D_{\rm L}$, scales with the time lag as $D_{\rm L} \propto (\Delta \tau)^2$ 
(see eq.\ (\ref{pfrelative2})). However, the scaling of $D_{\rm L}$ was found to be 
shallower than $\propto (\Delta \tau)^2$ at $\Delta \tau \gsim 0.1 \tau_\eta$ 
(Fig.\ \ref{temporalstructure} in Appendix C). This explains why the scaling of $w_{\rm f}^{\prime}$ 
is slower than $\propto St$ for $St \gsim 0.1$. %As mentioned in \S 4.1,  
%an exact  $(\Delta \tau)^2$ scaling of $D_{\rm L}$ may appear at $\Delta \tau < 0.1 \tau_\eta$. 
%In that case, one may be able 
To verify the linear scaling of  $w_{\rm f}^{\prime}$,  one needs to  include 
smaller particles with $St \ll 0.1$. 
The data point at $St =0.1$ is significantly larger than our model 
prediction with a bi-exponential $\Phi_{\rm L}$, %which makes it questionable whether the expected linear 
%scaling would really recover at $St \ll 0.1$. 
%%Reconsider
%One reason for the discrepancy is that the bi-exponential fitting curve in Fig.\ \ref{temporalstructure} underestimates for the 
%Lagrangian structure function $D_{\rm L}$ at small time lag $\Delta \tau$. %The agreement of the model prediction and the data for 
%$w_{\rm f}^{\prime}$ may improve if a better fitting function 
%for $D_{\rm L}$ at small $\Delta \tau$ is adopted. 
and the reason may be that the trajectory computation for 
the smallest particles in our simulation is the least accurate. 
This is because for these small particles the ratio of the simulation time step to the friction time is 
the largest, $\simeq 1/10$, which only marginally 
meets the requirement for accurate trajectory integration (see \S 4). To order to   
improve the accuracy  for the $St \lsim 0.1$ particles, we will adopt a smaller time
step for the  trajectory integration of the smallest particles in future simulations.  
%the numerical accuracy in the trajectory integration is less accurate for smaller particles, 
%as the ratio of the integration time step to the friction timescale is smaller. 
%A future simulation with a smaller time step may provide a correction to the data point at $St=0.1$. 
%Reconsider: With a more accurate simulation including more particles with smaller $St$, the expected linear scaling may be confirmed below 
%$St=0.1$.  

%%\begin{figure}[t]
%%\centerline{\includegraphics[height=2.4in]{f2b.eps}}
%\caption{The 3D rms particle-flow relative velocity at distances $r=0$ (circles), $\frac{1}{2}$ (triangles) 
%%and $1 \eta$ (diamonds). 
%The dash and solid lines are model predictions, eqs.\ (\ref{pfrelativesingle}) and (\ref{pfrelativebi}), 
%using single- and bi- exponential forms for $\Phi_{\rm L}$ (and hence for $\Phi_1$), 
%respectively. In both lines, $T_{\rm L}$ is taken to be 15 $\tau_{\eta}$. For the 
%bi-exponential case, the parameter $z$ is set to 0.3. %The dashed line considerably overestimates $w^{\prime f}$ for $St \lsim 3$ particles.   
%The dotted line segment denotes a $St^{1/2}$ scaling.
%Right panel: the 3D rm particle relative velocity at distance $r=0$ (circles), $\frac{1}{2}$ (triangles) and $1 \eta$ (diamonds). 
%Lines are our model predictions. 
%using bi-exponential $\Phi_1$ with $z=0.3$ and $T_{\rm L} =14.4 \tau_\eta$. 
%The data points (circles) and the solid line for $r=0$ have been shown in Fig.\ \ref{gaspartrms}.}
%\label{gaspartrms2} 
%\end{figure}

\section{The Particle Relative Velocity}

To compute the particle relative velocity, we search pairs of particles from all the different species at given distances, 
$r=1$, $\frac{1}{2}$,  and $\frac{1}{4} \eta$.
For each particle (1), we count particles (2) in a distance shell 
$[r-dr/2, r+dr/2]$. The shell thickness $dr$ $(\ll r)$ is chosen such that the 
number of particle pairs available is of the order of $\gsim 10^4$,  sufficient for  the measurement of the rms relative velocity at an accuracy level of 
a few percent.
On the other hand, at $r\lsim \frac{1}{8} \eta$, the measurement accuracy becomes poor due to 
 the limited number of particle particles available at small distances (Paper I). We thus limit our measurements to $r\gsim \frac{1}{4} \eta$. 
Future simulations including a much larger number of particles will allow accurate measurements at $r\lsim \frac{1}{8} \eta$, and 
help resolve the collision statistics at smaller scales toward the particle size.
%set to $0.08r$, $0.08r$, and $0.16 r$ and for $r=1$, $\frac{1}{2}$, and $\frac{1}{4}\eta$, respectively. 
%The number of particle pairs at $r=\frac{1}{4}\eta$ with $dr= 0.16 r$ is of the order of $10^4$, 
%which is sufficient for  an accurate measurement of the rms relative velocity.  %Smaller $r$ values are desirable, but that would require including a larger 
%number of particles to ensure accurate statistics.  
%We find that the number of pairs at $r=\frac{1}{8}\eta$ is already limited and is too small to study the high-order statistics such as the PDF tails. 
%We will only measure the first-order moment of the relative velocity (and the collision kernel) 
%at this distance. To increase the number of pairs,  we set $\delta r = 0.32 r$ for $r=\frac{1}{8}\eta$.   
We calculate the 3D amplitude, $|\bs w|$, of the relative velocity 
and its radial component, $w_{\rm r} = {\bs w} \cdot {\bs r}/r$, for each pair 
and average over all pairs to obtain the rms values, $\langle w^2 \rangle^{1/2}$, and $\langle w_{\rm r}^2\rangle^{1/2}$. The rms speed 
in a tangential direction is calculated as $\langle w_{\rm r}^2\rangle^{1/2} = ( \langle w^2 \rangle - \langle w_{\rm r}^2\rangle)^{1/2}/\sqrt{2} $. 

 \begin{table}[h]
\begin{center}
\caption{Properties of the simulated flow and the model parameters}
\label{tbl-1}
\vspace{3mm}
\begin{tabular}{rl|rl}
\hline
\hline
$Re$ & 1000 &$C_{\rm K}$ \tablenote{Used in Eq.\ (\ref{sll}) for the longitudinal structure function.}  &  2             \\
$u'/u_\eta$ &  7 & $C_{\rm Kn}$ \tablenote{Used in Eq.\ (\ref{snn}) for the transverse structure function.} &  2.5       \\
$L/\eta$  & 140  & $C_{\rm T}$  \tablenote{Used  in Eq.\ (\ref{Tr}) for the eddy timescale as a function of length scale.}    & 0.4          \\
$T_{\rm eddy}/\tau_\eta$ &  $19.2$  &$\tau_{\rm c}$   \tablenote{Assumed transition time from the ballistic separation to the Richardson behavior.} & $-(\tau_{\rm p1} + \tau_{\rm p2})/2$ \\
%$\tau_{\rm T} $ & $4.3 \tau_\eta$&\\
$T_{\rm L}/\tau_\eta$ &  $14.4$& $g$    \tablenote{Adopted Richardson constants that best fit $\langle w^2 \rangle^{1/2}$ at different $r$.} &  $1.6 (r=1\eta) $         \\
$\tau_{\rm T} /\tau_\eta$ & $4.3$    &                                               &  $1.3(r=\frac{1}{2}\eta)$     \\
$z$  & 0.3 &                                              &  $1.0(r=\frac{1}{4}\eta)$     \\
\hline 
\hline
%\vspace{-1.3cm}
%\tablenotetext{(1)}{Used in Eq.\ (\ref{sll}) for the longitudinal structure function.}
%\tablenotetext{(2)}{Used in Eq.\ (\ref{snn}) for the transverse structure function.}
%\tablenotetext{(3)}{Used  in Eq.\ (\ref{Tr}) for the eddy timescale as a function of length scale.}
%\tablenotetext{(4)}{Assumed transition time from the ballistic separation to the Richardson behavior.}
%\tablenotetext{(5)}{Adopted Richardson constants that best fit $\langle w^2 \rangle^{1/2}$ at different $r$. }
\end{tabular}
\end{center}
\end{table}
  
%\begin{table}
%\begin{center}
%\caption{Properties of the simulated flow and the model parameters}
%\label{tbl-1}
%\vspace{3mm}
%\begin{tabular}{cc}
%\tableline\tableline
%$Re$ & 1000 \\
%$u'/u_\eta$ &  7 \\ 
%$L/\eta$  & 140  \\
%$T_{\rm L}/\tau_\eta$ &  $14.4$ \\
%$\tau_{\rm T} $ & $4.3 \tau_\eta$&\\
%$z$  & 0.3 \\
%$C_{\rm K}${\tablenotemark{(1)}}   &  2             \\
%$C_{\rm Kn}${\tablenotemark{(2)}} &  2.5       \\
%$C_{\rm T}${\tablenotemark{(3)}}    & 0.4          \\
%$\tau_{\rm c}${\tablenotemark{(4)}} & $-(\tau_{\rm p1} + \tau_{\rm p2})/2$ \\
%$g${\tablenotemark{(5)}} &  $1.6 (r=1\eta) $         \\
% &  1.3$(r=\frac{1}{2}\eta)$     \\
% &  1.0$(r=\frac{1}{4}\eta)$     \\
%\tableline \tableline
%\vspace{-1.3cm}
%\tablenotetext{(1)}{Used in Eq.\ (\ref{sll}) for the longitudinal structure function.}
%\tablenotetext{(2)}{Used in Eq.\ (\ref{snn}) for the transverse structure function.}
%\tablenotetext{(3)}{Used  in Eq.\ (\ref{Tr}) for the eddy timescale.}
%\tablenotetext{(4)}{Assumed transition time from the ballistic separation to the Richardson behavior.}
%\tablenotetext{(5)}{Adopted Richardson constants that best fit $\langle w^2 \rangle^{1/2}$ at different $r$. }
%\end{tabular}
%\end{center}
%\end{table}

In PP10, we showed that our model prediction for the rms relative velocity 
is in good agreement with the simulation data of Zhou et al.\ (2001) for 
the bidisperse case. Here we test the model more systematically using our simulation (at 
higher resolution than Zhou et al.). To compare with the data, we first fix the Stokes number, 
$St_1$, of particles (1), and examine the relative 
velocity as a function of $St_2$. In \S 6,2, we will study 
the rms relative velocity at fixed Stokes number ratios, $f \equiv St_{\ell} /St_{h}$. 

In Table I, we summarize the properties of the simulated flow and the parameters  adopted in our model. The measurements and choices 
of the parameters are presented in details in Paper I, to which we refer the reader for details. For example, as mentioned 
earlier, $T_{\rm L}$ and $z$ are obtained by fitting eq.\ (\ref{biexponential}) to the measured Lagrangian correlation function, 
$\Phi_{\rm L}$, while $C_{\rm K}$ and $C_{\rm Kn}$ are measured from the structure functions of the flow. 
The choice of $\tau_{\rm c}$ is explained in Appendix B, and a detailed discussion on 
the selection of $g$ values at different $r$ is given in \S 6.3.

\subsection{The rms relative velocity at fixed $St_1$}

\begin{figure}[h]
\centerline{\includegraphics[width=1.1\columnwidth]{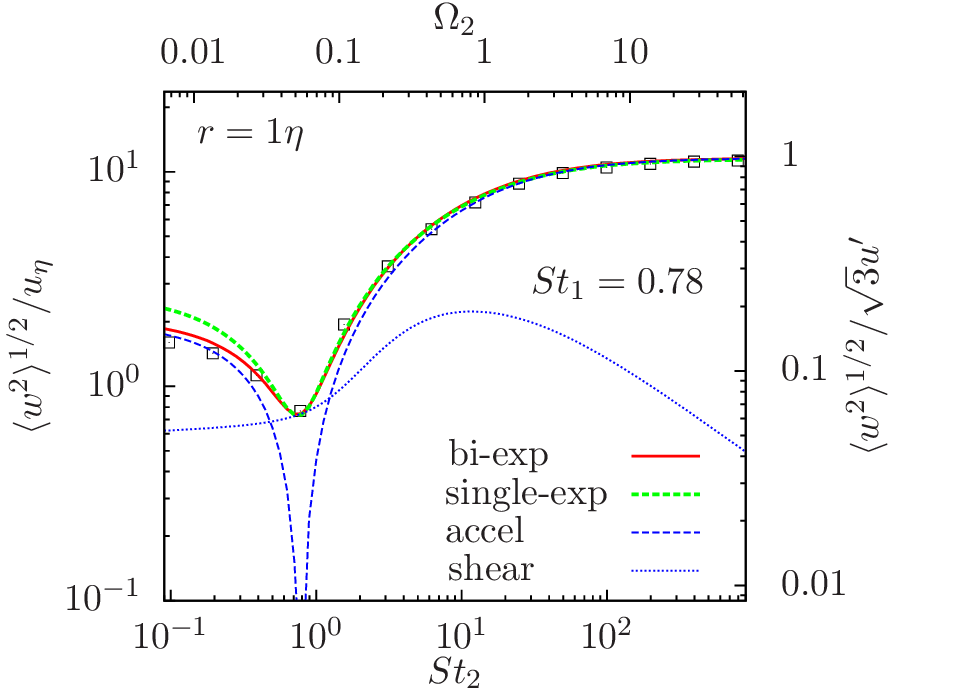}}
\caption{3D rms relative velocity as a function of $St_2$ for $St_1= 0.78$. %The data points 
%are simulation results measured at $r=1\eta$. 
Red and green lines are our model predictions 
using bi- and single- exponential temporal correlation functions ($\Phi_1$ and $\Phi_2$), 
respectively.  A two-phase separation is adopted for the generalized shear contribution. 
The ballistic phase is assumed to last for $(\tau_{\rm p1} + \tau_{\rm p2})/2$, 
and $g$ is set to $1.6$ for the Richardson phase. The blue dashed and dotted lines correspond to 
the contributions by the generalized acceleration ($\sqrt{3A}$) and shear ($\sqrt{{\mathcal S}_{ii}}$) 
terms in our model with bi-exponential temporal correlations.}
\label{3drms3} 
\end{figure}

In Fig.\ \ref{3drms3}, we plot the 3D rms relative velocity, $\langle w^2 \rangle^{1/2}$, 
for $St_1 = 0.78$. The data points are the simulation result for a particle 
distance of $r=1\eta$. Clearly,  $\langle w^2 \rangle^{1/2}$ shows a dip at $St_2 =St_1$. 
As discussed in PP10, the correlation between the velocities of equal-size  particles is 
stronger than that between different ones.  A higher velocity correlation corresponds to a 
smaller relative velocity,  leading to the formation of a dip at $St_2 \simeq St_1$.  
%which decreases with increasing velocity correlation.  

The solid red line is the prediction of our model using a bi-exponential form for both the 
1-particle and the 2-particle temporal trajectory correlations, $\Phi_1$ and $\Phi_2$.  %along the trajectory (or trajectories) of one or two particles. 
As explained above, we set the Lagrangian correlation timescale to $14.4 \tau_\eta$, and 
the parameter $z$ to $0.3$. %in both $\Phi_1$ and $\Phi_2$. 
%which are obtained  from the direct measurement of the Lagrangian correlation function, $\Phi_{\rm L}$ 
%(Paper I; see also Appendix B). 
For the calculation of the shear contribution, 
%we made use of eq.\ (\ref{Tr})  for the timescale, $T(\ell)$, 
%as a function of the eddy size, $\ell$, in $\Phi_2$, and eqs.\ (\ref{sll}) and (\ref{snn}) 
%for the spatial structure functions of the flow. 
we adopt a two-phase backward-in-time  separation behavior  consisting  of an initial  ballistic phase  
and a Richardson phase. The motivation and the justification for the assumed behavior can be found  
Paper I and Appendix B of the current paper.  The ballistic phase is assumed to last for a duration of 
$(\tau_{\rm p1} +\tau_{\rm p2})/2$ for particles of different sizes,  and then  connect to  a Richardson 
phase with a Richardson constant  of $g  = 1.6$.    
The same separation is used in all our model predictions for $r=1\eta$ 
in the rest of the paper. 
%The ballistic phase is assumed to l 
%As mentioned earlier, the exact separation behavior for particles of different 
%sizes is unknown, and the adopted behavior is simply a assumption. 
When $St_2 =St_1$, the assumed separation behavior reduces to that 
of identical particle pairs discussed in Paper I. %with a ballistic separation for a duration of $\tau_{\rm p}$ 
%and a Richardson phase 
The choice of  $g=1.6$ is because, as shown in Paper I, it gives a successful fit to the 
simulation data for the relative velocity of equal-size particles at $r=1\eta$ . As mentioned  earlier, a 
difference for the pair separation of different particles 
%the generalized shear term has a difference 
from the case of equal-size particles is that the separation rate in the ballistic phase has a contribution 
from the generalized acceleration term. The red line is in good agreement with the data, supporting the physical 
picture of our model for the bidipserse case. The dip center in the red line corresponds to 
our prediction for the monodisperse case with $St =0.78$. 
We point out that the separation behavior we used for particles of different sizes is largely an educated guess. 
However, the exact behavior turns out to be unimportant for very different particles as the main contribution 
to their relative velocity is the generalized acceleration term (see below).

The blue dashed and dotted lines in Fig.\ \ref{3drms3} correspond to the generalized 
acceleration ($\sqrt{3A}$) and shear ($\sqrt{\mathcal{S}_{ii}}$) terms in our model with 
bi-exponential temporal correlations. As discussed in \S 3, the acceleration term vanishes for 
identical particles, and this is responsible for the dip of $\langle w^2 \rangle^{1/2}$ at 
$St_2\simeq St_1$.  It increases the relative velocity on both sides of the dip.
The generalized shear term first increases with $St_2$, and then 
turns over and decreases at large $St_2$. 
This can be understood from %eq.\ (\ref{sii}) for $\mathcal{S}_{ii}$ or 
the approximate equation (\ref{approxshear}) for the shear term, ${\bs w}_{\rm s}$. 
%allows the contribution from larger $|\tau'|$ to the double integral in eq.\ (\ref{sii}). This 
%results in an increase of $\mathcal{S}_{ii}$, because $S_{ll}$, $S_{nn}$, and $T(R)$ in the 
%integrand all increase with the particle separation, $R$, and hence with increasing $|\tau'|$. 
%Alternatively, this increase of $\mathcal{S}_{ii}$ could be viewed 
The primary distance, $R_{\rm p}$, in eq.\ (\ref{approxshear}) is mainly controlled 
by the smaller particle, and, for $St_2 <St_1$, it increases with increasing $St_2$. 
The increase of $R_{\rm p}$ and/or $T_{\rm p}$ leads to the increase of the shear 
term for $St_2$ below $St_1$. 
%and/or the timescale $T_{\rm p}$ in eq.\ (\ref{approxshear}), which 
On the other hand,  as $St_2$ exceeds $St_1$, particle (1) becomes the smaller particle, 
%sets in and affects the range of $\tau'$ that may contribute to the double integral.  
%$\Phi_2$ tends to limit the range of the larger particle's memory that 
%contributes to the double integral to be within a timescale of $\simeq T_{\rm p}$ 
%from the memory time, $\simeq - \tau_{p1}$, of the smaller particle.  
and, at sufficiently large $St_2$,  both $R_{\rm p}$ and  $T_{\rm p}$ become 
independent of $St_2$ and approach constants.
After that,  eq.\ (\ref{approxshear}) indicates that the shear contribution, 
${\bs w}_{\rm s}$ would finally decrease as $St_2^{-1/2}$, as seen in Fig.\ \ref{3drms3}.  
This decrease corresponds to the reduction in  the range of the memory of particle (2) 
that can contribute to $\mathcal{S}_{ii}$ by the $\Phi_2$ term in eq.\ (\ref{sii}).

%\footnote{Following the same reasoning as in \S 3.2.4 of  Paper I,  when the particle separation, 
%$R(-\tau_{\rm p1}, -\tau_{\rm p2})$, at $\tau=-\tau_{\rm p1}$ and  $\tau'=-\tau_{\rm p2}$ is larger than the integral scale of 
%the flow, the $\Phi_2$ term would provide a factor, $T_{\rm L}/\tau_{\rm p2}$, which tends to cause $\mathcal{S}_{ii}$ 
%to decrease with increasing $St_2$.  See a detailed discussion on the role of $\Phi_2$ in Paper I.}
%Therefore, $\mathcal{S}_{ii}$ finally drops  with increasing $St_2$.  

In the green dashed line, we used a single-exponential form for both $\Phi_1$ and $\Phi_2$. 
The green dashed line is higher than the red solid line, and overestimates the data points at 
small $St_2$ ($\lsim 0.2$). We find that the generalized shear contribution is insensitive to 
the function form of $\Phi_2$, and replacing  $\Phi_2$ with a bi-exponential form does not 
result in a significant difference in the shear contribution\footnote{Similar to the monodisperse 
case in Paper I, when integrated, the dependence of the double integral in eq.\ (\ref{sii}) 
of the generalized shear term on $\Phi_2$ condenses to a dependence just on the timescale 
$T(R)$ in $\Phi_2$.}. The difference between the red and green lines at small $St_2$ is 
mainly due to the generalized acceleration term. The green line makes use of  eq.\ (\ref{accelsolutionsingle}) 
for the acceleration contribution, which ignores the $z^2/2$ terms in eq.\ (\ref{accelsolution}) 
based on the bi-exponential $\Phi_1$. In our simulated flow, $z^2/2=0.045$, and we have $\Omega  \lsim 0.07$ 
for particles with $St  \lsim 1$. Therefore, for these particles, the $z^2/2$ terms in eq.\ (\ref{accelsolution}) 
are not negligible. 
%Using $T_{\rm L}=14.4 \tau_\eta$ and $z=0.3$, this condition corresponds to $St_{1, 2} \ll 0.65$ 
%in our flow. 
Physically, for the small particles with $St\lsim 1$, the generalized acceleration term has a 
dependence on the local flow acceleration, $a$, which is not accounted for by the 
single-exponential temporal correlation. As discussed in \S 3.2.1, for these particles, 
adopting a bi-exponential $\Phi_1$ is needed for an accurate estimate of the acceleration contribution. 
%to the relative velocity 
%if both particles have $\Omega \lsim 3 z^2$ (corresponding to $St \lsim 4$ in our flow).
%as it gives a prediction in better agreement with the simulation data.   

We also attempted to test the S-T prediction, $3a^2 (\tau_{\rm p2} -\tau_{\rm p1})^2+\frac{\bar{\epsilon}}{3\nu}r^2$, 
for the 3D variance against our data for small particles with $St_{1,2}  \lsim 1$. The measured value of 
the acceleration variance $a^2$ in our flow is $4.6 \tau_\eta^{-2}$ (see  Paper I or Appendix C). 
We aimed at specifically testing the validity of the acceleration term, $3a^2 (\tau_{\rm p2} -\tau_{\rm p1})^2$. 
Thus, at a given $St_1$, we replaced 
the shear term, $\frac{\bar{\epsilon}}{3\nu}r^2$, in the S-T formula by the measured value 
of the monodisperse variance, $\langle w^2 \rangle_{\rm mono}$, at $St_2=St_1$.  
Comparing $(3a^2 (\tau_{\rm p2} -\tau_{\rm p1})^2 +  \langle w^2 \rangle_{\rm mono}^2)^{1/2}$ with 
the data, we find that the acceleration term in the S-T prediction works well only for the smallest two 
particles in our simulation, i.e., for $St=0.1$ and $St=0.19$ particles.  As mentioned in \S 3.2.1,  
the generalized acceleration contribution can be approximated by $3 a^2 (\tau_{\rm p2} -\tau_{\rm p1})^2$ 
only if $\Omega_{1,2} \lsim 0.08z^2$. With $T_{\rm L}=14.4$ and $z=0.3$ in our flow, 
this condition corresponds to $St_{1, 2} \lsim 0.1$, which is met only by the two smallest 
particles in our simulation. By including smaller particles, one may further test 
the S-T prediction for the bidisperse case with $St_{1, 2} \ll 0.1$, 
where the acceleration term is expected to be determined completely  by 
the flow acceleration, $a$. 
%It was observed earlier in the particle-flow velocity 
%that at $St\simeq 0.1$...   

\begin{figure*}[t]
\includegraphics[height=2.8in]{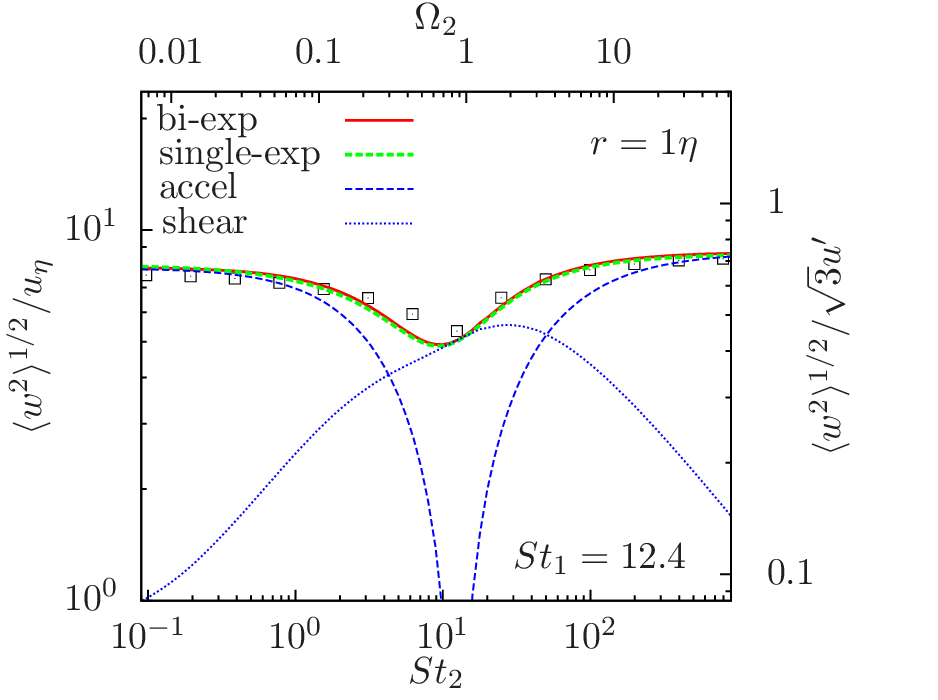}
\includegraphics[height=2.8in]{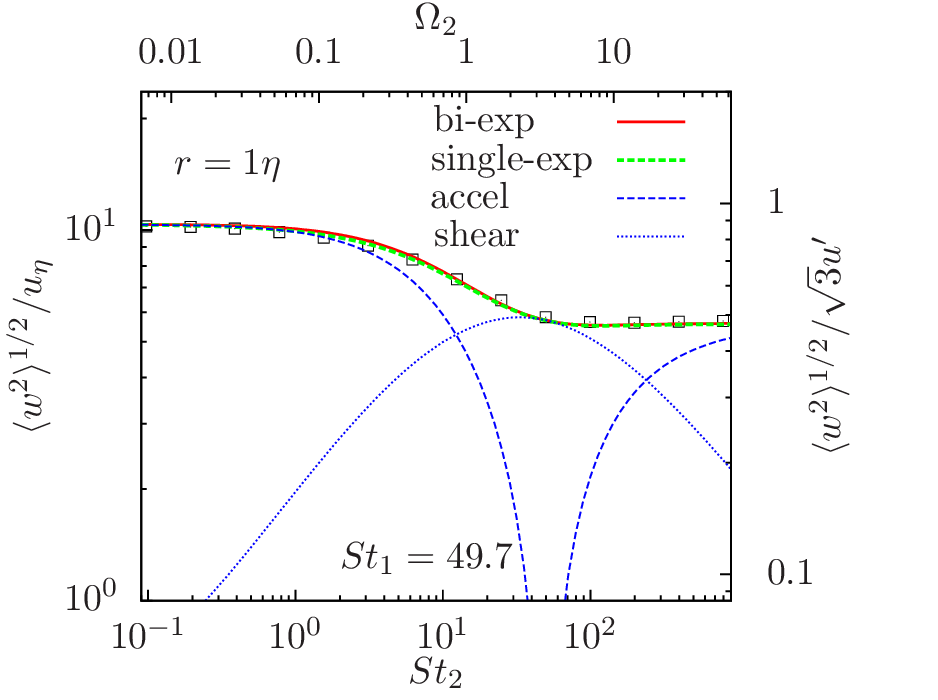}
\caption{Same as Fig.\ \ref{3drms3}, but for $St_1= 12.4$ (left panel) and 
$St_1 =49.7$ (right panel). See caption of Fig.\ \ref{3drms3} for details.}
\label{3drms79} 
\end{figure*}

Fig.\ \ref{3drms79} shows the simulation results and our model predictions for $St_1 = 12.4$ (left panel) 
and $St_1=49.7$ (right panel). The two panels are plotted in the same way as Fig.\ \ref{3drms3}. 
The assumptions and parameters used in the model predictions are also the same as for the 
$St_1=0.78$ case in Fig.\ \ref{3drms3}. For $St_1 = 12.4$ and $St_1 = 49.7$, there is no significant difference 
between the predictions with single- or bi- exponential temporal correlation functions. 
As mentioned in \S 3.2.1, if either of the two particles has $\Omega \gsim 3 z^2$ 
(as is the case for $St_1 = 12.4$ and $St_1 = 49.7$), eqs.\ (\ref{accelsolutionsingle}) 
and (\ref{accelsolution}) for the generalized acceleration term, derived 
from single- or bi-exponential $\Phi_{1}$, respectively, are close to each other. 
Together with the fact that the shear contribution is insensitive to the form of $\Phi_{2}$, 
this explains the coincidence of the red and green lines in both panels 
of Fig.\ \ref{3drms79}. 

For $St_1 = 12.4$, the rms relative velocity changes only 
slightly with $St_2$. The variation is less than $\simeq 40 \%$. 
This is of particular interest as $St_1 = 12.4$ corresponds to a friction time close to the Lagrangian correlation 
time, $T_{\rm L}$($=14.4 \tau_\eta$), of our flow. The small variation of $\langle w^2 \rangle$ for $\tau_{\rm p1}\simeq T_{\rm L}$ 
can be understood by considering three interesting limits: $St_2 \to 0$ (toward the left Y-axis), 
$St_2=St_1$ and $St_2 \to \infty$ (the right Y-axis). In the $St_2 \to 0$ limit, $\langle w^2 \rangle^{1/2}$ is essentially the particle-flow 
rms relative speed, $w_{\rm f}^{\prime}$, of particles (1) (see \S 3.2.1). 
Then, using eq.\ (\ref{pfrelativesingle}) (or eq.\ (\ref{pfrelativebi})) for $w_{\rm f}^{\prime}$, 
we have $\langle w^2 \rangle^{1/2} = u'[3\Omega_1/(1+\Omega_1)]^{1/2} \simeq \sqrt{\frac{3}{2}}u'$ 
for $\tau_{\rm p1}\simeq T_{\rm L}$ and $St_2\to 0$. In the opposite limit 
$St_2 \to \infty$, the velocity of particles (2) is negligible, and its relative velocity with respect to particles (1) 
is essentially the 1-particle velocity, $v'$, of particles (1). Using 
eq.\ (4) (or (6)) in Paper I for $v'$,  we find that $\langle w^2 \rangle^{1/2}  = u'[3/(1+\Omega_1)]^{1/2} \simeq \sqrt{\frac{3}{2}} u'$ 
for $\tau_{\rm p1}\simeq T_{\rm L}$ and $St_2\to \infty$. This  is the same as the estimate for the $St_2\to 0$ limit. 
Finally, for the monodisperse case with $St_2=St_1$, we have shown in Paper I 
that the rms relative speed of identical particles with $\tau_{\rm p}\simeq T_{\rm L}$ is about 
half the flow rms velocity, i.e.,  $\langle w^2 \rangle^{1/2} \simeq \frac{\sqrt{3} }{2} u'$, which is 
smaller by a factor of $\sqrt{2}$ than the estimated values for the two extreme limits above. 
As seen in the left panel of Fig.\ \ref{3drms79}, the rms relative speed 
at the dip ($St_2 =St_1$) is indeed smaller than at the left and right Y-axes by 30-40\%. 
The fact that the difference of the three limits is within $\lsim 40\%$ 
explains the small variation of $\langle w^2 \rangle^{1/2}$ in the whole $St_2$ range. The 
discussion here is general and not limited to our simulation. We expect that, 
for $\tau_{\rm p1} =T_{\rm L}$ in any flow, we have $\langle w^2 \rangle^{1/2}\simeq$  $\sqrt{\frac{3}{2}} u'$, $\frac{\sqrt{3} }{2} u'$ and $\sqrt{\frac{3}{2}} u'$  
at $St_2 \to 0$, $St_2=St_1$ and $St_2 \to \infty$,  respectively. 
%, at $\tau_{\rm p1} = \tau_{\rm p2} =T_{\rm L}$,

In the right panel of Fig.\ \ref{3drms79}, no dip exists in the data or the model prediction for 
$St_1 =49.7$. Instead, $\langle w^2 \rangle^{1/2}$ decreases monotonically 
as $St_2$ increases from 0.1 to $795$. We find that 
the disappearance of dips actually starts at $St_1 = 24.9$ (or  $\tau_{\rm p1} = 1.7 T_{\rm L}$). 
Recall that the dip formation for smaller $St_1$ is due to the tighter velocity 
correlation between particles of similar sizes. But 
if $\tau_{\rm p}$ is considerably larger than  $T_{\rm L}$, the particle velocities are not 
significantly correlated even for two particles of exactly the same size (see PP10). 
Therefore, dips are not expected in the $\langle w^2 \rangle^{1/2}$ vs.\ $St_2$ curve 
if $\tau_{\rm p1} \gsim T_{\rm L}$.
%\footnote{Another perspective to understand the absence of dips for $\tau_{\rm p1} \gg T_{\rm L}$
%is to compare two limiting cases: identical particles with $\tau_{\rm p2} =  \tau_{\rm p1}$, 
%and the $\tau_{\rm p2} \to \infty$ limit.  In the first case, 
%$\langle w^2 \rangle$ is estimated by $(6T_{\rm L}/\tau_{\rm p1})^{1/2} u'$ (see eq.\ 12 of Paper I), 
%%with $\tau_{\rm p1} \gg T_{\rm L}$ and $\tau_{\rm p2} \to \infty$, 
%while, for the second limit, we have $\langle w^2 \rangle \to (3T_{\rm L}/\tau_{\rm p1})^{1/2} u'$ 
%from eq.\ (\ref{largeparticles}). As the latter is twice smaller than the monodisperse case,
%no dip would form at $St_2 =St_1$ for $\tau_{\rm p1} \gg T_{\rm L}$.}.
%%($St_{1} \gsim 24.9$) 

The fitting quality of our model around the dip for $St_1 = 12.4$ is not as satisfactory as 
the $St_1=0.78$ and $St_1 = 49.7$ cases. The predicted curve overestimates 
the width of the dip.  Furthermore,  rather than exactly at $St_2 =St_1$, 
the dip center in the prediction is located at $St_2$ slightly below $St_1$.  
The same is found for $St_1 =3.11$ and $St_1 =6.21$.  Apparently, this mismatch 
of the dip center is due to the decrease of the generalized shear term 
with decreasing $St_2$ at $St_2 \le St_1$. The poorer fit of our model around the 
dips for $St_1 =3.11$, $6.21$, and $St_1=12.4$ than for the 
$St_1 \lsim 1$ and $\tau_{\rm p1} \gsim T_{\rm L}$ cases 
suggests that the assumptions of the model are the least accurate for 
inertial-range particles of similar size. The following simplifying assumptions 
may be responsible for this lower accuracy.  

First, as discussed in Appendix A, we ignored a term named $C_{ij}$ 
in our general formulation for the particle structure function, $S_{{\rm p}ij}$, in the 
bidisperse case. Second, in \S 3.2.1 we neglected the anti-symmetric term $B_{ij}^{-}$, 
when evaluating $\mathcal{A}_{ij}$. Both $C_{ij}$ and  $B_{ij}^{-}$ 
are related to the asymmetry in the flow velocity statistics along trajectories 
of the two  particles. They are not exactly zero for different particles, and it 
is possible that they make non-negligible contributions for inertial-range particles of slightly different sizes. 
%\footnote{Both $C_{ij}$ and  $B_{ij}^{-}$ are 
%expected to increase as the friction time difference of the two particles 
%increases.  However, this does not exclude the possibility that 
%their contributions {\it relative} to the generalized acceleration or 
%shear terms are the largest for particles of slightly different sizes
%and make non-negligible contributions only 
%for particles of similar sizes.}. 
Third, the assumptions in our model for the generalized shear term may be less 
accurate in the bidisperse case. One example is the assumed separation 
behavior of particle pairs. As pointed out in Appendix B,  how particle pairs of different 
sizes exactly separate backward in time is largely unknown. A direct study of the pair separation 
of slightly different particles in the inertial range may help improve the 
model prediction. Theoretically, an improvement of our model for 
inertial-range particles of similar sizes accounting for all the possibilities listed above 
is of significant interest, as it may further refine the understanding of the physics.
%as we are interested in inertial-range dust particles in protoplanetary disks.  
On the other hand, from a practical point of view, it may not 
be particularly useful to develop a high-accuracy model for the 
rms relative velocity, which, as mentioned earlier, is not directly 
applicable to the collision statistics (see Paper I). 
It could be more convenient to simply use the collision 
statistics measured from the simulation data. 
Overall, our model is in good agreement with the data despite the small mismatch for 
inertial-range particles of slightly different sizes, and, in general, it provides a successful 
physical picture for the particle relative speed in the bidisperse case. 

\begin{figure}[h]
\centerline{\includegraphics[width=1.15\columnwidth]{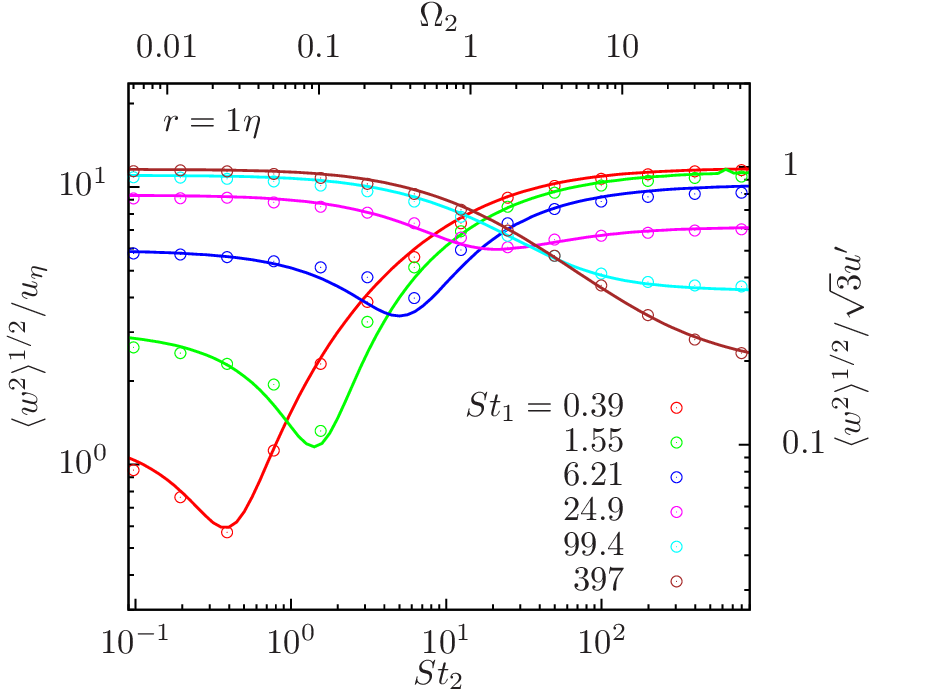}}
\caption{The 3D rms relative velocity for different values of $St_1$. Data points are simulation 
results at $r=1\eta$. Lines are our model predictions using bi-exponential 
correlation functions, $\Phi_1$ and $\Phi_2$. %A two-phase separation with $g=1.6$ for the Richardson phase 
%is adopted for the shear term.
}
\label{3drmsall} 
\end{figure}

From Figs.\ \ref{3drms3} and  \ref{3drms79}, we see that  the generalized shear term 
is important only for particles of similar sizes, and
the acceleration term starts to dominate 
if the Stokes numbers differ by a factor of $\gsim 4$. We stress that the 
interesting limits, $St_2 \to 0$, $St_2=St_1$ and $St_2 \to \infty$, are 
very useful delimiters for the behavior of the relative velocity at a given $St_1$. 

We summarize the general picture for the bidisperse relative velocity in Fig.\  \ref{3drmsall}, which shows $\langle w^2 \rangle^{1/2}$ at $r=1\eta$ 
for six values of $St_1$ ranging from 0.39 to 397.  Lines are the model predictions using bi-exponential $\Phi_1$ and $\Phi_2$ 
and the same separation behavior as in Figs.\ \ref{3drms3} and \ref{3drms79}. 
There is a clear difference in the behavior of $\langle w^2 \rangle^{1/2}$ for $\tau_{\rm p1} \lsim T_{\rm L}$ and $\tau_{\rm p1} \gsim T_{\rm L}$. 
For $\tau_{\rm p1} \lsim T_{\rm L}$,  a dip exists at $St_2=St_1$, and, away from the dip, $\langle w^2 \rangle^{1/2}$ increases 
toward $St_2 \to 0$ and $St_2 \to \infty$. On the other hand, for $\tau_{\rm p1} \gsim T_{\rm L}$, there are no dips, 
and the relative speed decreases monotonically with increasing $St_2$.  In the $St_2 \to 0$  limit, i.e., 
on the left Y-axis, $\langle w^2 \rangle^{1/2}$ corresponds to the particle-flow relative velocity for 
particles (1) and increases with increasing $St_1$ (see eq.\ (\ref{pfrelativebi}) and Fig.\ \ref{gaspartrms}). 
In the other limit $St _2\to \infty$ (the right Y-axis), the relative velocity decreases monotonically 
with increasing $St_1$, corresponding to the decrease of the 1-particle relative velocity of 
particles (1) with increasing $\tau_{\rm p1}$ (see eq.\ (6) and Fig.\ 5 of Paper I). 
%Again this is equivalent to the black dotted line (or Fig.\ X of paper I). 
For $\tau_{\rm p2} \simeq T_{\rm L} $(i.e., $St_2 \simeq 14.4 $), the change of $\langle w^2 \rangle^{1/2}$ 
with $St_1$  in the vertical direction is slight, corresponding to the small variation 
of $\langle w^2 \rangle^{1/2}$ as a function of $St_2$ in the horizontal direction for 
$St_1 = 12.4$ (see the left panel of Fig.\ \ref{3drms79}). 

\subsection{The rms relative velocity at fixed Stokes ratios}

\begin{figure}[h]
\centerline{\includegraphics[width=1.15\columnwidth]{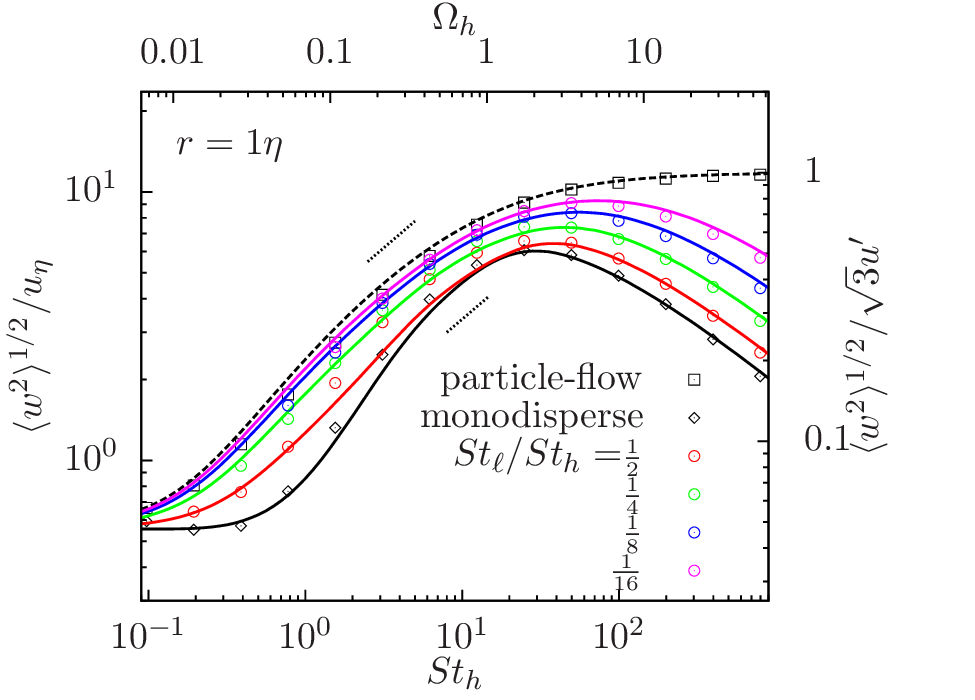}}
\caption{The 3D rms relative velocity at $r=1\eta$ for particle pairs with fixed Stokes 
ratios, $f\equiv St_{\ell}/St_{h} = \frac{1}{2}$ (red), $\frac{1}{4}$ (green), $\frac{1}{8}$ (blue) and $\frac{1}{16}$ 
(magenta). 
Black diamonds and squares correspond to the monodisperse case ($f=1$) 
and the particle-flow relative velocity ($f=0$). The lines show our model predictions. 
%using bi-exponential temporal correlation functions. 
%and a two-phase separation behavior with a Richardson constant 
%of $g=1.6$. 
The two black dotted line segments correspond to a $St_{h}^{1/2}$ scaling. 
%The dash and solid lines are model predictions, eqs.\ (\ref{pfrelativesingle}) and (\ref{pfrelativebi}), 
%using single- and bi- exponential forms for $\Phi_{\rm L}$ (and hence for $\Phi_1$), 
%respectively. In both lines, $T_{\rm L}$ is taken to be 15 $\tau_{\eta}$. For the 
%bi-exponential case, the parameter $z$ is set to 0.3. %The dashed line considerably overestimates $w^{\prime f}$ for $St \lsim 3$ particles.   
%The dotted line segment denotes a $St^{1/2}$ scaling.
%Right panel: the 3D rm particle relative velocity at distance $r=0$ (circles), $\frac{1}{2}$ (triangles) and $1 \eta$ (diamonds). 
}
\label{fixedratio} 
\end{figure}

%We have fixed the Stokes number of one particle, and examined the relative velocity 
%as a function of the other Stokes number. 
Instead of fixing one of the Stokes numbers, it may also be convenient to 
analyze the relative velocity of particle pairs with a fixed Stokes number ratio, 
$f$. As a reminder, $f$ is defined as the ratio of the lower Stokes number, 
$St_{\ell}$ ($\equiv \min (St_1, St_2)$), to the higher one, 
$St_{h}$ ($\equiv \max (St_1, St_2)$), so that $0 \le f \le 1$. 
In Fig.\ \ref{fixedratio}, we show the 3D rms relative speed at $r=1\eta$ as 
a function of $St_{h}$ for different values of $f$. The monodisperse case 
(black diamonds) corresponds to $f=1$. As discussed in Paper I,  for small identical 
particles at $St \lsim 1$, the relative velocity at $r=1\eta$ is mainly determined by the 
local flow velocity difference and is essentially independent of $St$.
% (see the shear terms in the S-T formula, eq.\ \ref{saffmanturner}). 
As $St$ increases above $1$, the particle memory of the {\it spatial} flow velocity difference in the past becomes important
and %together with the particle separation backward in time, it causes 
the relative velocity starts to increase.  
%This can be seen from the approximate equation (\ref{approxshear}) for ${\bs w}_{\rm s}$.  
%For inertial-range particles, the primary distance $R_{\rm p}$ and the timescale $T_{\rm p}$ 
%increase with $St$, leading to the increase of ${\bs w}_{\rm s}$. 
A $St^{1/2}$ scaling (dotted line segment) was predicted  for inertial-range particles by various models.  
For large particles with $\tau_{\rm p} \gg T_{\rm L}$, the memory/correlation time of the flow at the largest scales 
is shorter than the particle memory, and the flow memory cutoff  
(by the $\Phi_2$ term) causes a decrease of the relative velocity with $St$ as $St^{-1/2}$.  
%(see also eq.\ (\ref{approxshear})).   

The black squares show the particle-flow relative velocity, $\langle w_{\rm f}^2\rangle^{1/2}$, 
corresponding to $f \to 0$.  For a consistent comparison, $\langle w_{\rm f}^2\rangle^{1/2}$ is
measured here at the same distance ($r=1\eta$) as in the particle-particle cases\footnote{Note that the particle-flow relative 
velocity, $w_{\rm f}^{\prime}$, shown in Fig.\ \ref{gaspartrms} is at zero distance.  
The data points for $w_{\rm f}^{\prime}$ at $r=0$ and $\langle w_{\rm f}^2\rangle^{1/2}$ at $r=1\eta$ 
are found to coincide at $St \gsim 1$. But at $St \lsim 1$, $\langle w_{\rm f}^2\rangle^{1/2}$ at $r\simeq 1\eta$ 
is slightly larger due to the contribution of the shear term.
The shear contribution to $\langle w_{\rm f}^2\rangle$ for the $St\lsim1$ particles is approximately $\frac{\bar{\epsilon}}{3\nu} r^2$.}. 
%The method 
%we used 
%To measure the particle-flow relative velocity at finite distances %is given in Appendix C (see Fig.\ \ref{gaspartrms2}).
%In the right panel, we show the particle-flow relative velocity at finite distances. 
We used the TSC interpolation to obtain the flow velocities at a separation $r$ 
from the position of each particle in the three orthogonal directions of the simulation grid. 
The black dotted line shows our model prediction.
%In each direction ${\bs e}_{i}$, the particle-flow relative velocity at $r$ is computed as ${\bs w}_{\rm f} = {\bs u}({\bs X}(t)+r{\bs e}_{i}, t)-{\bs v}(t)$. 
%We averaged the statistics measured from the three directions.  
%Fig.\ \ref{gaspartrms2} plots the 3D rms particle-flow relative velocity at $r=0$ (circles), $\frac{1}{2}$ (triangles) 
%and $1 \eta$ (diamonds). 
To compute the prediction, we set one of the friction times 
in eq.\ (\ref{sii}) to zero, which reduces the equation to a single integral. We used 
exactly the same assumptions and parameters as in the predictions for the particle-particle
relative velocity in the bidisperse case. The prediction  is in good agreement 
with the data points. 

%For example, for the particle tracer separation, we adopt the same two-phase behavior. We used bi-exponential 
%temporal correlations, $\Phi_1$ and $\Phi_2$, with $z=0.3$ and $T_{\rm L} =14.4\eta$ (i.e., the  same as in the 
%solid line in Fig.\ \ref{gaspartrms}; see \S5).}
%The same parameters were used  for $\Phi_2$ in the generalized shear term. 
%The prediction can be well approximated by $\langle w_{\rm f} ^2 \rangle = 3A + \frac{\bar{\epsilon}}{3\nu} r^2$, 
%where the second term corresponds to the shear contribution. The shear term causes the particle-flow relative velocity to increase with 
%$r$ for small particles. 
%As $St$ increases, the acceleration term increases, and the particle-flow relative 
%velocity becomes $r-$independent, when $3A$ dominates. 
%Note that, at $r=\frac{1}{2}\eta$, the particle-flow relative velocity converges for all  particles (with $St\gsim 0.1$) in 
%our simulation.

\begin{figure*}[t]
\includegraphics[height=2.6in]{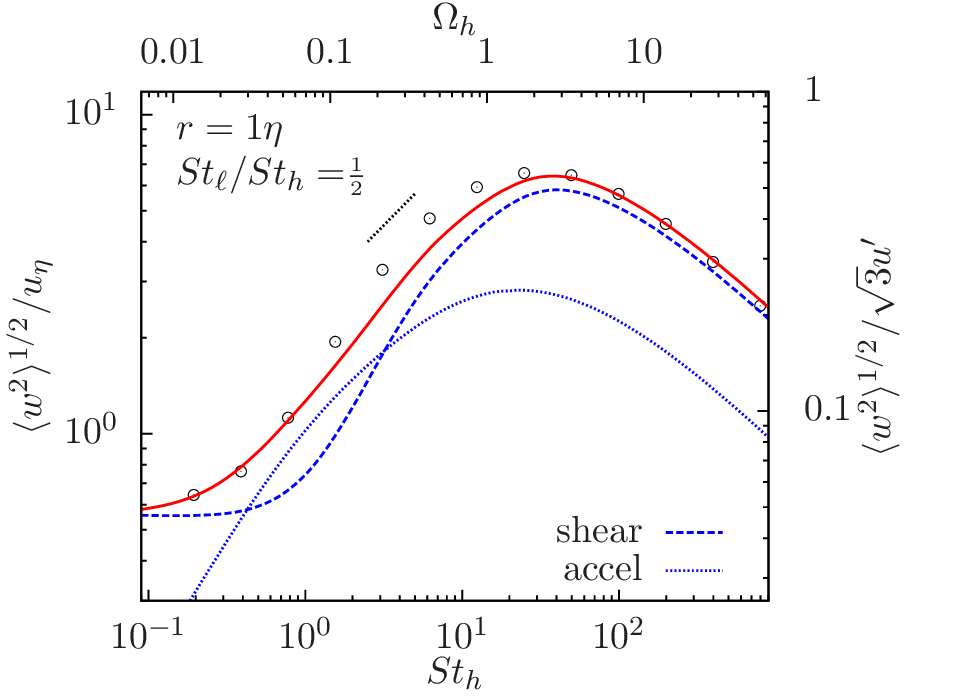}
\includegraphics[height=2.6in]{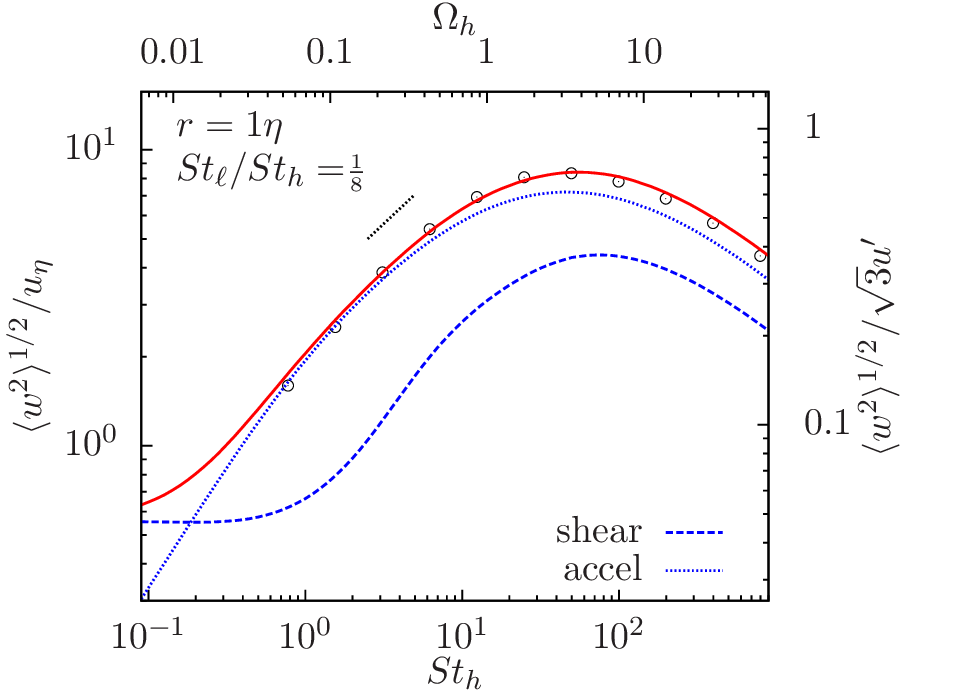}
\caption{The 3D rms relative velocity at Stokes ratios  $f=\frac{1}{2}$ (left panel)
and $\frac{1}{8}$ (right panel). Solid lines are the same model predictions as 
in Fig.\ \ref{fixedratio}. Dashed and dotted lines are the generalized 
shear and acceleration contributions, respectively. 
Dotted line segments denote $St_{h}^{1/2}$ scaling.}
\label{fixedratio28} 
\end{figure*}

The color data points show the simulation results for $\frac{1}{16} \le f \le \frac{1}{2}$,
which all lie in between the particle-flow relative velocity ($f=0$) and 
the monodisperse case ($f=1$). Due to the contribution of the generalized acceleration term, the relative 
velocity increases as $f$ decreases.  At a given $St_{h}$, the increase of 
$\langle w_{\rm f}^2\rangle^{1/2}$ with decreasing $f$ corresponds to the increase of the data points or 
lines in Fig.\ \ref{3drmsall} toward the left Y-axis, i.e., at $St_2 \le St_1$. The color lines in the figure are our model 
predictions using bi-exponential temporal correlation functions and a two-phase separation behavior with 
a Richardson constant of $g=1.6$ (i.e., the same as the lines 
in Figs.\ \ref{3drms3}, \ref{3drms79}, and \ref{3drmsall}). %{\bf The same separation is adopted in all our model predictions in the rest of the paper for 
%$r=1\eta$}
The model prediction matches reasonably well 
the data. As $St_{h} \to 0$, all the data points and 
lines appear to converge to the same value. At any given $f$, 
both particles become tracer particles as 
$St_{h} \to 0$. Therefore, in the $St_{h} \to 0$ limit, 
the relative velocity always approaches the spatial flow 
velocity difference across the particle distance, 
$r$ (see the shear term in the S-T formula). 

Like the monodisperse case, $\langle w^2 \rangle^{1/2}$ first increases with $St_{h}$, reaches 
a maximum and finally decreases.  We find that in general the peak of $\langle w^2 \rangle^{1/2}$ lies in the 
range $T_{\rm L} \lsim \tau_{\rm ph} \lsim T_{\rm L}/f$ (or equivalently $fT_{\rm L} \lsim \tau_{\rm pl} \lsim T_{\rm L}$). 
In this range,  the variation of $\langle w^2 \rangle^{1/2}$ is small. The value of $St_{h}$ at which $\langle w^2 \rangle^{1/2}$ peaks 
increases with decreasing $f$.  For $St_{\ell} \gg  T_{\rm L}$ ($St_{h} \gg  T_{\rm L}/f$), $\langle w^2 \rangle^{1/2}$ is 
expected to decrease as $St_{h}^{-1/2}$ (see eq.\ (\ref{largeparticles}) and the discussion below).  

The two dotted line segments denote a $St_{h}^{1/2}$ scaling for $St_{h}$ 
in the inertial range. Such a scaling has been predicted for inertial-range particles of equal sizes 
($f=1$) by various models (see Paper I). The same scaling was 
also predicted by our model for the relative velocity, $w_{\rm f}$, between the flow and inertial-range particles (i.e., for $f=0$; see \S 2 and 5). 
This suggests the possibility of  a universal $St_{h}^{1/2}$ scaling in the inertial range 
for any $f$ between 0 and 1. Using eq.\ (\ref{approxaccel}) and the assumption $\Delta {\bs u}_{\rm T} \simeq \Delta {\bs u}_{\rm L}$, we 
see that, for a fixed $f$, the acceleration term, ${\bs w}_{\rm a}$, would scale 
as $St_{h}^{1/2}$ for $\tau_{\rm ph}$ in the inertial range.  
%
%(the linear range is wider for smaller $f$), 
This could also be shown from eq.\ (\ref{accelsolution}): at a given $f$, 
the acceleration contribution, $A$, to the relative velocity 
variance goes linearly with $\Omega_{h}$ if  $z^2/2 \ll \Omega_{h} \ll 1$. 
Therefore, the $St_{h}^{1/2}$ scaling is expected for 
$f\lsim\frac{1}{4}$, where the acceleration effect dominates.  Considering the high probability 
that the shear term, ${\bs w}_{\rm s}$, in the monodisperse case 
scales as $St^{1/2}$ in the inertial range, it may be generally true that the 
scaling applies to any $f$.  However, to observe a convincing $St_{h}^{1/2}$ 
scaling, the flow must have an extended inertial range. Higher-resolution simulations are 
needed to verify if a $St_{h}^{1/2}$ scaling holds in general for any values of $f$ in 
the range $0 \le f \le1$.

In Fig.\ \ref{fixedratio28}, we show the rms relative velocity for $f=\frac{1}{2}$ (left panel) 
and $f=\frac{1}{8}$ (right panel) in more details. The data points and the solid lines are 
the same as the corresponding ones in Fig.\ \ref{fixedratio}. 
The discrepancy between the data and our model prediction is largest for $f=\frac{1}{2}$.  
%and, in particular, for $St_{h}$ in the inertial-range, $1.55\le St_{h} \le 12.4$ (the left panel of Fig.\ \ref{fixedratio28}). 
In this case, our model underestimates the data points by $\simeq 20\%$ for $St_{h} = 3.11$ and $6.21$, 
and by  $\simeq 15\%$ for $St_{\rm 2} = 1.55$ and $12.4$.  On the other hand,  the model 
prediction agrees with the data quite well for smaller ($St_{h} \lsim 1$) or larger ($\tau_{\rm ph} \gsim T_{\rm L}$) particles. 
A detailed discussion on the possible reasons for the discrepancy for inertial-range particles 
of similar sizes  was  given in \S 6.1. The agreement improves with decreasing $f$, 
and at $f =\frac{1}{4}$ the discrepancy is $< 10\%$.

The dashed and dotted lines in Fig.\ \ref{fixedratio28} plot the generalized shear and acceleration contributions in our model 
predictions. At $f=\frac{1}{2}$, the relative velocity is mainly contributed by the shear term, except for 
$0.4 \le St_{h} \le 3$.  The contribution of the acceleration term increases with decreasing $f$, 
and, at $f \le \frac{1}{8}$, the rms relative velocity is dominated by the acceleration contribution.

The behaviors of the generalized acceleration and shear contributions as a function of $St_{h}$ can be 
understood from eqs.\ (\ref{approxaccel}) and (\ref{approxshear}) for ${\bs w}_{\rm a}$ and ${\bs w}_{\rm s}$, respectively. 
Eq.\ (\ref{approxaccel}) indicates that, at a fixed $f$,  
${\bs w}_{\rm a}$ increases with $St_{h}$ for $\tau_{\rm ph} \lsim T_{\rm L}$ 
because $\Delta_{\rm T} {\bs u}(\tau_{\rm ph})$ increases with $\tau_{\rm ph}$.  
Due to the $(1+ \Omega_{\ell})^{-1/2}$ (or $(1+ f\Omega_{h})^{-1/2}$) factor 
in eq.\ (\ref{approxaccel}), %its contribution to the rms relative velocity 
${\bs w}_{\rm a}$ decreases as $St_{h}^{-1/2}$ %(or $ \Omega_{h}^{-1/2}$) 
at $\tau_{\rm ph} \gg T_{\rm L}/f$.  
These are indeed observed in the dotted lines in Fig.\ \ref{fixedratio28}.  
The acceleration term is roughly constant for $T_{\rm L} \lsim \tau_{\rm ph} \lsim T_{\rm L}/f$ because in this range 
the amplitude of $\Delta_{\rm T} {\bs u}(\tau_{\rm ph})$ is $\simeq u'$  
and $ \Omega_{\ell} \lsim 1$. 
In order to understand the shear contribution, we calculate the rms of ${\bs w}_{\rm s}$ 
from eq.\ (\ref{approxshear}) using the method described 
in Footnote 4 to estimate the primary distance, $R_{\rm p}$, and the timescale, $T_{\rm p}$. 
%at the scale $R_{\rm p}$. 
The calculation shows that the rms of ${\bs w}_{\rm s}$ 
is consistent with $\sqrt{\mathcal {S}_{ii}}$ computed from the double integral equation (\ref{sii}), 
supporting the validity of eq.\ (\ref{approxshear}) as an approximate estimate for the shear term. 
At small $St_{h}$, both $R_{\rm p}$ and $T_{\rm p}$ in eq.\ (\ref{approxshear}) increase with $St_{h}$, 
and thus the shear contribution increases. As $St_{h}$ 
keeps increasing, $T_{\rm p}$ finally reaches the maximum value ($T_{\rm L}$), when the friction time, $\tau_{\rm p,l}$, of the smaller particle 
increases up to $\simeq T_{\rm L}$. At  $\tau_{\rm p,l} \gsim T_{\rm L}$,  
$T_{\rm p}$ stays constant ($=T_{\rm L}$), and the $[T_{\rm p}/(T_{\rm p}+T_{\rm p,h})]^{1/2}$ 
term in eq.\ (\ref{approxshear}) causes ${\bs w}_{\rm s}$ to decrease with  $St_{h}$. 
The condition $\tau_{\rm p,l} \gsim T_{\rm L}$ corresponds to $\Omega_{\ell} \gsim 1$ or 
$\Omega_{h} \gsim 1/f$. This explains why the peak of the shear contribution occurs at larger $St_{h}$ 
for smaller $f$.  Similar to the acceleration term,  the shear contribution, $\sqrt{S_{ii}}$,
decreases as $\propto St_{h}^{-1/2}$ at $\Omega_{h} \gg 1/f$.  

%In summary, if one Stokes number is fixed, the bidisperse relative velocity as a function of the other Stokes number 
%has three interesting limits:... 

\subsection{Dependence on the particle distance}

\begin{figure*}[t]
\includegraphics[height=2.7in]{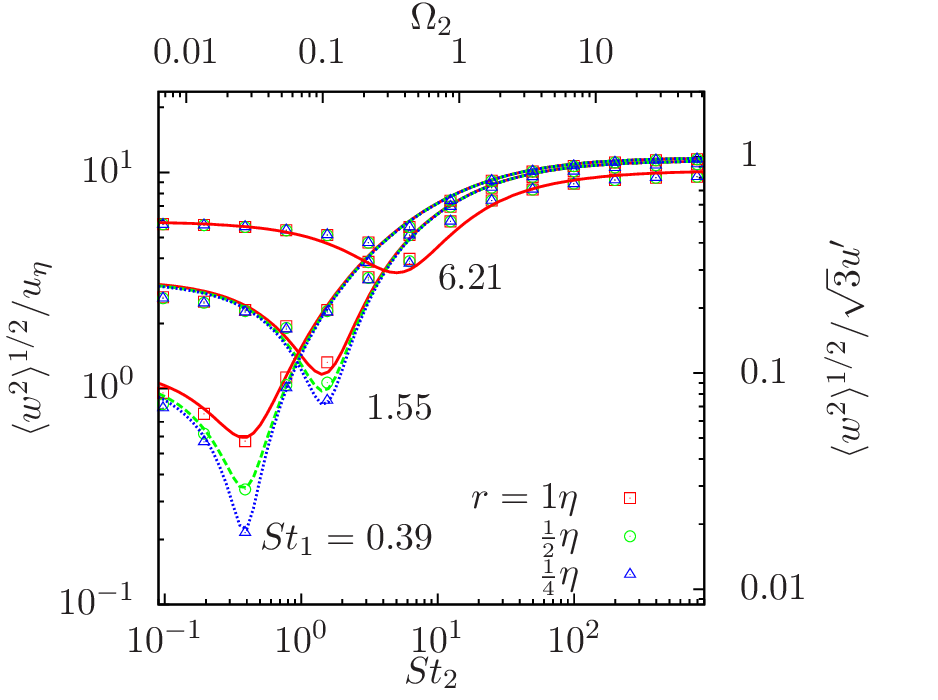}
\includegraphics[height=2.7in]{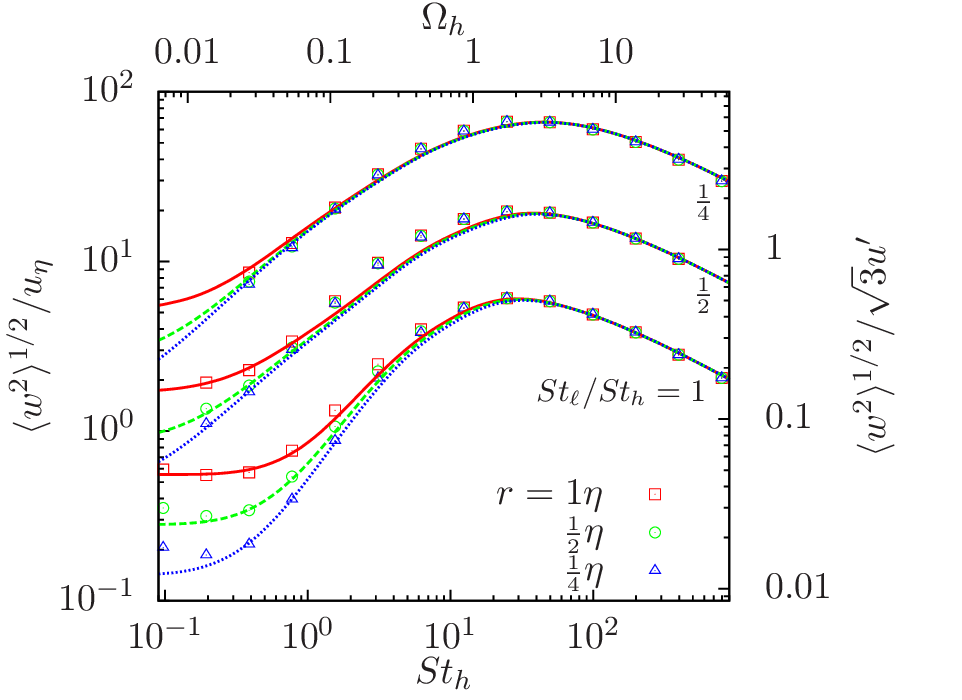}
\caption{Dependence of 3D rms relative velocity on the particle distance, $r$. %Red, green and blue data points 
%correspond to $r=1$, $\frac{1}{2}$, and $\frac{1}{4}\eta$, respectively. 
Left and right panels plot results at fixed $St_1$ and fixed Stokes ratios, respectively. 
In both panels, lines are our model predictions using bi-exponential $\Phi_1$ and $\Phi_2$ and 
a two-phase separation behavior with $g=1.6$, $1.3$ and $1.0$ for $r=1$, $\frac{1}{2}$, and $\frac{1}{4}\eta$, respectively. 
For clarity,  the data points and lines in the right panel 
for $f = \frac{1}{2}$ and $\frac{1}{4}$ are shifted upward by a factor of 3 and 9, respectively.
}
\label{3drmsscale} 
\end{figure*}

In Fig.\ \ref{3drmsscale}, we show the 3D rms relative 
velocity at different particle distances. The left and right panels plot $\langle w^2\rangle^{1/2}$ 
at fixed values of $St_1$ and fixed Stokes ratios $f$, respectively. The solid, 
dashed and dotted lines are our model predictions for $r=1$, $\frac{1}{2}$, and $\frac{1}{4} \eta$ 
using bi-exponential temporal correlations and a two-phase behavior for the 
particle separation backward in time.  %as in Fig.\ \ref{3drmsall}. 
In the Richardson phase,  $g$ is set to 1.6, 1.3 and 1.0 for $r=1$, $\frac{1}{2}$, and $\frac{1}{4} \eta$, 
respectively. These values of $g$ are the same as those used in Paper I that best fit 
the rms relative velocity of equal-size particles at the corresponding distances.  
{As discussed in Paper I, our choice of smaller Richardson constant $g$ at smaller $r$ 
is based on the observation that the directly measured value of $g$ for the backward-in-time pair 
separation of tracer particles in our simulated flow decreases with decreasing initial 
distance $r$. This is likely due to the limited inertial range of the simulated flow, and  $g$ may reach 
a universal constant  value if  the flow Reynolds number is  sufficiently larger (Paper I). 
It would be convenient if one can set $g$ to a fixed constant in our model. 
However, the measured $g$ for tracer particles suggests otherwise, and allowing $g$ to change
 with $r$ for inertial particles in our flow appears to be a more natural choice, as fixing $g$ to a single 
 value may leave the misleading impression 
that  $g$ is strictly constant with $r$ in our simulated flow. We thus pursued best fits of 
our model prediction to the simulation data by varying $g$ with $r$. The fact that the best-fit $g$ 
decreases with $r$ is consistent with the direct measurement of $g$ for tracer particles. 
In protoplanetary turbulence, the Reynolds  number is much larger  than in our simulation, 
and one may expect a constant $g$ at different values of $r$. However, 
this expectation and the exact value of $g$ in flows with much larger $Re$ 
needs to be examined by numerical simulations  at considerably higher resolution. }     

%However, there is no sufficient 
%evidence that $g$ is a constant in our simulated flow. 
%However, in the figures shown here, we chose to let $g$ change 
%with $r$, as suggested by the measured $g$ for tracer particles in our flow.

%For clarity,  the data points and lines  for $f = \frac{1}{2}$ and $\frac{1}{4}$ in 
%the right panel are shifted upward by a factor of 3 and 9, respectively.
%As discussed in Paper I,  the best-fit value of $g$ and its decrease with 
%decreasing $r$ is consistent with that measured from the backward separation of tracer particles in our 
%simulated flow. 

In the left panel of Fig.\ \ref{3drmsscale}, we see that,  for $St_1<6.2$, the dips at $St_2= St_1$ become deeper with 
decreasing $r$. 
%As $St_1$ increases above $6.2$ , the depth of the dip becomes independent of $r$. 
This corresponds to the $r-$dependence of the relative velocity in the monodisperse case (see the $f=1$ case in the right panel). 
For small equal-size particles in the S-T limit, the relative velocity decreases with 
decreasing $r$, as it is determined largely by the local flow velocity 
difference across $r$.  As $St$ increases, the particle memory of the spatial flow velocity 
difference in the past provides larger contribution, and the $r-$dependence 
becomes weaker. At $St \gsim 6.2$, the particle memory time is significant, and 
the particle separation at a friction time ago is insensitive to its initial value, $r$. 
This explains the $r-$independence of the dip for $St_1 = 6.2$ in the left panel, 
as well as the $r-$independence of the $f=1$ case at $St \gsim 6.2$ in the right panel. 

In the bidisperse case, the generalized acceleration term is independent 
of the particle distance (see \S 3.2 and Appendix A), and its presence reduces the $r-$dependence of $\langle w^2 \rangle^{1/2}$. 
In the left panel of Fig.\ \ref{3drmsscale}, the relative speed is less dependent on $r$, 
as $St_2$ moves away from the dip center. Clearly, toward both sides of the dips, the contribution from the 
generalized acceleration term increases. %Once the acceleration effect dominates, the relative velocity is $r-$independent. 
For $St_1 \gsim 6.2$, the generalized shear term is already independent of $r$ for $St_2$ around $St_1$, 
and this suggests that, if one of the Stokes numbers is larger than $6.2$,  
$\langle w^2 \rangle^{1/2}$ is $r-$independent.  

The same behavior is seen in the right panel: the $r-$dependence of $\langle w^2 \rangle^{1/2}$ becomes 
weaker as the Stokes ratio $f$ decreases. %At a given $f$, there is a minimum value for the larger Stokes number, 
%$St_{h}$, above which the rms relative velocity is independent of $r$. For example, for $f=\frac{1}{2}$, 
%$\langle w^2 \rangle^{1/2}$ is independent of $r$ at $St_{h} \gsim 0.78$ for $r$ below $\frac{1}{2} \eta$.
%This minimum  value for $St_{h}$ decreases with decreasing $f$.  
%In fact, if the ratio of the two Stokes numbers is larger than 2, the rms relative speed already converges at $r=\frac{1}{4}\eta$ (see Fig.\  \ref{3drmsscale}). 
The weaker $r-$dependence in the bidisperse case makes the evaluation of the collision statistics much easier than in the 
monodisperse case. As discussed in the Introduction (see also Paper I),  
dust particles in protoplanetary disks are nearly point-like particles, and in principle one needs to extrapolate the measured statistics to 
the $r \to 0$ limit  before applying it to dust particle collisions. In Paper I, we 
found that, for equal-size particles with $St \lsim 1$,  the extrapolation is very 
challenging because their relative velocity has a significant $r-$dependence 
that may persist to very small $r$. In the bidisperse case, the relative velocity of 
small particles would converge once the $r-$independent acceleration contribution 
dominates, making it considerably easier to directly compute 
the collision statistics from the numerical simulations.

\begin{figure*}[t]
\includegraphics[height=2.7in]{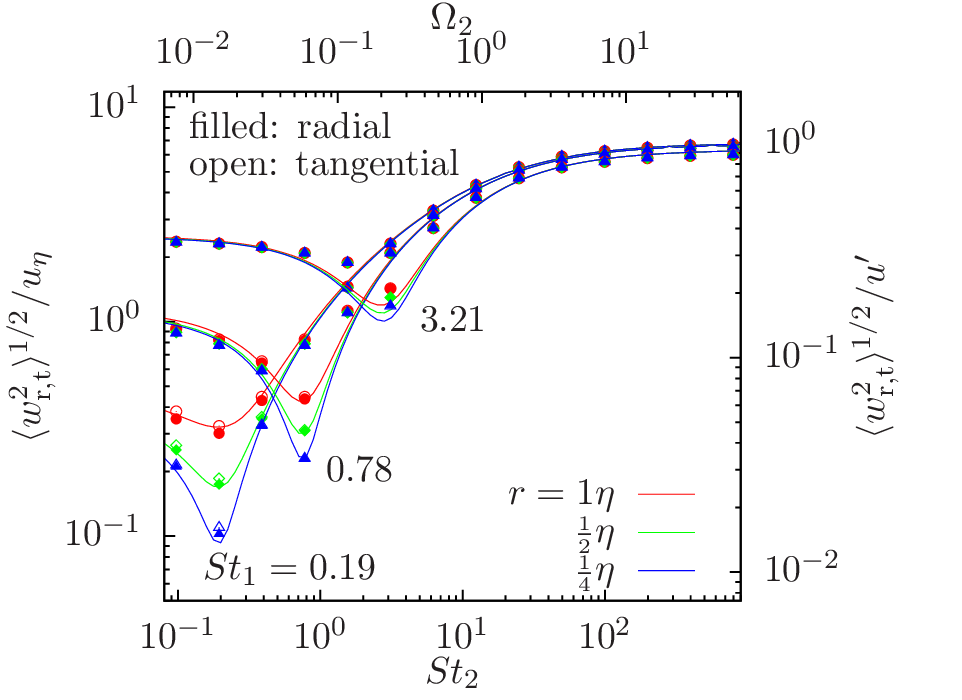}
\includegraphics[height=2.7in]{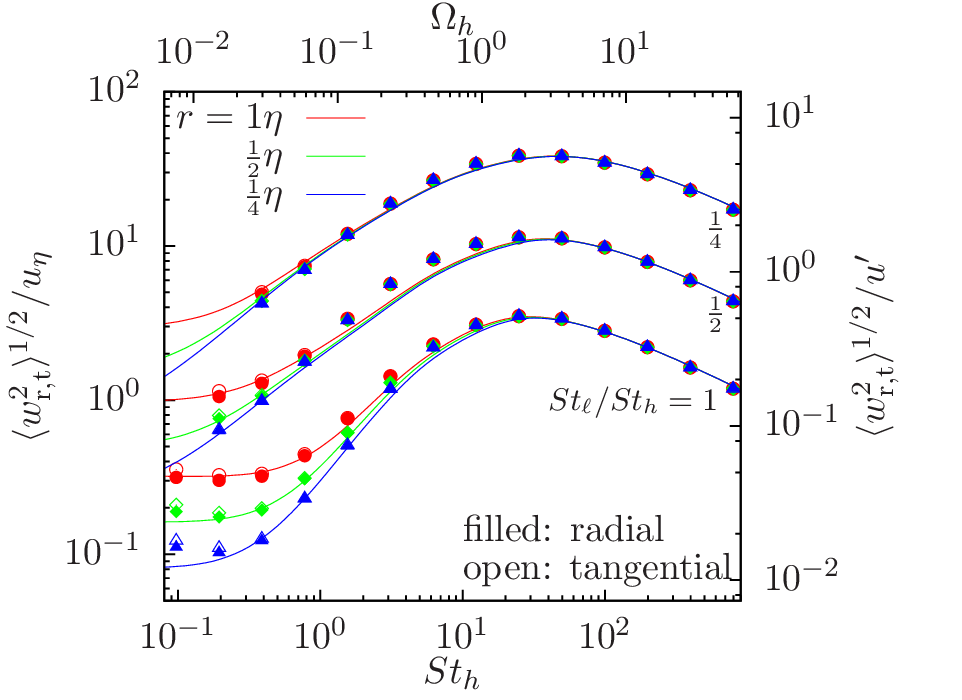}
\caption{Radial (filled circles) and tangential (open circles) rms relative speeds at different $r$. 
Left and right panels show results at fixed $St_1$ and fixed Stokes ratios, 
respectively.  The lines are our model predictions using the same parameters as in Fig.\ \ref{3drmsscale}. %bi-exponential $\Phi_1$ and $\Phi_2$
%and a two-phase separation with $g=1.6$, $1.3$ and $1.0$ 
%for the Richardson phase for $r=1$, $\frac{1}{2}$, and $\frac{1}{4}\eta$, respectively. 
Eq.\ (\ref{randomdirection}) is adopted for %the angular average of the trajectory structure tensor, 
$\langle S_{ij} \rangle_{\rm ang}$, which predicts $\langle w_{\rm r}^2 \rangle = \langle w_{\rm t}^2 \rangle = \frac{1}{3} \langle w^2 \rangle$.
For clarity, the data points and lines in the right panel for $f = \frac{1}{2}$ and $\frac{1}{4}$ are shifted upward by a factor of 3 and 9, 
respectively.}
\label{rt} 
\end{figure*}

We point out that, in principle, the Richardson constant $g$ is not a free parameter, as it is 
physically controlled by the properties of the turbulent flow and the particle dynamics. However, 
the exact value of $g$ for inertial particles is largely unknown, and its behavior with decreasing $r$ and 
the flow Reynolds number, $Re$, is currently an open question. We thus treated $g$ 
essentially as a free parameter in our model prediction. Our assumption concerning particle 
separation will be refined and improved with the help of future numerical studies that directly 
investigate the backward-in-time separation 
of inertial particles. Due to the uncertainty of $g$, it is useful to examine the dependence of our model 
prediction on $g$. The $g-$dependence of our model for equal-size particles was shown in the left 
panel of  Fig.\ 2  in PP10. It was found in PP10 that the dependence on $g$ is rather weak, and, for 
inertial-range particles, the model prediction for $\langle w^2\rangle^{1/2}$ scales with $g$ as $g^{1/3}$.  
Note that the right panel of Fig.\ 2 in PP10 assumed a different separation behavior and a different Reynolds number 
than in the current work. Repeating the same analysis of $g$-dependence 
with the separation behavior adopted in this work and $Re$ in our simulated flow, 
we found a similar weak dependence on $g$: The predicted $\langle w^2\rangle^{1/2}$ decreases 
only by $\lsim 20\%$ for $ 0.8 \lsim St \lsim 10$, as $g$ decreases by each factor of 2, consistent with the 
$g^{1/3}$ dependence. This weak $g$-dependence, to some degree, alleviates the problem 
of uncertainties in $g$ and its behavior with $r$ and $Re$. In order to estimate 
the effects of the uncertainties in $g$, we also attempted to compare our model 
with $g$ fixed at a single value of $1.6$ against simulation results for all $r$. It turns out that, due to the weak $g-$dependence, 
our model with $g=1.6$ is in acceptable agreement with the data for all 
$r \gsim \frac{1}{4}\eta$.  For example, for equal-size particles, changing $g$ from the best value of 1.0  for $r=\frac{1}{4} \eta$ 
to1.6 only leads to a $\lsim 15\%$ overestimate for the $St$ 
range $0.4 \lsim St \lsim 3$. Also $g$ does not affect the acceleration term, and thus 
the $g-$dependence is even weaker for particle of different sizes.
%if the acceleration term makes a considerable contribution. 
If the Stokes numbers differ by a factor of $\gsim 2$,  our model prediction 
is barely affected as $g$ is changes from 1 (or 1.3)  to 1.6.  We emphasize that 
fixing $g$ to a single value here is for an illustration, as 
there is no evidence that $g$ is constant in our flow. As mentioned earlier, $g$ may be constant 
with $r$ in realistic flows with much larger $Re$. 

%because the relative velocity of there particles does not converge at $r=\frac{1}{4}\eta$ for small particles.  
%In fact,  with $St \gsim 1$.
%In other words, if a particle has a Stokes number larger than $6.2$, its relative velocity 
%with any other particle is $r-$independent.   

\subsection{The radial and tangential rms relative speeds}

In Fig.\ \ref{rt}, we show the rms relative speeds in the radial 
($\langle w_{\rm r}^2 \rangle^{1/2}$; filled symbols) and tangential 
($\langle w_{\rm t}^2 \rangle^{1/2}$; open symbols) directions  
at $r=1$, $\frac{1}{2}$ and $\frac{1}{4}\eta$.  In the left panel, we fix 
$St_1$ at $0.19$, $0.78$ and $3.21$, while the right panel shows 
fixed Stokes ratios, $f=1$, $\frac{1}{2}$ and $\frac{1}{4}$. 
Results on the radial and tangential relative speeds in the monodisperse 
case (the $f=1$ case in the right panel of Fig.\ \ref{rt}) were already discussed in Paper I.
%identical particles. 
The S-T formula 
%, eq.\ (\ref{saffmanturner}), predicts that the shear contribution to the relative 
%speed variance ($\langle w_{\rm t}^2 \rangle$) in the tangential direction is twice larger than 
%that ($\langle w_{\rm t}^2 \rangle$) in the radial direction. This suggests 
predicts that the tangential rms is larger than the radial 
one by $\sqrt{2}$ for identical particles with $St \ll 1$ (see eq.\ (\ref{saffmanturner})). This factor originates 
from the difference in the longitudinal and tangental structure functions 
of incompressible turbulence. However, in Paper I we found that, at $\frac{1}{4} \eta \le r \le 1\eta$, 
the tangential-to-radial rms ratio is about 1.1 for the smallest particles ($St=0.1$) in our simulation, 
and it decreases to unity at $St \simeq 1$, above which $\langle w_{\rm t}^2 \rangle^{1/2}$ and  $\langle w_{\rm r}^2 \rangle^{1/2}$ 
are equal (see the $f=1$ data in the right panel). 
One reason for the near equality of $\langle w_{\rm t}^2 \rangle^{1/2}$ and $\langle w_{\rm r}^2 \rangle^{1/2}$ 
for the $St \lsim 1$ particles is the deviation of their trajectories 
from the flow elements, which randomizes the direction of the relative velocity 
with respect to the particle separation, ${\bs r}$. 
%leading to $\langle w_{\rm r}^2 \rangle \simeq  \langle w_{\rm t}^2 \rangle$
%for all particles with $St \gsim 0.1$.  Another 
For larger particles, the effect of the particle memory of the 
spatial flow velocity difference in the past and the stochastic particle 
separation backward in time also tends to equalize $\langle w_{\rm r}^2 \rangle^{1/2}$ 
and $\langle w_{\rm t}^2 \rangle^{1/2}$. %This latter effect is stronger for larger $St$ and/or at smaller $r$.}. 
%This is also observed from the open and filled data points at dip centers in Fig.\ \ref{rt}. 
%This ratio is significantly smaller than the S-T prediction. 
%We concluded in Paper I that eq.\ (\ref{randomdirection}) for the angular average of $S_{\rm ij}$ is preferred as it predicts 
%$\langle w_{\rm r}^2 \rangle = \langle w_{\rm t}^2 \rangle$.

In the bidisperse case, the acceleration contributions to the radial and tangential components are 
equal both in the S-T formula (eq.\ (\ref{saffmanturner})) and in our model prediction 
(see eq.\ (\ref{accelform})). Therefore, the tangential-to-radial ratio for different particles is 
expected to be closer to unity than the monodisperse case. In left panel of Fig.\ \ref{rt},  
we see that $\langle w_{\rm t}^2 \rangle^{1/2}$ is slightly larger than $\langle w_{\rm r}^2 \rangle^{1/2}$ at the dip 
center for $St_1=0.19$. As $St_2$ moves away from $St_1=0.19$,  the acceleration contribution increases,  
leading to  a decrease in the difference between $\langle w_{\rm r}^2 \rangle^{1/2}$ and $\langle w_{\rm t}^2 \rangle^{1/2}$. 
%which 
%tends to equalize $\langle w_{\rm r}^2 \rangle^{1/2}$ and $\langle w_{\rm t}^2 \rangle$. 
There is also a general trend 
for $\langle w_{\rm r}^2 \rangle^{1/2}$ and $\langle w_{\rm t}^2 \rangle$ to equalize 
as $r$ decreases. For the bidisperse case, 
this corresponds to a relative increase in the acceleration contribution, 
as the shear term decreases with decreasing $r$ for small particles in the S-T limit.
%with $St_{1,2} \lsim 1$ 
The same behavior is seen in the right panel.
%This explains why the tangential and radial radial speeds for $St_1= 0.19$ become closer as $r$ decreases (see Fig.\ \ref{rt}).
%to radial difference with decreasing $r$. 
If one Stokes number is larger  than $\simeq 1$, the tangential-to-radial ratio is unity, since 
$\langle w_{\rm r}^2 \rangle^{1/2}$ and $\langle w_{\rm t}^2 \rangle^{1/2}$ are already equal in the monodispese case with $St \gsim1$.
%, and the acceleration contribution also 
%tends to equalize $\langle w_{\rm t}^2 \rangle^{1/2}$ and $\langle w_{\rm r}^2 \rangle^{1/2}$ 
%Since the radial and tangential rms relative speeds are close for identical particles.  

Due to the almost equality of the radial and tangential rms speeds for all Stokes pairs 
in our simulation, we adopt  eq.\ (\ref{randomdirection}) for the angular average of the 
trajectory structure tensor, $\langle S_{{\rm T}ij} \rangle_{\rm ang}$, which predicts 
$\langle w_{\rm r}^2 \rangle = \langle w_{\rm t}^2 \rangle = \frac{1}{3} \langle w^2\rangle$. 
The lines in Fig.\ \ref{rt} are our model predictions from this equation\footnote{As discussed in Appendix B, 
using eq.\ (26) in Paper I for  $\langle S_{ij} \rangle_{\rm ang}$ 
would recover the S-T prediction for small particles, which, however, gives poorer fits to the data.}, 
which are in good agreement with the data points. Again, we adopted bi-exponential $\Phi_1$ and $\Phi_2$ 
and a two-phase separation behavior with $g=1.6$, $1.3$ and $1.0$ for  $r=1$, $\frac{1}{2}$, and $\frac{1}{4}\eta$, 
respectively. The $r-$dependence of $\langle w_{\rm r}^2 \rangle^{1/2}$ and $\langle w_{\rm t}^2 \rangle^{1/2}$ 
is similar that of the 3D rms shown in Fig.\ \ref{3drmsscale}.
\\

To summarize \S 6.3 and 6.4, we found that the interesting 
features of the rms relative velocity for equal-size particles discussed in Paper I  
become weaker in the bisdisperse case. The generalized acceleration term is rather featureless: 
It is independent of the particle distance, and provides equal contributions to the radial and tangential
components of the relative velocity.  
As the acceleration contribution increases with increasing Stokes number difference, both the $r-$dependence and the tangential-to-radial ratio decrease.
In Appendix D, we follow Paper I to split particle pairs at given distances into two 
groups with negative and positive radial relative 
speed, corresponding to particles approaching and separating from each other, respectively.  
%examine the difference in the relative velocities of particle pairs that are 
%approaching or separating from each other, compare 
%Splitting particle pairs into these two groups 
This division is of interest because only approaching particle 
pairs may lead to collisions. Paper I found that, 
for equal-size particles with $St \lsim 6$, approaching pairs have larger relative speed than separating ones.  
In the bidisperse case, the acceleration contribution is independent of the relative motions of the two 
particles as it depends only on the flow velocity statistics along individual particle trajectories.  
Therefore, as shown in Appendix D, the asymmetry between approaching and separating pairs is 
weaker in the bidisperse case than in the monodisperse one.

\section{Summary and Conclusions}

We have investigated the relative velocity of inertial particles  suspended in turbulent 
flows, extending our earlier work on equal-size particles (Pan \& Padoan 2013; Paper I)  to  
the general bidisperse case for different particles of arbitrary sizes.  
We have made use of the same numerical simulation presented in 
Paper I, which evolved 14 species of inertial particles in a simulated turbulent flow.
% covering the entire scale range of the  
The particle friction time, $\tau_{\rm p}$, ranges from $0.1\tau_\eta$ ($St =0.1$) to $54T_{\rm L}$ ($St=795$), 
with $\tau_\eta$ and $T_{\rm L}$ the Kolmogorov timescale and the Lagrangian correlation 
time of the flow, respectively.  We computed the rms relative velocity, 
$\langle w^2 \rangle ^{1/2}$, for all Stokes number pairs $(St_1, St_2)$ available in the simulation, 
and tested the PP10 model for the general bidisperse case. 
%We have also conducted a systematic analysis of the probability distribution function (PDF) of the relative velocity 
%as a function of the Stokes number pair.  
Here we list our main conclusions.

\begin{enumerate}

\item As a special bidisperse case, we examined the relative velocity, ${\bs w}_{\rm f}$, 
between inertial particles and the local flow velocity.  We showed that 
${\bs w}_{\rm f}$ can be roughly estimated as the {\it temporal} 
flow velocity difference, $\Delta {\bs u}_{\rm T}(\Delta \tau)$, 
along the particle trajectory at a time lag, $\Delta \tau$, close 
to the particle friction time $\tau_{\rm p}$. A simple model is developed for the rms of ${\bs w}_{\rm f}$, assuming 
that the temporal flow velocity correlation on the particle trajectory can be 
approximated by the Lagrangian correlation function, $\Phi_{\rm L}$. 
Adopting a bi-exponential form for $\Phi_{\rm L}$, our model 
is in good agreement with the simulation data. In particular, 
it predicts that the rms of  ${\bs w}_{\rm f}$ increases linearly with $St$ for $St\ll1$,  scales as $St^{1/2}$  
in the inertial range, and finally approaches the flow rms velocity for $\tau_{\rm p} \gg T_{\rm L}$.
 %and approaches constant for $\tau_{\rm p} \gg T_{\rm L}$. 
%The PDF shape of  ${\bs w}_{\rm f}$ becomes thinner with increasing $St$, 
%and approaches Gaussian for $\tau_{\rm p} \gsim 3.5-7 T_{\rm L}$. This behavior 
%corresponds to the thinning trend of the PDF of $\Delta {\bs u}_{\rm T}$
%with increasing time lag as inferred from the PDFs of the 
%Lagrangian and Eulerian temporal flow velocity differences. 
The particle-flow relative velocity is an interesting delimiter that 
helps confine the relative velocity behavior in the general bidisperse case.  

\item We introduced the general formulation of PP10 for 
the relative velocity of different particles of arbitrary sizes. The formulation 
shows that the relative velocity variance is contributed by two terms, named as the 
generalized acceleration and shear terms because they reduce to the acceleration 
and shear terms in the Saffman-Turner formula 
for small particles with $St \ll 1$.  The generalized acceleration term originates from 
different responses of particles of different sizes to the flow velocities. 
We established an approximate relation between the generalized acceleration 
term and the {\it temporal} flow velocity difference, $\Delta {\bs u}_{\rm T}$, 
along the trajectory of the larger particle.  
On the other hand, the generalized shear term represents the contribution from the 
particles' memory of the {\it spatial} flow velocity difference, $\Delta {\bs u}$, across the distance 
of the two particles at given times in the past. %(see detailed discussions in Paper I). 
An analytical expression is derived for the generalized acceleration term, while the generalized shear 
term is modeled in a similar way as  the mondispserse model presented in Paper I, 
accounting for the combined effects of the particle memory and 
the separation of particle pairs backward in time. For equal-size particles, 
the acceleration term vanishes, and only the shear term contributes. 

\item Using our simulation, we computed the rms relative velocity, $\langle w^2 \rangle ^{1/2}$, between particles 
of any different sizes. We first examined  $\langle w^2 \rangle ^{1/2}$ as a function of $St_2$ at fixed values of $St_1$.  
If $\tau_{\rm p1} \lsim T_{\rm L}$, the relative velocity shows a dip around 
$St_2 \simeq St_1$, indicating that the velocities of nearby particles of similar sizes 
have a tighter correlation than particles of different sizes. The dip disappears for 
$\tau_{\rm p1} \gsim T_{\rm L}$. The generalized shear term dominates the 
contribution to the rms relative velocity for particles of similar sizes,
while the acceleration term dominates if the Stokes numbers differ by more 
than a factor of $\simeq 4$. %In the limits $St_2 \to 0$ and $St_2 \to \infty$,  $\langle w^2 \rangle ^{1/2}$ 
%approaches the particle-flow relative velocity and the 1-particle velocity of particles (1), respectively. 

Defining the ratio, $f\equiv  St_{\ell}/St_{h}$, between the small ($St_{\ell}$) and large ($St_{h}$) Stokes numbers, 
we also considered $\langle w^2 \rangle ^{1/2}$
as a function of $St_{h}$ at fixed values of $0\le f\le 1$. 
The limits $f\to0$ and $f\to1$ correspond to the particle-flow relative velocity 
and the monodisperse case, respectively. At a fixed $f$, 
$\langle w^2 \rangle ^{1/2}$ increases with $St_{h}$ for $ \tau_{\rm ph} \lsim T_{\rm L}$, 
stays roughly constant for $T_{\rm L} \lsim \tau_{\rm ph} \lsim T_{\rm L}/f$ (or equivalently 
$f T_{\rm L}  \lsim \tau_{\rm p l} \lsim T_{\rm L}$), and finally decreases as $St_{h}^{-1/2}$ for 
$\tau_{h} \gg T_{\rm L}/f$. For any value of $f$, a $St_{h}^{1/2}$ scaling is 
predicted, if the larger friction time, $\tau_{\rm p,h}$, is within the inertial range of the 
flow. This $St_{h}^{1/2}$ scaling will have to be verified in future simulations with higher 
resolutions. At a given $St_{h}$,  $\langle w^2 \rangle ^{1/2}$ increases with decreasing $f$ due to the 
increase of the acceleration contribution, which starts to dominate at $f\lsim 1/4$.  

The generalized acceleration contribution is independent of the distance, $r$, %or the relative motions of the two particles, and provides equal contributions to 
%the radial and tangential components of the relative velocity, 
and thus reduces the $r-$dependence of  the relative velocity 
between small, different  particles, making it easier to achieve 
numerical convergence for the collision statistics of point-like particles at $r \to 0$. 

The prediction of the PP10 model is in good agreement with 
the simulation data. The largest discrepancy occurs for $f=\frac{1}{2}$ and $St_{h}$ 
in the inertial range, where the model underestimates the rms 
relative velocity by 15-20\%. At other values of $f$, the discrepancy between our model and the simulation
is $<10\%$.  This confirms the validity of the physical picture revealed by our model. 
%between different particles than equal-size particles in the Saffman-Turner limit.   

%the difference in the radial and tangential rms relative speeds, and the asymmetry between 
%approaching and separating pairs found in Paper I for identical particles all decrease when 

\end{enumerate}

%Despite extensive conclusions drawn in the current work for the particle relative 
%velocity in the general bidisperse case, 
%e will focus on Future developments will be made to investigate more practical measures for the study 
%to dust particle collisions in protoplaneraty turbulence.  
We emphasize that the theoretical modeling of the rms relative velocity is 
important for understanding the fundamental physics, even though its practical use is 
limited. In future work, we will focus on establishing statistical measures or 
tools that can be applied to model dust particle collisions in protoplanetary turbulence. 
We have started an effort in an ongoing paper (Pan \& Padoan 2014) 
to explore the collision kernel %and the average collision velocity as a function of the Stokes number pair  
in the general bidisperse case, accounting for  turbulence-induced collision velocity 
and the effect of turbulent clustering.  In the next paper of this series, we will 
systematically examine the probability distribution of the collision velocity, which is needed 
to determine the fractions of collisions leading to sticking, bouncing or fragmentation. 
Due to the limited resolution, the simulated flow in the current work has only a short inertial range, 
and our model prediction for particles in the inertial range remains to be tested and 
validated.  %Dust particles of sub-millimeter to meter size in protoplanetary turbulence  belong 
%to the inertial range, and modeling the collisions of these particles is crucial to assessing the viability of the 
%planetesimal formation mechanism by particle coagulation.  
Future simulations at higher resolutions are being planned to obtain accurate 
measurements for  the collision statistics of  inertial-range particles.  

\acknowledgements 
Resources supporting this work were provided by the NASA High-End Computing (HEC) Program through the NASA Advanced Supercomputing (NAS) 
Division at Ames Research Center,and by the Port d'Informaci— Cient'fica (PIC), Spain, maintained by a collaboration of the 
Institut de F'sica d'Altes Energies (IFAE) and the Centro de Investigaciones EnergŽticas, Medioambientales y
Tecnol—gicas (CIEMAT).
LP is supported by a Clay Fellowship at Harvard-Smithsonian Center for Astrophysics. 
PP acknowledges support by the FP7-PEOPLE- 2010-RG grant PIRG07-GA-2010-261359.

\appendix

\section{A: The Formulation of Pan \& Padoan (2010)}

We review the general PP10 formulation for particles of different sizes.  It follows from eq.\ (\ref{particlestructure}) that the particle 
velocity structure tensor, $S_{{\rm p}ij}$, has four terms, $\left\langle v^{(1)}_i v^{(1)}_j \right\rangle$, $-\left\langle v^{(1)}_i v^{(2)}_j \right\rangle$, $-\left\langle v^{(2)}_i v^{(1)}_j \right\rangle$, and $\left\langle v^{(2)}_i v^{(2)}_j \right\rangle$. Inserting the formal solution, eq.\ (\ref{formalsolution}), for $\bs{v}^{(1)}$ and/or  $\bs{v}^{(2)}$ 
into each term yields a double integral. The terms $\left \langle v^{(1)}_i v^{(1)}_j \right \rangle$ and $\left \langle v^{(2)}_i v^{(2)}_j \right \rangle$ 
correspond to the 1-particle velocity variances, which have been evaluated in \S 2 of Paper I. For particle (1), $\left \langle v^{(1)}_i v^{(1)}_j \right \rangle$ at $t=0$ 
can be calculated as,  
\begin{equation}
%\left \langle v^{(1)}_i v^{(1)}_j \right \rangle  =  \int_{-\infty}^0  \frac {d\tau}{\tau_{p1}} \int_{-\infty}^0 \frac {d\tau'}{\tau_{p1}}
% \left \langle u_i \left({\bs X}^{(1)} (\tau), \tau\right) u_j \left({\bs X}^{(1)}(\tau'),\tau'\right) \right \rangle \exp \left( \frac {\tau}{\tau_{p1}} %\right) \exp \left (\frac{\tau'}{\tau_{p1}}\right)        
\left \langle v^{(1)}_i v^{(1)}_j \right \rangle  =  \int_{-\infty}^0  \frac {d\tau}{\tau_{\rm p1}} \int_{-\infty}^0 \frac {d\tau'}{\tau_{\rm p1}}
 B^{(1)}_{ij}(\tau, \tau') \exp \left( \frac {\tau}{\tau_{\rm p1}} \right) \exp \left (\frac{\tau'}{\tau_{\rm p1}}\right),  
%\label{eq1}
\end{equation}
where $B^{(1)}_{ij} = \left (\equiv \left \langle u_i^{(1)} (\tau) u_j^{(1)}(\tau') \right \rangle \right)$ is the trajectory correlation tensor of particle (1) (see \S 2).
%(see \S 2 and Paper I).
A similar equation can be derived for $\left \langle v^{(2)}_i v^{(2)}_j \right\rangle$, which depends on $B^{(2)}_{ij}$. %of particle (2).
%$\left(\equiv \left \langle u_i^{(2)} (\tau) u_j^{(2)}(\tau') \right \rangle \right)$. 
%Note that we have replaced the lower and upper limits in the formal solution, eq.\ (\ref{formalsolution}) by $0$ and $\infty$. This is because we are mainly 
%interested in the case where the particle dynamics has fully relaxed and the velocity statistics reached a steady state. 
%The former allows us to set $t_0 \to -\infty$ and the latter means that.  

The cross terms, $\left \langle v^{(1)}_i v^{(2)}_j \right\rangle$ and $\left\langle v^{(2)}_i v^{(1)}_j \right\rangle$, involve the 
memories of the flow velocities by both particles.  %For particles of different sizes, the exponential memory 
%cutoffs are different.  
For the first cross term, we have,
\begin{equation}
%\left \langle v^{(1)}_i v^{(1)}_j \right \rangle  =  \int_{-\infty}^0  \frac {d\tau}{\tau_{p1}} \int_{-\infty}^0 \frac {d\tau'}{\tau_{p1}}
% \left \langle u_i \left({\bs X}^{(1)} (\tau), \tau\right) u_j \left({\bs X}^{(1)}(\tau'),\tau'\right) \right \rangle \exp \left( \frac {\tau}{\tau_{p1}} %\right) \exp \left (\frac{\tau'}{\tau_{p1}}\right)        
\left \langle v^{(1)}_i v^{(2)}_j \right \rangle  =  \int_{-\infty}^0  \frac {d\tau}{\tau_{\rm p1}} \int_{-\infty}^0 \frac {d\tau'}{\tau_{\rm p2}}
\left \langle u_i^{(1)} (\tau) u_j^{(2)}(\tau') \right \rangle  \exp \left( \frac {\tau}{\tau_{\rm p1}} \right) \exp \left (\frac{\tau'}{\tau_{\rm p2}}\right),  
%\label{eq1}
\end{equation}
where $\left \langle u_i^{(1)} (\tau) u_j^{(2)}(\tau') \right \rangle$ is the correlation of the flow velocities seen by particles (1) and (2) at $\tau$ and $\tau'$, respectively.  
A similar integral equation exists for $\left\langle v^{(2)}_i v^{(1)}_j \right\rangle$, 
which contains $\left \langle u_j^{(1)} (\tau) u_i^{(2)}(\tau') \right \rangle$. 
The sum of $\left \langle u_i^{(1)} (\tau) u_j^{(2)}(\tau') \right \rangle$ 
and $\left \langle u_j^{(1)} (\tau) u_i^{(2)}(\tau') \right \rangle$ can be written as,  
\begin{equation}
%\left \langle v^{(1)}_i v^{(1)}_j \right \rangle  =  \int_{-\infty}^0  \frac {d\tau}{\tau_{p1}} \int_{-\infty}^0 \frac {d\tau'}{\tau_{p1}}
% \left \langle u_i \left({\bs X}^{(1)} (\tau), \tau\right) u_j \left({\bs X}^{(1)}(\tau'),\tau'\right) \right \rangle \exp \left( \frac {\tau}{\tau_{p1}} %\right) \exp \left (\frac{\tau'}{\tau_{p1}}\right)        
 \left \langle u_i^{(1)} (\tau) u_j^{(2)}(\tau') \right \rangle  +  \left \langle u_j^{(1)} (\tau) u_i^{(2)}(\tau') \right \rangle  
= -S_{{\rm T}ij} ({\bs r}, \tau, \tau')  +  B_{ij}^{(1)} (\tau, \tau')+  B_{ij}^{(2)} (\tau, \tau')  - C_{ij}({\bs r}, \tau, \tau'), 
\label{identity}
\end{equation}
where $S_{{\rm T}ij} ({\bs r}, \tau, \tau')$ is named as the trajectory structure tensor. It is defined as, 
\begin{equation}
S_{{\rm T}ij} ({\bs r}, \tau, \tau') = \left \langle \left[u_i^{(1)} (\tau) - u_i^{(2)} (\tau) \right] \left[ u_j^{(1)} (\tau') - u_j^{(2)} (\tau') \right]\right \rangle, 
\label{trajectstructuretensor}
\end{equation} 
which represents the correlation of the flow velocity differences seen by the two particles 
at two times. It depends on the particle separation ${\bs r}$ at $t=0$ through the constraint ${\bs X}^{(2)} (0) -  {\bs X}^{(1)} (0) = {\bs r}$. 
%We modeled $S_{{\rm T}ij} $ in detail in Paper I for the monodisperse case. 

The last term in eq.\ (\ref{identity}) is defined as $C_{ij} ({\bs r}, \tau, \tau') \equiv \left \langle u_j^{(1)} (\tau') u_i^{(2)} (\tau) \right \rangle - \left \langle u_j^{(1)} (\tau) u_i^{(2)} (\tau') \right \rangle$.  %A possible dependence on ${\bs r}$ is explicitly indicated. 
Since $C_{ij}$ is anti-symmetric under the exchange of $\tau$ and $\tau'$, it is easy to 
see that ${\Large \int} \frac{d \tau}{\tau_{\rm p1}} {\Large \int} \frac{d\tau'}{\tau_{\rm p2}} C_{ij} \exp(\frac{\tau}{\tau_{\rm p1}}) \exp(\frac{\tau'}{\tau_{\rm p2}})$ is 
zero for equal-size particles with $\tau_{\rm p1} = \tau_{\rm p2}$.  It can also been shown that 
$C_{ij} = 0$ if both particles are small with $St_{1,2} \ll 1$. On the other hand, the term is not expected to 
exactly vanish for particles of arbitrarily different sizes.  %This is because the temporal statistics 
%of the flow velocity along the trajectories of different particles may be different. 
%For example, a small particle would follow the Lagrangian 
%trajectory, while the flow velocity seen by a very large particle may be closer to Eulerian.  
%This may lead to 
For example,  given the particle distance at $\tau$,  $\left \langle u_j^{(1)} (\tau') u_i^{(2)} (\tau) \right \rangle$ 
depends on the trajectory  of particle (1), or more precisely, the flow velocity decorrelation along its trajectory 
from $\tau$ to $\tau'$, while $\left \langle u_j^{(1)} (\tau) u_i^{(2)} (\tau') \right \rangle$ is controlled by the trajectory of 
particle (2). Therefore, a difference may exist between $\left \langle u_j^{(1)} (\tau') u_i^{(2)} (\tau) \right \rangle$ 
and $\left \langle u_j^{(1)} (\tau) u_i^{(2)} (\tau') \right \rangle$ due to the different 
temporal statistics along the trajectories of the two particles. However, Paper I showed that the Lagrangian 
and Eulerian temporal correlation functions in our simulated flow are close to each 
other, meaning that the decorrelation of the flow velocity along a  trajectory 
of a small particle in the $St\ll 1$ limit may be similar to that for a 
large particle with $\tau_{\rm p} \gg T_{\rm L}$. 
%This result was used to justify the assumption 
%made in Paper I  that the trajectory correlation function ($B_{{\rm}ij}$) for particles of any 
%size can be approximated by the Lagrangian correlation function. 
%Such an assumption was made in Paper I for the monodisperse case. 
%This assumption will also be adopted throughout our model for the bidisperse case.  
Based on this extreme case, one could assume  that, qualitatively, 
$\left \langle u_j^{(1)} (\tau') u_i^{(2)} (\tau) \right \rangle
%$ and $
\simeq \left \langle u_j^{(1)} (\tau) u_i^{(2)} (\tau') \right \rangle$  for  particles of any different sizes. 
%are approximately equal.  
We thus neglect $C_{ij}$ in our model, even though the quantitative accuracy of the assumption is unclear.
%Second, for identical particles, the trajectories of the two particles are statistically equivalent, 
%and we expect $\left \langle u_j^{(1)} (\tau') u_i^{(2)} (\tau) \right \rangle 
The $C_{ij}$ term was ignored in PP10, where it was found 
that, without $C_{ij}$, the model prediction is in good agreement 
with the simulation results of Zhou et al.\ (2001) for the bidisperse case. 

Adding the four terms of $S_{{\rm p}ij}$ together, rearranging the integrals using eq.\ (\ref{identity}), 
and neglecting $C_{ij}$, the particle structure tensor can be written as two terms, 
\begin{equation}
S_{{\rm p}ij} ({\bs r}) = \mathcal{A}_{ij} +\mathcal{S}_{ij} {\rm}, 
\label{pp10form}
\end{equation}
where $\mathcal{A}_{ij}$ and $\mathcal{S}_{ij}$, named the generalized acceleration and shear terms, respectively (PP10),  
reduce to the acceleration and shear terms in the S-T limit, eq.\ (\ref{saffmanturner}), for $St_{1,2} \ll 1$, respectively (see \S 3.1 and \S 3.2.1).  
%The tensor $\mathcal{T}$ is given by 
%\begin{equation}
% \mathcal{T}_{ij} =  \int_{-\infty}^0  \frac {d\tau}{\tau_{\rm p1}}\int_{-\infty}^0 \frac{d\tau'}{\tau_{\rm p2}}C^{(1)(2)}_{ij} ({\bs r}; \tau, \tau') \exp \left(\frac{\tau}{\tau_{\rm p1}} \right) \exp \left(\frac{\tau'}{\tau_{\rm p2}}\right)
%\end{equation}

The generalized acceleration term is given by,  
\begin{gather}
\vspace {2mm}
\mathcal{A}_{ij} = \hspace{2mm} \int_{-\infty}^0  \frac {d\tau}{\tau_{\rm p1}} \int_{-\infty}^0 \frac {d\tau'}{\tau_{\rm p1}} B^{(1)}_{{\rm}ij}(\tau, \tau') \exp \left (\frac {\tau}{\tau_{\rm p1}} \right) \exp \left(\frac{\tau'}{\tau_{\rm p1}}\right)  \hspace{2.5cm} \notag \\
\hspace{1.cm} -\int_{-\infty}^0  \frac {d\tau}{\tau_{\rm p1}} \int_{-\infty}^0 \frac {d\tau'}{\tau_{\rm p2}} \left[B^{(1)}_{{\rm}ij}(\tau, \tau') + B^{(2)}_{{\rm}ij}(\tau, \tau')\right]\exp \left(\frac{\tau}{\tau_{\rm p1}}\right) \exp \left(\frac{\tau'}{\tau_{\rm p2}}\right)  \notag \\ 
%\hspace{0.9cm} {\displaystyle -\int_{-\infty}^0  \frac {d\tau}{\tau_{p1}} \int_{-\infty}^0 \frac {d\tau'}{\tau_{p2}} \left( B^{(1)}_{{\rm T}ij}(\tau, \tau') + B^{(2)}_{{\rm T}ij}(\tau, \tau') \right) \exp \left (\frac {\tau}{\tau_{p1}} \right) \exp \left(\frac{\tau'}{\tau_{p2}}\right) }\\ %\vspace{-4mm}
\hspace{0.cm} + \int_{-\infty}^0  \frac {d\tau}{\tau_{\rm p2}} \int_{-\infty}^0 \frac{d\tau'}{\tau_{\rm p2}}  B^{(2)}_{{\rm}ij}(\tau, \tau')  
\exp \left(\frac{\tau}{\tau_{\rm p2}}\right) \exp \left(\frac{\tau'}{\tau_{\rm p2}}\right). \hspace{1.5cm} 
\label{accel}
\end{gather}
Clearly, $\mathcal{A}_{ij}$ vanishes if $\tau_{\rm p1} = \tau_{\rm p2}$. $\mathcal{A}_{ij}$ is independent 
of ${\bs r}$, as it depends only on the flow velocity statistics ($B^{(1)}_{{\rm}ij}$ and $B^{(2)}_{{\rm}ij}$) along {\it individual} trajectories of the two particles. 

The generalized shear term reads, 
\begin{equation}            
\mathcal{S}_{ij} = \int_{-\infty}^0  \frac {d\tau}{\tau_{\rm p1}}\int_{-\infty}^0 \frac{d\tau'}{\tau_{\rm p2}}
S_{{\rm T}ij} ({\bs r}; \tau, \tau') \exp \left(\frac{\tau}{\tau_{\rm p1}} \right) \exp \left(\frac{\tau'}{\tau_{\rm p2}}\right), 
\label{shear}
\end{equation}
which represents the contribution from the particles' memory of the flow velocity difference they saw in the past. 
In the limit $\tau_{\rm p1}, \tau_{\rm p2} \to 0$, %(or $St_{1,2} \ll 1$),  
the exponential cutoffs can be viewed as delta functions, and we thus have $\mathcal{S}_{ij} \to S_{{\rm T}ij} ({\bs r}; 0, 0) = S_{ij} ({\bs r})$ 
with $S_{ij} ({\bs r})$ the flow structure tensor (PP10 and Paper I). Eq.\ (\ref{shear}) thus reproduces the shear terms in the S-T limit (eq.\ (\ref{saffmanturner})).  

\section{B: Modeling the Generalized Shear Term}

%In particular, it predicts that the tangential rms relative speed  ($\langle w_{\rm t}^2 \rangle^{1/2}$) is larger than the radial one ($\langle w_{\rm r}^2 \rangle^{1/2}$) 
%by a factor of $\sqrt{2}$ in the S-T limit (see the shear terms in eq.\ (\ref{saffmanturner})). However, this was not confirmed by our simulation 
%data for equal-size particles with $St \gsim 0.1$. 
 
We list the assumptions made to calculate eq.\ (\ref{sij}) for the generalized shear term, $\mathcal{S}_{ij}$.
In an isotropic flow,  the Eulerian structure tensor $S_{ij} ({\bs R})$ in eq.\ (\ref{sij}) is written as 
$S_{\rm nn}(R) \delta_{ij}  + [S_{\rm ll}(R)- S_{\rm nn}(R)]\frac{R_i R_j} {R^2}$, with $S_{\rm ll}$ and $S_{\rm nn}$ the 
longitudinal and transverse structure functions. To evaluate the angular average of $
S_{ij} ({\bs R})$, we assume that the direction of ${\bs R}$ is  isotropic and random, resulting in
%\begin{equation}
$\langle S_{ij} ({\bs R}) \rangle_{\rm ang} = \frac{1}{3}\left[S_{\rm ll}(R) + 2 S_{\rm nn} (R) \right] \delta_{ij}$ (which is eq.\ (\ref{randomdirection}) in the text). 
%\label{randomdirection2}
%\end{equation} 
This assumption suggests the shear contributions for the radial and tangential rms relative speeds are 
equal for all particles. %In the monodisperse case, this second assumption was found to be in better agreement with simulation data 
%(Paper I).  
%With eq.\ (\ref{randomdirection}) for $\langle S_{ij} ({\bs R})\rangle_{\rm ang}$, 
%the radial and tangential relative velocity variances are equal, $\langle w^2_{\rm r} \rangle = \langle w^2_{\rm t} \rangle = \langle w^2 \rangle/3$. 
%In this case, the radial and tangential rms  values are immediately known once $\langle w^2\rangle$ 
%is solved. 
%The difference in the two assumptions for $\langle S_{ij} ({\bs R})\rangle_{\rm ang}$ is 
%that eq.\ (\ref{randomdirection1}) reproduces the shear terms in eq.\ (\ref{saffmanturner}) for 
%$\langle w_{\rm r}^2\rangle$ and $\langle w_{\rm t}^2\rangle$ in the S-T limit, while, 
%and, 
Under this assumption, the shear terms In the S-T limit ($St_{1,2} \ll 1$) are given by $\frac{\bar{\epsilon}}{9 \nu} r^2$ 
for both $\langle w_{\rm r}^2\rangle$ and $\langle w_{\rm t}^2\rangle$. %The other approximation for $\langle S_{ij} ({\bs R})\rangle_{\rm ang}$ given in Appendix A (eq.\ (\ref{randomdirection1})) recovers 
In PP10 and Paper I, another assumption was adopted for the angular average (see eq.\ (26) in paper I), 
%where %the direction of the separation change ${\bs R} -{\bs r}$ rom the initial value, 
%${\bs r}$, (rather than ${\bs R}$) is taken to be isotropic (see eq.\  (XX) in Paper I) 
which exactly reproduces the shear terms %$ \frac{1}{15} \frac {\bar{\epsilon}}{\nu} r^2$ and $\frac{2}{15} \frac {\bar{\epsilon}}{\nu} r^2$, for 
for the radial and tangential variances in the S-T formula (eq.\ (\ref{saffmanturner})) 
%predicts that  $\langle w_{\rm t}^2\rangle = 2 \langle w_{\rm t}^2\rangle$ for 
%identical particles
for $St_{1,2} \ll 1$. This second assumption is, however, in poorer agreement with our simulation data for particles with $St \gsim 0.1$. 
We mainly consider eq.\ (\ref{randomdirection}) for $\langle S_{ij} ({\bs R})\rangle_{\rm ang}$ in this paper.

%%This leads to,
%%\begin{equation}
%%\langle S_{ij} ({\bs R}) \rangle_{\rm ang} = \left[ \left(\frac{1}{3}-\frac{r^2}{3R^2}\right) S_{\rm ll}(R) + \left( \frac{2}{3} +\frac{r^2 }{3R^2} \right) 
%%S_{\rm nn} (R) \right] \delta_{ij}
%%+ \left[S_{\rm ll}(R)  - S_{\rm nn}(R)\right]\frac{r_i r_j}{R^2},
%%\label{randomdirection1}
%%\end{equation} 
%%which approaches the flow structure tensor $S_{ij} ({\bs r})$ in the limit $R\to r$. 
%to reproduce the S-T prediction for identical particles with $St \ll 1$ (Paper I). 
%\begin{gather}
%\langle w_{\rm r}^2 \rangle =A + \int_{-\infty}^0 \frac {d\tau} {\tau_{\rm p1}}  \int_{-\infty}^0 \frac {d\tau'} {\tau_{\rm p2}} \left[ \left( \frac{1}{3} + \frac{2r^2}{3R^2}\right) S_{\rm ll}(R) + \left(  \frac{2}{3}- \frac{2r^2}{3R^2} \right) 
%S_{\rm nn} (R) \right] \times \notag \\
%\hspace{4.5cm}\Phi_2 \big(\tau' -\tau, R \big)  \exp \left(\frac{\tau}{\tau_{\rm p1}}\right) \exp\left( \frac{\tau'}{\tau_{\rm p2}}\right).
%\label{wr}
%\end{gather} 
%Due to the dependence of $R$ on the 3D variance, $\langle w^2\rangle$, in the 
%ballistic phase,  one needs to solve $\langle w^2\rangle$ before integrating eq.\ (\ref{wr}). 
%The tangential variance can be obtained by $\langle w^2_{\rm t}\rangle= ( \langle w^2\rangle -  \langle w^2_{\rm r} \rangle)/2$.   
%We will mainly use eq.\ (\ref{randomdirection}), which is in better agreement with the simulation data. 

In Paper I we considered both a single and  a bi-exponential form for the temporal correlation function $\Phi_2$ in eq.\ (\ref{sij}), i.e, 
%of the flow velocity differences along the trajectories of the two particles,
\begin{gather}
\Phi_{2}(\Delta \tau, R) = \exp \left(-\frac{- |\Delta \tau|}{T(R) }  \right),
\label{exponential2}
\end{gather}
and, \begin{gather}
\Phi_{2}(\Delta \tau, R)= {\frac{1}{2 \sqrt {1-2z^2} }} \vast\{ \big(1 + \sqrt{1-2z^2}\big) \exp \Bigg[-\frac{2 |\Delta \tau|}{ \big(1+ \sqrt{1-2z^2}\big) T(R) }  \Bigg] - \notag\\
\hspace{4.7cm} \big(1-\sqrt{1-2z^2}\big)   
 \exp \left [ {- \frac{2 |\Delta \tau|}{ \big(1 - \sqrt{1-2z^2} \big) T(R)}}  \right] \vast\}, 
\label{biexponential2}
\end{gather}
where the correlation time $T(R)$ corresponds to the eddy turnover time at the scale $R$. 
%Eq.\ (\ref{biexponential2}) is in the same form as eq.\ (\ref{biexponential}) for 
%$\Phi_{\rm L}$ with $T(R)$  replacing $T_{\rm L}$.  

%The parameter $z$ is the same as that in eq.\ (\ref{biexponential}), 
%and will be set to be a constant, $0.3$, as measured in our  simulated flow (Paper I).  

The typical particle distance, $R$, between $\tau$ and $\tau'$ was approximated by 
\begin{equation} 
R (\tau, \tau')= [d(\tau) d(\tau')]^{1/2},  
\label{distance}
\end{equation}
where $d(\tau)$ and $d(\tau')$ are the rms particle distances, at $\tau$ and $\tau'$.
It was motived by the fact that $S_{{\rm T} ij}$ is zero if either $d(\tau)$ or $d(\tau')$ 
is zero (PP10). 
The backward  separation of inertial particle pairs has not been 
explored until recently (Bragg, Ireland, \& Collins 2014).  The simulation results of  Bec et al.\ (2010) for the forward-in-time dispersion of equal-size 
particles were used %pair dispersion of inertial particles  we used the simulation results of  for the 
as a guide for the assumption for the backward separation (PP10). We adopted a two-phase behavior. 
%as a function of time.
In the first phase, the particle pairs separate ballistically, %with $d(\tau)$ increasing linearly with time, 
\begin{equation}
d^2 (\tau) = r^2 +\langle w^2 \rangle \tau^2, 
\label{ballistic}
\end{equation}
where $\langle w^2 \rangle$ is the 3D relative velocity variance. The second phase follows the Richardson's law,    
\begin{equation}
d^2 (\tau) \simeq g \bar{\epsilon} |\tau|^3,
\label{richardson}
\end{equation}
where $g$ is the Richardson constant (Paper I).  For equal-size particles, the transition 
between the two phases was assumed to occur at a friction time or so, $\tau \simeq -\tau_{\rm p}$, 
based on the results of Bec et al.\ (2010). We thus connect the behaviors at a transition time, 
$\tau_{\rm c} = -\tau_{\rm p}$.    
%for $St \gsim 3$ particles. 
A number of uncertainties in the assumed two-phase behavior were pointed out and 
justified in Paper I for equal-size particles.  In a recent study, Bragg et al.\  (2014) 
examined both forward and backward-in-time separation behaviors of inertial 
particle pairs, and very briefly mentioned possible problem in our assumption 
based on the forward separation behavior. Apparently, their claim of 
the potential accuracy problem in our model prediction due to the 
adopted separation behavior is discussed in a forthcoming paper that has not 
yet been published. We will consider/address their comments in future works.      
%(see Paper I for details). 
%For example, the Richardson behavior is expected to end once $d$ is larger than the 
%integral scale, $L$, of the flow. However, it was shown in Paper I that our model prediction is 
%insensitive to the separation behavior after the particle distance exceeds $L$, and this 
%justifies using the Richardson law at $d \gsim L$. Also, for small particles with $St \lsim 3$, 
%the ballistic separation for a friction timescale or so does not bring the particle distance to inertial-range scales, and there is an intermediate 
%phase before the transition to the Richardson behavior. This complexity was neglected. %as the exact behavior in the intermediate phase is unknown. 
%We refer the reader to Paper I for a detailed discussion on these issue.
%Finally, Bec et al. (2010) only considered initial distances above the Kolmogorov scale,  
%while we are interested in small distances ($r \ll \eta $) for dust particle collisions. 
%A direct study of the backward separation
%for all particles at $r\ll\eta$ may considerably improve the model.  
We will adopt the same two-phase behavior for the bidisperse case. 
An additional uncertainty here is that it is not clear how long the ballistic phase 
lasts, as the two friction times are different. Following PP10, 
we assume that the transition occurs at $\tau_{\rm c} = -(\tau_{\rm p1} +\tau_{\rm p2})/2$. 
This may be questionable for particles of very different sizes. It, 
however, does not present a severe problem because, in that case, 
it is the generalized acceleration term that gives the dominant contribution
%rather than the shear term 
to the relative velocity (see \S 6.1). For the Richardson constant, $g$, we will adopt the values 
used in Paper I that best fit the data for identical particles.

%However, this prediction was not confirmed by 
%our monodisperse result even for the smallest particles ($St \simeq 0.1$) 
%in our simulation (Paper I). 

To calculate the generalized shear term, we also need the flow structure functions, $S_{\rm ll}$ and $S_{\rm nn}$, 
as functions of the length scale, $\ell$ (see eq.\ \ref{sii} in \S 3.3).  Following Paper I, we adopt the following connecting 
formulas, 
\begin{equation}
S_{\rm ll}  (\ell)= 2 u'^2 \left[ 1- \exp \left( - \frac{(\ell/ \eta)}{(15C_{\rm K})^{3/4}} \right) \right]^{4/3}  
\left [\frac{(\ell/ \eta)^4}{(\ell/ \eta)^4 + (2 u'^2 /C_{\rm K} u_\eta^2)^{6} } \right]^{1/6},  
\label{sll}
\end{equation}
%and,
\begin{equation}
S_{\rm nn} (\ell)= 2  u'^2 \left[ 1- \exp \left( - \frac{(\ell/ \eta)^{4/3}}{(15C_{\rm Kn}/2)} \right) \right] 
\left [\frac{(\ell/ \eta)^4}{(\ell/ \eta)^4 + (2u'^2 /C_{\rm Kn} u_\eta^2 )^{6} } \right]^{1/6}.   
\label{snn}
\end{equation}
%where the Kolmogorov velocity $u_\eta$ is defined as $u_\eta \equiv (\nu \bar{\epsilon})^{1/4}$.
With $C_{\rm K } =2$ and $C_{\rm Kn} =2.5$, %correctly reproduce the velocity scalings in different scale ranges and 
the two formulas fit well the measured structure functions  in our simulated flow.
%A comparison of these formulas to the measured structure functions shows that the best-fit parameters 
%We found that $C_{\rm K } =2$ and $C_{\rm kn} =2.5$ in our simulated flow. 
The correlation timescale, $T$, in eqs.\ (\ref{exponential2}) and (\ref{biexponential2}) for $\Phi_2$ 
is also obtained by a connecting formula (Zaichik et al.\ 2006),    
\begin{equation}
T (\ell)  = T_{\rm L} \left[ 1- \exp \left( -   \left(\frac{C_{\rm T}}{\sqrt{5}}\right)^{3/2} 
(\ell/ \eta) \right) \right]^{-2/3} \left [\frac{(\ell/ \eta)^4}{(\ell/ \eta)^4 +  (T_{\rm L}/( C_{\rm T} \tau_\eta))^{6} } \right]^{1/6}, 
\label{Tr}
\end{equation}
with the parameter  $C_{\rm T}$ set to $0.4$ (Paper I). %The connecting formulas complete our model for the 
%generalized shear term. 

\section{C: Lagrangian and Eulerian Temporal Structure Functions}

As discussed in \S 2 and \S 3, the particle-flow relative velocity 
and the generalized acceleration term in the bidisperse 
case depend on the temporal flow velocity difference, $\Delta_{\rm T}{\bs u}(\Delta \tau)$, 
along particle trajectories. As small particles more or less follow Lagrangian 
tracers, it is helpful to examine the flow velocity difference 
$\Delta { \bs u}_{\rm L} (\Delta \tau)\equiv  {\bs u} ({\bs X}_{\rm L}(t+\Delta \tau),  t+\Delta \tau)- 
{\bs u} ({\bs X}_{\rm L}(t),  t)$ as a function of the time lag, $\Delta \tau$, along Lagrangian trajectories ${\bs X}_{\rm L}(t)$. 
We also consider the Eulerian temporal velocity difference 
$\Delta { \bs u}_{\rm E} (\Delta \tau) \equiv  {\bs u} ({\bs x}, t+\Delta \tau) - {\bs u} ({\bs x}, t)$ 
at fixed points, ${\bs x}$, as the flow velocity seen by large 
particles with $\tau_{\rm p} \gsim T_{\rm L}$ may be better described as Eulerian (Paper I).  %Here we show the Lagrangian and 
%Eulerian structure functions in our simulated flow. 
%The PDFs of $\Delta_{\rm L}{\bs u} $ and $\Delta_{\rm E}{\bs u}$ as a function of $\Delta \tau$
%are given in Appendix B. 
%The behavior of the shape of  these PDFs as a function of $\Delta \tau$ provides interesting 
%clues to understand the PDFs of the particle-flow relative velocity (\S5) and 
%of the particle relative velocity in the bidisperse case (\S 6). 

%The Eulerian temporal velocity difference $\Delta_{\rm E} {\bs u}$ is relevant for the study of large particles with 
%$\tau_{\rm p} \gg T_{\rm L}$, which move 
%much slower than the flow velocity.  

\begin{figure}[t]
\centerline{\includegraphics[width=0.55\columnwidth]{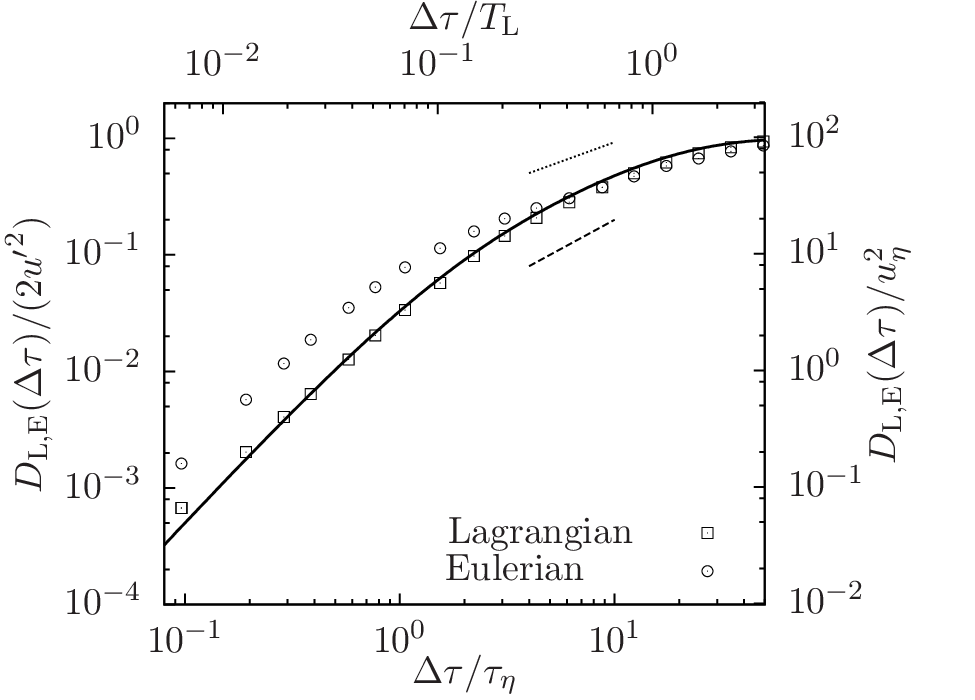}}
\caption{Lagrangian (squares) and Eulerian (circles) temporal structure 
functions. %The time lag is normalized to $\tau_\eta$ and $T_{\rm L}$ on 
%the bottom and top X-axises, while 
The left and right Y-axises normalize these functions to $2u'^2$ and $u_{\eta}^2$, respectively. The solid 
line corresponds to $1-\Phi_{\rm L} (\Delta \tau)$ using a bi-exponential 
$\Phi_{\rm L}$ (eq.\ (\ref{biexponential})) with $z=0.3$ and $T_{\rm L} = 15 \tau_\eta$. 
The dashed and dotted line segments denote a linear and 
a $(\Delta \tau)^{2/3}$ scaling, respectively, for $\Delta \tau$ in the inertial range.}
\label{temporalstructure} 
\end{figure}

We define Lagrangian/Eulerian structure tensors 
as $D_{{\rm L, E} ij} (\Delta \tau)\equiv \langle \Delta { \bs u}_{{\rm L, E}i}(\Delta \tau) \Delta {\bs u}_{{\rm L, E}j}(\Delta \tau) \rangle$, 
and by isotropy $D_{{\rm L, E} ij} (\Delta \tau)= D_{\rm L, E} (\Delta \tau) \delta_{ij}$. 
By definition, $D_{\rm L, E} = 2 u'^2 (1- \Phi_{\rm L, E})$, 
where $\Phi_{\rm L, E}$ are the temporal correlation functions of Lagrangian and Eulerian velocities 
(see Paper I).  Fig.\ \ref{temporalstructure} 
shows $D_{\rm L}$ (squares) and $D_{\rm E}$ (circles) as a function of 
$\Delta \tau$. The figure is equivalent to Figure 2 in Paper I for $\Phi_{\rm L, E}$. The solid line %, corresponding to the 
%solid line in Figure 2 of Paper I,  
plots $(1-\Phi_{\rm L})$ using a bi-exponential $\Phi_{\rm L}$ (eq.\ (\ref{biexponential})) with $z=0.3$ and $T_{\rm L} =15 \tau_\eta$, 
consistent with the values obtained in Figure 2 of Paper I. 
Except at the smallest $\Delta \tau$, the line provides a good fit to the data.
%measured $D_{\rm L}$. 
%and we thus adopt %eq.\ (\ref{biexponential}) with $z=0.3$ and $T_{\rm L} =15 \tau_\eta$ 
%this best fit in the calculation of  our model predictions. 
The measured values of $z$ and $T_{\rm L}$ indicate a Taylor micro timescale of  $4.3 \tau_\eta$ and a 1-D 
rms acceleration, $a$, of  $2.2 \tau_\eta^{-1}$. 
%A similar estimate was given in Paper I from the fit to 
%the Lagrangian correlation function.  

The dashed and dotted line segments denote a linear and a $(\Delta \tau)^{2/3}$ scaling. 
In the inertial range,  $D_{\rm L}(\Delta \tau)$ is expected to be $\simeq \bar{\epsilon} \Delta \tau$ 
from the Lagrangian version of Kolmogorov's similarity theory (Monin \& Yaglom 1975), 
while the $(\Delta \tau)^{2/3}$ scaling for $D_{\rm E}(\Delta \tau)$ follows from the random 
Taylor hypothesis that connects the Eulerian temporal and spatial structure functions (Tennekes 1975).
%Due to the low numerical resolution, 
The inertial range of our flow is short especially for the Lagrangian structure function. A linear scaling is 
barely seen in $D_{\rm L}$. At $\Delta \tau$ below the Taylor micro timescale $\tau_{\rm T}\simeq 4.3 \tau_\eta$, 
$D_{\rm L}$ is expected to be $a^2 (\Delta \tau)^2$, but we see 
the slope of the square data points at the smallest $\Delta \tau$ ($\simeq0.1 \tau_\eta$) 
in the figure is slightly shallower than $(\Delta \tau)^2$. 
%As a consequence, the measured $D_{\rm L}$ does not allows an accurate evaluation 
%of the rms flow acceleration, $a$. The value of $a$ given above 
%should be taken as an approximate estimate. 
An exact $(\Delta \tau)^2$ scaling may appear when one extends the measurement to $\Delta \tau$ 
below $0.1\tau_\eta$.  
%Or it is also possible that the trajectory integration of tracer particles in our simulation is not sufficiently accurate at a timescale as 
%small as $0.1\tau_\eta$.  
We note that $D_{\rm L}$ merges with $D_{\rm E}$ at $\Delta \tau \simeq 8 \tau_\eta$, 
suggesting that we can use $D_{\rm L}$ in our model for particles with 
$\tau_{\rm p} \gsim T_{\rm L} \simeq 15 \tau_\eta$, even though 
trajectories of these large particles would significantly deviate from  tracer 
particles. In other words, this provides a justification for approximating $\Delta_{\rm T} {\bs u}$ 
%along the particle trajectories 
by $\Delta_{\rm L} {\bs u}$ for all particles.
% but also for particles with $St \gsim 8$. 

\section{D: Approaching and separating particle pairs}

\begin{figure*}[t]
\includegraphics[height=2.7in]{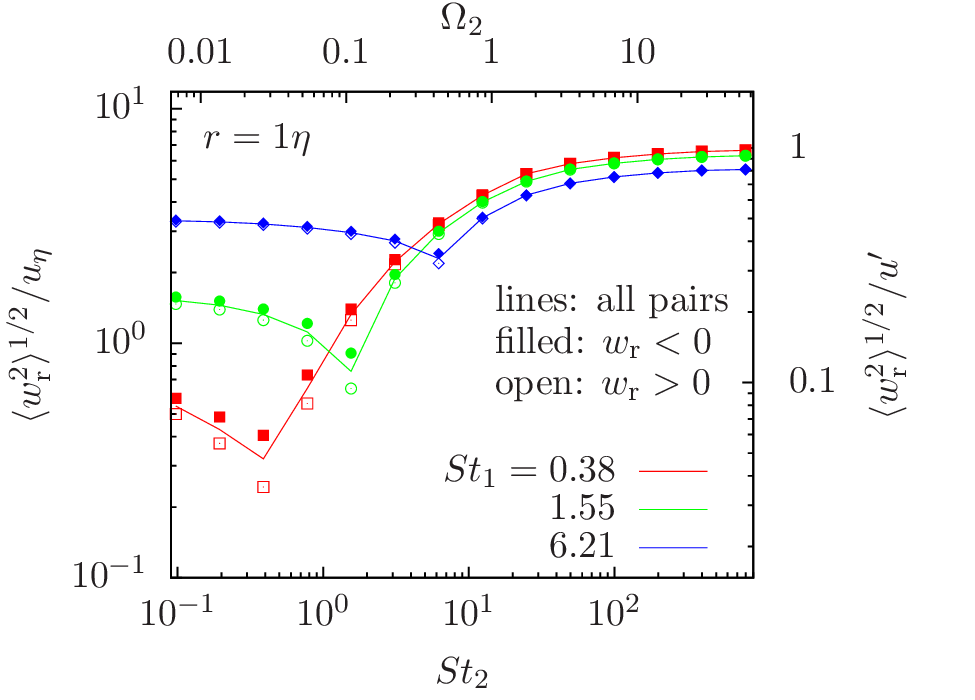}
\includegraphics[height=2.7in]{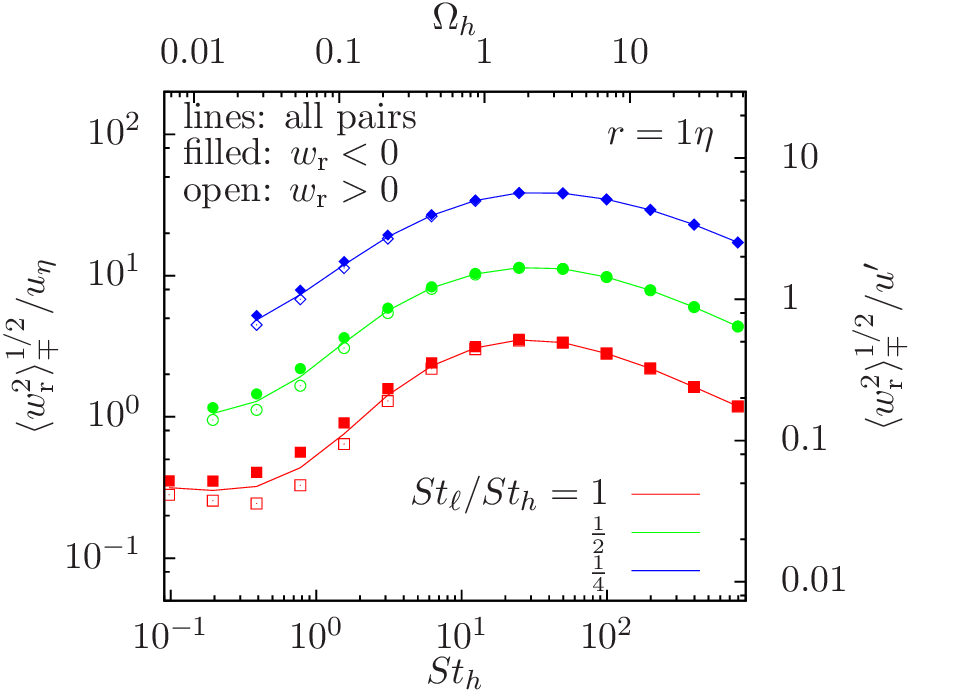}
\caption{Radial rms relative speeds for approaching (filled symbols) and 
separating (open symbols) pairs at $r=1 \eta$. 
Solid lines are the overall radial rms accounting for all pairs in 
both groups. Left  and right panels show results at fixed $St_1$ and fixed Stokes ratios. respectively.
For clarity, the data for $f=\frac{1}{2}$ and $\frac{1}{4}$ in the right panel 
are shifted upward by a factor of 3 and 9, respectively. }
%{\bf  The normalizations of the particle friction time can be converted 
%using $\Omega = St/14.4$ and $\Omega_{\rm eddy} =St/19.2$}.}
\label{radialpm} 
\end{figure*}

As in Paper I, we split particle pairs at given distances into two 
groups with negative ($w_{\rm r}<0$) and positive ($w_{\rm r}>0$) radial relative 
speed, corresponding to particle pairs approaching and separating from each other, respectively. 
We named them as minus and plus groups. 
%Only the pairs in the minus group each other and may 
%lead to collisions. 
Although only the minus group is relevant for collisions, it is of 
theoretical interest to compare the two groups. 
Fig.\ \ref{radialpm} plots the radial rms relative speeds for particle pairs in the minus ($\langle w_{\rm r}^2 \rangle_-^{1/2}$; filled symbols) and plus 
($\langle w_{\rm r}^2 \rangle_+^{1/2}$; open symbols) groups at a distance of $r=1\eta$. 
%The left and right panels show results 
%at fixed $St_1$ and fixed Stokes ratios, respectively.  %at $f=1$, $\frac{1}{2}$ and $\frac{1}{4}$ 

In Paper I, we showed that, for identical particles with $St \lsim 6.2$, the 
rms relative speed in the minus group is larger than in the plus group. 
For $St\ll1$ particles, one expects the relative speed to inherit such an 
asymmetry from the flow (%The PDF of the longitudinal flow velocity difference 
%has an intrinsic skewness toward negative values 
see Appendix B of Paper I). As $St$ increases, the asymmetry is first 
amplified and then decreases at $St \gsim 0.4$ (see the $f=1$ data in 
the right panel). The amplification is due to the fact that approaching pairs 
come from a larger distance in the near past than separating ones. 
This tends to make the relative speed in the minus group larger 
because the relative velocity of identical particles is determined by 
their memory of the flow velocity difference in the past, which 
is larger at larger particle separation. For larger particles with $St  \gsim 0.4$, 
separating pairs would move past each other within a friction time in the past, 
and their separation then starts to increase backward in time. 
Consequently, the difference in the particle distance at a friction time ago for 
minus and plus pairs decreases, causing the asymmetry to decrease. 
It finally disappears for $St \gsim 6.2$, as the amplitude of the 
particle separation at a friction time ago becomes insensitive to $r$ or the condition in the near past (Paper I).
%The asymmetry behavior for identical particles can be seen from the dip centers in 
%Fig.\ \ref{radialpm}. 

The left panel of Fig.\ \ref{radialpm} shows that, for each $St_1$, 
the difference between the two groups is largest for identical particles, and decreases as $St_2$ 
moves away from the dip center. This is because the contribution of the generalized acceleration term 
does not depend on whether the particles are approaching or separating. 
The acceleration term is determined by the flow velocity statistics along the individual 
trajectories of the two particles (see \S 3.2.1 and Fig.\ \ref{cartoonaccel}), and thus independent 
of their relative motions. 
%and does not cause an asymmetry in. 
Therefore, the asymmetry  of the plus and minus groups decreases when the acceleration 
contribution increases. This is also seen in the right panel where the asymmetry 
is weaker at smaller $f$. 
If one Stokes number is larger than $6.21$, the asymmetry disappears for any 
bidisperse case.
As mentioned in Paper I, the asymmetry in $\langle w_{\rm r}^2 \rangle_-^{1/2}$ and $\langle w_{\rm r}^2 \rangle_+^{1/2}$ is 
related to the spatial clustering of the particles. A smaller difference in 
$\langle w_{\rm r}^2 \rangle_-^{1/2}$ and $\langle w_{\rm r}^2 \rangle_+^{1/2}$ 
for particles of different sizes suggests a weaker clustering in the bidisperse case (Pan et al.\ 2011). 

The asymmetry tends to decrease with decreasing $r$, and 
the decrease is  faster for the bidisperse case than the monodisperse 
case. %Again this is because the contribution from the acceleration effect becomes relatively larger 
%if the shear term decreases with decreasing $r$. 
The asymmetry would disappear at sufficiently small $r$ once the acceleration term dominates.  
%ratio is larger than 2.  
%On the other hand, for identical particles, the difference in $\langle w_{\rm r}^2 \rangle_-^{1/2}$ 
%and $\langle w_{\rm r}^2 \rangle_+^{1/2}$ for $St \lsim 1$ decreases 
%slightly as $r$ decreases from $\frac{1}{2}$ to $\frac{1}{4} \eta$. 
%%In the monodisperse case, 
%but the ratio, $\langle w_{\rm r}^2 \rangle_-^{1/2}/\langle w_{\rm r}^2 \rangle_+^{1/2}$ is largely unchanged for $St \lsim 0.49$ as $r$ decreases to $\frac{1}{4}\eta$. 
%it remains to be checked whether the asymmetry for small identical particles with $St \lsim 1$
%completely vanishes as $r\to 0$ (see details in Paper I). 
We also examined the tangential ($\langle w_{\rm t}^2 \rangle_{\mp}^{1/2}$) 
and the 3D ($\langle w^2 \rangle_{\mp}^{1/2}$) rms relative speeds in the minus 
and plus groups, and found a similar asymmetry as in the case of the radial component.  
%We point 
%out that $\langle w_{\rm t}^2 \rangle_{\mp}^{1/2}$ and $\langle w^2 \rangle_{\mp}^{1/2}$
%correspond to the widths of 
%PDFs, $P(w_{\rm t}|w_{\rm r}\lessgtr 0; St_1, St_2)$ and $P(|{\bs w}||w_{\rm r}\lessgtr 0; St_1, St_2)$, 
%conditioned on the sign of $w_{\rm r}$. These conditional PDFs  will be discussed in \S 7.
Paper I showed that the radial and tangential rms relative speeds of approaching 
pairs are about equal (i.e., $\langle w_{\rm r}^2 \rangle_-^{1/2} \simeq \langle w_{\rm t}^2 \rangle_-^{1/2}$) 
for all particles in our simulation. The same is true for the bidisperse case, as the acceleration term 
has the effect of equalizing the radial and tangential components. 

\end{document}